\documentclass[11pt]{article}
\usepackage{amsfonts,amsmath,amssymb}
\usepackage{enumerate}
\usepackage{hyperref}
\usepackage{bbm}
\usepackage{nicefrac}
\usepackage[all]{xy}
\usepackage{graphicx}
\usepackage{bm}

\usepackage{slashed}

\usepackage{upgreek}

\usepackage{booktabs}

\newcommand{\beq}{\begin{equation}}
\newcommand{\eeq}{\end{equation}}

\newcommand{\be}{\begin{equation}}
\newcommand{\ee}{\end{equation}}

\newcommand{\bea}{\begin{eqnarray}}
\newcommand{\eea}{\end{eqnarray}}

\newcommand{\bes}{\begin{subequations}}
\newcommand{\ees}{\end{subequations}}

\newcommand{\cN}{{\cal N}}

\def\sst#1{{\scriptscriptstyle #1}}

\def\0{{\sst{(0)}}}
\def\1{{\sst{(1)}}}
\def\2{{\sst{(2)}}}
\def\3{{\sst{(3)}}}
\def\4{{\sst{(4)}}}
\def\5{{\sst{(5)}}}
\def\6{{\sst{(6)}}}
\def\7{{\sst{(7)}}}
\def\8{{\sst{(8)}}}

\def\tA{\widetilde A}

\usepackage{multirow}
\usepackage{rotating}

\allowdisplaybreaks

\newcommand{\tmu}{\tilde{\mu}}

\begin{document}

\makeatletter
\renewcommand{\theequation}{\thesection.\arabic{equation}}
\@addtoreset{equation}{section}
\makeatother

\begin{titlepage}

\begin{flushright}
IFT-UAM/CSIC-18-048
\end{flushright}

\vspace{5pt}

   \begin{center}
   \baselineskip=16pt

   \begin{Large}\textbf{
The geometry of ${\cal N} =3$ AdS$_4$ in massive IIA  
}

   \end{Large}

   		
\vspace{25pt}
		
{\large  G.~Bruno De Luca,$^{1}$ Gabriele Lo Monaco,$^{1}$ Niall T. Macpherson,$^{2}$  \\[5pt]
Alessandro Tomasiello$^{1}$ and  Oscar Varela$^{3,4,5}$}
		
\vspace{30pt}

	\begin{small}

	  {\it $^{1}$ Dipartimento di Fisica, Universit\`a di Milano-Bicocca, 20126 Milan, Italy \\
	  INFN, sezione di Milano-Bicocca, 20126 Milan, Italy}

	\vspace{15pt}
	
	  {\it $^{2}$ SISSA International School for Advanced Studies \\
	  and INFN, sezione di Trieste, 34136 Trieste, Italy}

	\vspace{15pt}
	
	{\it $^{3}$  Department of Physics, Utah State University, Logan, UT 84322, USA}

	\vspace{15pt}
          
   {\it $^{4}$ Departmento de F\'\i sica Te\'orica and Instituto de F\'\i sica Te\'orica UAM/CSIC , \\
   Universidad Aut\'onoma de Madrid, Cantoblanco, 28049 Madrid, Spain}

	\vspace{15pt}
          
   {\it $^{5}$ Max-Planck-Institut f\"ur Gravitationsphysik (Albert-Einstein-Institut), \\Am M\"uhlenberg 1, D-14476 Potsdam, Germany}

	\end{small}

\vskip 50pt

\end{center}

\begin{center}
\textbf{Abstract}
\end{center}

\begin{quote}

The geometry of the ${\cal N} = 3$, SO(4)--invariant, AdS$_4$ solution of massive type IIA supergravity that uplifts from the ${\cal N} = 3 $ vacuum of $D=4$ ${\cal N} = 8$ dyonic ISO(7) supergravity is investigated. Firstly, a $D=4$, SO(4)--invariant restricted duality hierarchy is constructed and used to uplift the entire, dynamical SO(4)--invariant sector to massive type IIA. The resulting consistent uplift formulae are used to obtain a new local expression for the ${\cal N} = 3 $ AdS$_4$ solution in massive IIA and analyse its geometry. Locally, the internal $S^6$ geometry corresponds to a warped fibration of $S^2$ and a hemisphere of $S^4$. This can be regarded as a warped generalisation of the usual twistor fibration geometry. Finally, the triplet of Killing spinors corresponding to the $\cN=3$ solution are constructed and shown to obey the massive type IIA Killing spinor equations.

\end{quote}

\vfill

\end{titlepage}

\tableofcontents


\section{Introduction}

Massive type IIA supergravity \cite{Romans:1985tz} admits a consistent truncation on the six-sphere to maximal supergravity in four dimensions with gauge group $\textrm{ISO}(7) = \textrm{SO}(7) \ltimes \mathbb{R}^7$ \cite{Guarino:2015jca,Guarino:2015vca}. The gauging is dyonic, in the sense of \cite{Dall'Agata:2012bb,Dall'Agata:2014ita} (see also \cite{Inverso:2015viq}). By virtue of the consistency of the truncation, all solutions of the four-dimensional theory uplift on $S^6$ to solutions of massive type IIA supergravity. In particular, the critical points (which can only be AdS) of the four-dimensional scalar potential give rise to supersymmetric and non-supersymmetric ten-dimensional solutions of the form $\textrm{AdS}_4 \times S^6$. This product is generically warped and the metric on $S^6$ displays an isometry group $G \subset \textrm{SO}(7)$ related to the residual symmetry within ISO(7) supergravity of the critical point it uplifts from. Using this technique, new massive type IIA solutions have been found \cite{Guarino:2015jca,Varela:2015uca,Pang:2015vna} and previously known ones \cite{Romans:1985tz,Behrndt:2004km,Lust:2008zd} have been recovered. Other supersymmetric AdS$_4$ solutions of massive type IIA supergravity have been recently found using other methods in \cite{Apruzzi:2015wna,Rota:2015aoa}. Previous constructions of supersymmetric AdS$_4$ solutions in massive type IIA supergravity include \cite{Lust:2004ig,Grana:2006kf,Tomasiello:2007eq,Koerber:2008rx,Petrini:2009ur,Lust:2009mb}.

In this paper, we investigate the ten-dimensional uplift of the $\cN=3$ SO(4)--invariant critical point of dyonic ISO(7) supergravity. This $D=4$ critical point was found in \cite{Gallerati:2014xra}. A local form of its massive type IIA uplift has already appeared in \cite{Pang:2015vna}. Here, we provide an alternate local form of this $\cN=3$ AdS$_4$ solution of massive IIA supergravity (equation (\ref{SO4SolN=3})) and discuss its geometric features. The internal space of the $\cN=3$ solution is topologically $S^6$, endowed with a geometry that can be locally regarded as an $S^2$ bundle over a half-$S^4$. This is a generalisation of the twistor bundle over a quaternionic-K\"ahler manifold of positive curvature, see {\it e.g.}~\cite{Cvetic:2002kj, Tomasiello:2007eq} for reviews. The twistor fibration allows one to engineer nearly-K\"ahler or half-flat geometries on six-manifolds $M_6$ of topology different than $S^6$, see {\it e.g.}~\cite{Tomasiello:2007eq}. In turn, a well known class of $\cN=1$ (direct) product solutions AdS$_4 \times M_6$ of massive IIA supergravity entails a nearly-K\"ahler \cite{Behrndt:2004km,Behrndt:2004mj} or a half-flat structure \cite{Lust:2004ig,Tomasiello:2007eq,Koerber:2008rx} on $M_6$. 

It is suggestive that this $\cN=3$ solution formally corresponds to an elaboration of these $\cN=1$ constructions. This is reminiscent of the situation for a well-known class of $D=11$ direct product solutions  involving  AdS$_4$ and a tri-Sasaki seven-manifold. Recall that the latter can be regarded as an $S^3$ bundle over a quaternionic-K\"ahler base, equipped with an Einstein metric on the total space. This class of solutions is $\cN=3$, see {\it e.g.}~\cite{Acharya:1998db}. On each tri-Sasaki manifold, a second Einstein metric can be obtained by squashing the $S^3$ fibers by a certain constant amount. The resulting $D=11$ AdS$_4$ solution is $\cN=1$, see \cite{Awada:1982pk,Acharya:1998db}. The analogy with these $\cN=3$ and $\cN=1$ solutions of $D=11$ supergravity should not be taken too far, though. The internal metric of the massive IIA $\cN=3$ solution is certainly not Einstein, unlike the $\cN=1$ nearly-K\"ahler solutions of \cite{Behrndt:2004km}. In the IIA $\cN=3$ solution, the $S^2$ fibers are squashed, not by a constant, but by a warping function of the $S^4$ hemisphere base.  The connection does not have definite duality properties, unlike in the usual twistor fibration. Finally, the $\cN=3$ solution involves a warped, rather than direct, product of AdS$_4$ and the internal topological $S^6$. Like in the $D=11$ tri-Sasaki case, though, the SO(3) R-symmetry acts on the fibers of the $\cN=3$ massive IIA solution.

The type IIA $\cN=3$ solution displays a local SO(4) symmetry, inherited from that preserved by the $\cN=3$ critical point of the $D=4$ supergravity. More generally, we construct in section \ref{sec:SO(4)-sector} the restricted, in the sense of \cite{Guarino:2015qaa}, duality hierarchy \cite{deWit:2008ta,Bergshoeff:2009ph} of $D=4$ ISO(7) supergravity that is invariant under this SO(4). This result is particularly useful, as it allows us to consistently embed the entire, dynamical SO(4)--invariant sector of the $D=4$ $\cN=8$ supergravity into massive type IIA. The explicit consistent uplift formulae are presented in section \ref{sec:SO4sectorinIIA}. These formulae give the ten-dimensional uplift of any SO(4)--invariant solution of the ISO(7) supergravity, including solutions with running scalars. The local and global features of this consistent embedding formulae are discussed at length, and generalisations are given. Section \ref{sec:urthertruncs} discusses further truncations. The truncation to the dynamical G$_2$--invariant sector of \cite{Guarino:2015vca} is recovered, and an example that illustrates the usefulness of the duality hierarchy approach is worked out. In section \ref{sec:AdSsolutions}, we turn our attention to the massive IIA uplift of solutions of the $D=4$ supergravity, focusing on vacuum solutions. In particular, a new local form of the $\cN=3$ AdS$_4$ solution in massive IIA is provided. Finally, in section \ref{sec:N=3susy}, the solution is demonstrated to indeed be $\cN=3$ by explicitly building the triplet of Killing spinors that it preserves.


\section{A $D=4$, SO(4)--invariant duality hierarchy} \label{sec:SO(4)-sector}


We are interested in the sector of $D=4$ $\cN=8$ dyonically-gauged ISO(7) supergravity \cite{Guarino:2015qaa} that retains the fields that are invariant under the SO(4) subgroup of ISO(7) defined by the embedding \cite{Gallerati:2014xra}
\begin{equation}
\label{embedding_SO4}
\textrm{SO}(7) \, \supset \, 
%
 %
  \textrm{SO}(3)^\prime \times \textrm{SO}(4)^\prime \,  \supset \, 
   \textrm{SO}(3)_{\textrm{d}} \times \textrm{SO}(3)_{\textrm{R}}  \, \equiv \, 
   \textrm{SO}(4) \ ,
\end{equation}
with $\textrm{SO}(4)^\prime \equiv \textrm{SO}(3)_{\textrm{L}} \times \textrm{SO}(3)_{\textrm{R}}$ and $ \textrm{SO}(3)_{\textrm{d}}$ the diagonal subgroup of $\textrm{SO}(3)^\prime \times \textrm{SO}(3)_{\textrm{L}}$. Equivalently, this SO(4) is also the maximal subgroup of the G$_2$ contained in SO(7),
\begin{equation}
\label{embedding_SO4_2}
\textrm{SO}(7) \, \supset \, 
 \textrm{G}_{2} \, \supset \, 
   \textrm{SO}(4) \ .
\end{equation}
The Lagrangian corresponding to this sector of the $\cN=8$ ISO(7) supergravity was given in \cite{Guarino:2015qaa}, and the vacuum structure was studied in detail there. Here, we complete the analysis of the SO(4)--invariant sector by determining the restricted, in the sense of \cite{Guarino:2015qaa}, duality hierarchy \cite{deWit:2008ta,Bergshoeff:2009ph} in this sector. 

The SO(4)--invariant sector of $\cN=8$ ISO(7) supergravity corresponds to an $\cN=1$ supergravity coupled to two chiral multiplets that parametrise a K\"ahler submanifold 
\begin{eqnarray} 
\label{ScalManN=1}
\frac{\textrm{SU}(1,1)}{\textrm{U}(1)} \times  \frac{\textrm{SU}(1,1)}{\textrm{U}(1)} \,   
\end{eqnarray}
of E$_{7(7)}/$SU(8). The $\textrm{SU}(1,1)^2$ in the numerator is the commutant of the SO(4) in (\ref{embedding_SO4}) or (\ref{embedding_SO4_2}) inside E$_{7(7)}$. According to table 2 of \cite{Guarino:2015qaa}, the SO(4)-singlets of the restricted, SL(7)--covariant tensor hierarchy considered therein give rise to one two-form and two three-form potentials in this sector. To summarise and fix the notation, the SO(4)--invariant, restricted duality hierarchy of $D=4$ $\cN=8$ supergravity contains the following real fields:
\begin{eqnarray} \label{fieldContentHierarchy}
\textrm{1 metric} & : & \quad ds_4^2 \; ,  \nonumber \\
\textrm{4 scalars} & : & \quad \varphi \; , \;  \chi \; , \; \phi \; , \;  \rho \; ,  \nonumber \\
\textrm{1 two-form} & : & \quad B \; ,
\nonumber \\
\textrm{2 three-forms} & : & \quad C^1 \; , \;  C^2 \; .
\end{eqnarray}
The embedding of the scalars into the $\cN=8$ E$_{7(7)}/$SU(8) manifold was discussed at length in \cite{Guarino:2015qaa}. In turn, the two- and three-form potentials in (\ref{fieldContentHierarchy}) are embedded into the SL(7)--covariant two-, ${\cal B}_I{}^J$, and three-forms, ${\cal C}^{IJ}$, defined in \cite{Guarino:2015qaa} via
\begin{eqnarray} \label{embeddingTH}
{\cal B}_i{}^j = \tfrac47 \, B \, \delta^j_i \; , \qquad 
{\cal B}_{\hat i}{}^{\hat j} = -\tfrac37 \, B \, \delta^{\hat j}_{\hat i} \; , \qquad 
{\cal C}^{ij}  = C^1 \, \delta^{ij} \; , \qquad 
{\cal C}^{\hat{i}\hat{j}}  = C^2 \, \delta^{\hat{i}\hat{j}} \; ,
\end{eqnarray}
and ${\cal B}_i{}^{\hat{j}} = {\cal B}_{\hat{j}}{}^i = {\cal C}^{i {\hat{j}}} =0$. Here, we have split the SL(7) indices $I, J = 1 , \ldots ,7$ as $I = (i , {\hat{i}})$, $i = 1,2,3$, ${\hat{i}}=0,1,2,3$, as in appendix \ref{subset:S6Geom}. 

In a conventional formulation, only the metric and the scalars in (\ref{fieldContentHierarchy}) enter the $D=4$ Lagrangian. This reads \cite{Guarino:2015qaa}
\begin{equation}
\label{L_SO4}
\mathcal{L}  = (R - V) \, \textrm{vol}_4  + \tfrac{6}{2} \left[ d\varphi \wedge * d\varphi + e^{2 \varphi} \, d\chi \wedge * d\chi \right] + \tfrac{1}{2} \left[ d\phi \wedge * d\phi + e^{2 \phi} \, d\rho \wedge * d\rho \right]  \ ,
\end{equation}
\begin{center}
\begin{table}[t]
\renewcommand{\arraystretch}{1.5}
\begin{center}
\begin{tabular}{cc|cccc|cc}
\noalign{\hrule height 1pt}
$\mathcal{N}$ & $G$ & $c^{-1/3} \, \chi$ & $c^{-1/3} \, e^{-\varphi}$ & $c^{-1/3} \, \rho$ & $c^{-1/3} \,  e^{-\phi}$ & $g^{-2} \, c^{1/3} \, V$ & ref. \\
\noalign{\hrule height 1pt}
$\mathcal{N}=3$ & $\textrm{SO}(4)$ & $\frac{1}{2^{4/3}}$ & $\frac{3^{1/2}}{2^{4/3}} $ & $-\frac{1}{2^{1/3}}$ & $\frac{3^{1/2}}{2^{1/3}}$ & $-\frac{2^{16/3}}{3^{1/2}} $ & \cite{Gallerati:2014xra} \\[5pt]
$\mathcal{N}=1$ & $\textrm{G}_{2}$ & $-\frac{1}{2^{7/3}} $ & $\frac{5^{1/2} \, 3^{1/2}}{2^{7/3}}$ & $-\frac{1}{2^{7/3}} $ & $\frac{5^{1/2} \, 3^{1/2}}{2^{7/3}}$ & $- \frac{2^{28/3} \, 3^{1/2}}{5^{5/2}} $ & \cite{Borghese:2012qm} \\[5pt]
\hline
$\mathcal{N}=0$ & $\textrm{SO}(7)_+$ & $0$ & $\frac{1}{5^{1/6}}$ & $0$ & $\frac{1}{5^{1/6}}$ & $-3 \, 5^{7/6}$ & \cite{DallAgata:2011aa} \\[5pt]
$\mathcal{N}=0$ & $\textrm{G}_{2}$ & $\frac{1}{2^{4/3}}$ & $\frac{3^{1/2}}{2^{4/3}}$ & $\frac{1}{2^{4/3}}$ & $\frac{3^{1/2}}{2^{4/3}}$ & $-\frac{2^{16/3}}{3^{1/2}}$ & \cite{Borghese:2012qm} \\[5pt]
$\mathcal{N}=0$ & $\textrm{SO}(4)$ & $0.412$ & $0.651$ & $0.068$ & $1.147 $ & $-23.513$ & \cite{Guarino:2015qaa} \\
\noalign{\hrule height 1pt}
\end{tabular}
\end{center}
\caption{\small{Critical points of the scalar potential (\ref{VSO4}), namely, of $\cN=8$ ISO(7)-dyonically-gauged supergravity with invariance equal or larger than the SO(4) subgroup of SO(7) defined in (\ref{embedding_SO4}). For each point we give the residual supersymmetry $\cN$ and bosonic symmetry $G$ within the full $\cN=8$ theory, its location, the cosmological constant $V$ and the reference where it was first found. We have employed the shorthand $c \equiv m/g$. All of these data are reproduced from \cite{Guarino:2015qaa}.}\normalsize}
\label{Table:SO4Points}
\end{table}
\end{center}
\noindent with the scalar potential given by \cite{Guarino:2015qaa}
\begin{equation}
\label{VSO4}
\begin{array}{lll}
V &=& \frac{1}{2} \, g^{2}  \, e^{-\phi } (1+e^{2 \varphi } \chi ^2)  
\left[ -24 \, e^{\varphi +\phi } - 8 \, e^{2 \phi }   +  e^{2 \varphi } \, \Big(-3+  (8 \chi ^2-3 \rho ^2) \, e^{2 \phi } \Big) \right. \\[4mm]
&+& \left.  e^{4 \varphi }  \, \chi ^2 \, \Big(  9 +  (3 \rho +4 \chi )^2  \, e^{2 \phi }   \Big)   \right]  
-     g m \, \chi ^2 \, (3 \rho + 4 \chi ) \,  e^{6 \varphi +\phi }
+  \frac{1}{2} \, m^2 \, e^{6 \varphi +\phi }  \ .
\end{array}
\end{equation}
The constants $g$ and $m$ are the electric and magnetic gauge couplings of the parent $\cN=8$ ISO(7) supergravity.

When $g m \neq 0$, the scalar potential (\ref{VSO4}) contains AdS critical points that spontaneously break the $\cN=8$ supersymmetry and ISO(7) gauge symmetry of the full $D=4$ supergravity to some supersymmetry $\cN$ and residual symmetry $G$. See table \ref{Table:SO4Points} for a summary. The $\cN=3$ SO(4)--invariant point manifests itself as non-supersymmetric within the subtruncation (\ref{L_SO4}), (\ref{VSO4}), see \cite{Guarino:2015qaa} for further details. All these critical points are inherent to the dyonic ISO(7) gauging and  disappear in the purely electric $g \neq 0, m=0$, or purely magnetic,  $g = 0, m \neq 0$ limits. Accordingly, these four-dimensional solutions naturally uplift to massive type IIA supergravity on $S^6$ and do not have direct counterparts in either massless IIA on $S^6$ or massive IIA on $T^6$. 

The three- and four-form field strengths of the SO(4)--invariant two-form, $B$, and three-form potentials, $C^1$, $C^2$, are
\begin{eqnarray} \label{eq:FieldStrengths}
H_\3 = d B -2g \, C^1 + 2g \, C^2 \; , 
\qquad H^1_\4 = d C^1 \; , 
\qquad H^2_\4 = d C^2 \; .
\end{eqnarray}
These expressions follow from the generic expressions given in (2.8), (2.9) of \cite{Guarino:2015qaa} evaluated on equation (\ref{embeddingTH}) above. These field strengths are subject to the Bianchi identities
\begin{eqnarray} \label{eq:D=4Bianchis}
dH_\3 = -2g \, H^1_\4 + 2g \, H^2_\4 \; , 
\qquad d H^1_\4 \equiv 0 \; , 
\qquad d H^2_\4 \equiv 0 \; .
\end{eqnarray}
These in turn correspond to the SO(4)--invariant truncation of the generic, $\cN=8$ SL(7)--covariant Bianchi identities given in (2.13) of \cite{Guarino:2015qaa}. 

Not all of the fields in the SO(4)--invariant, restricted tensor hierarchy (\ref{fieldContentHierarchy}) carry independent degrees of freedom: the field strengths of the form potentials are subject to duality relations, see  \cite{Bergshoeff:2009ph,Guarino:2015qaa} for a generic discussion. Particularising the SL(7)-covariant duality equations (2.17), (2.18) of \cite{Guarino:2015qaa}  to the present case, we find the following duality relations obeyed by the SO(4)--invariant field strengths:
{\setlength\arraycolsep{2pt}
\begin{eqnarray} \label{eq:FieldStrengthDuality}
H_\3 &=&  * \Big( d\phi - e^{2\phi} \rho \, d\rho - d\varphi  + e^{2\varphi} \chi \, d\chi \Big) \; ,   \nonumber \\[10pt]
H^1_\4 &=& \Big[ g \,  e^{\varphi}  \big(1+e^{2\varphi} \chi^2 \big)  \Big(  4 - 4 \, e^{\phi+3 \varphi } \rho \chi^3  + e^{\varphi - \phi } \big(1- 3 e^{2\varphi} \chi^2 \big) \big(1+  e^{2\phi} \rho^2 \big)   \Big)  \nonumber \\[4pt]
&& \quad + m\, e^{\phi + 6\varphi } \rho \chi^2 \Big] \, \textrm{vol}_4 \; ,  \nonumber \\[10pt]
H^2_\4 &=&  \Big[ g \, \big(1+e^{2\varphi} \chi^2 \big)  \Big( 3 e^{\varphi}  - 3 \, e^{\phi+4 \varphi } \rho \chi^3  +6 e^{\phi} \big(1+e^{2\varphi} \chi^2 \big)   -4 e^{\phi} \big(1+e^{2\varphi} \chi^2 \big)^2    \Big) \nonumber \\[4pt]
&& \quad + m\, e^{\phi + 6\varphi } \chi^3 \Big]  \, \textrm{vol}_4   \; . 
\end{eqnarray}
}The Bianchi identities (\ref{eq:D=4Bianchis}), combined with the duality relations (\ref{eq:FieldStrengthDuality}), reproduce the scalar equations of motion that follow from the Lagrangian (\ref{L_SO4}), (\ref{VSO4}).

Even though it does not play a critical role in the IIA uplift, it is nevertheless useful to consider the SL(7)--singlet four-form field strength whose duality relation was given in (2.25) of  \cite{Guarino:2015qaa}. In the SO(4)--invariant case at hand, this duality relation reads
\begin{eqnarray} \label{eq:H4DualityAdditional}
\tilde{H}_\4 &=& e^{\phi+6\varphi}  \Big[ g \, \chi^2 \big( 3\rho +4 \chi  \big) - m \,  \Big] \, \textrm{vol}_4 \; .
\end{eqnarray}
Using (\ref{eq:FieldStrengthDuality}), (\ref{eq:H4DualityAdditional}), the scalar potential (\ref{VSO4}) can be checked to be related to the four-form field strengths $H^1_\4$, $H^2_\4$ and $\tilde{H}_\4$ through
\begin{equation} 
\label{DualitySU3H4V}
g \, ( 3 H^{1}_{\4} + 4 H^{2}_{\4} ) + m \, \tilde{H}_{\4} = -2 \, V \, \textrm{vol}_{4} \ .
\end{equation}
This is the SO(4)--invariant counterpart of the full $\cN=8$ expressions (2.28), (2.29) of \cite{Guarino:2015qaa}. At any of the critical points of the scalar potential (\ref{VSO4}), that were summarised in table \ref{Table:SO4Points} above, these four-form field strengths turn out to obey
\begin{equation}
\label{EOM_SU3}
g \, (3 H^{1}_{\4}|_0 + 4 \, H^{2}_{\4}|_0 ) +7 m \, \tilde{H}_{\4}|_0 =  0  \ , \qquad 
H^1_\4 |_0 = H^2_\4 |_0 \ , 
\end{equation}
where $|_0$ denote evaluation at a critical point. 

We conclude by recovering two interesting sectors of $D=4$ $\cN=8$ ISO(7) supergravity from the SO(4)--invariant sector. Firstly, according to the branching rule (\ref{embedding_SO4}), the $\textrm{SO}(3)^\prime \times \textrm{SO}(4)^\prime$--invariant sector is contained in the SO(4) sector. This is recovered by setting the pseudoscalars to zero, 
\begin{equation} \label{eq:SO4toSO3pSO4p}
\chi = \rho = 0 \; , 
\end{equation}
while retaining all other fields in the duality hierarchy (\ref{fieldContentHierarchy}). The $\textrm{SO}(3)^\prime \times \textrm{SO}(4)^\prime$--invariant Lagrangian, tensor field strengths, Bianchi identities and duality relations follow by letting $\chi = \rho = 0 $ in the expressions above. Secondly, as discussed in \cite{Guarino:2015qaa}, the G$_2$--invariant sector can be also recovered from the SO(4)--sector. This is apparent from the branching (\ref{embedding_SO4_2}). The G$_2$--invariant sector is recovered from the SO(4)--invariant sector through the identifications
\begin{equation} \label{eq:SO4toG2}
\varphi = \phi \; , \qquad 
\chi = \rho \; , \qquad  
B=0 \; , \qquad
C^1 = C^2 \equiv C \; ,
\end{equation}
along with $H_{\3} = 0 $ and $H^{1}_{\4} = H^{2}_{\4} \equiv H_{\4} $. These identifications bring the Lagrangian and duality relations to their G$_2$--invariant counterparts, given in section 4 of \cite{Guarino:2015qaa}.

\section{\mbox{Truncation from type IIA supergravity}} \label{sec:SO4sectorinIIA}

We are now ready to give the complete, non-linear embedding of the dynamical SO(4)--invariant sector of $D=4$ $\cN=8$ ISO(7) supergravity into massive type IIA. As discussed in \cite{Guarino:2015vca}, the embedding of the full $\cN=8$ theory is naturally expressed, at the level of the IIA metric, dilaton and form potentials, in terms of the restricted, SL(7)-duality hierarchy introduced in \cite{Guarino:2015qaa}. Accordingly, the complete IIA embedding of the SO(4)--invariant sector is naturally written in terms of the tensor hierarchy discussed in section \ref{sec:SO(4)-sector}.

\subsection{Consistent embedding formulae}
\label{subsec:SU3UpliftSubsec}


The SO(4)--invariant consistent embedding formulae can be obtained by particularising the $\cN=8$ formulae  given in (3.12), (3.13) of \cite{Guarino:2015vca} (see also \cite{Guarino:2015jca}) to the case at hand. It is a matter of simple algebra to find the embedding of the two- and three-form potentials of the $D=4$ tensor hierarchy (\ref{fieldContentHierarchy}) into their $D=10$ counterparts, using their $\cN=8$ embedding (\ref{embeddingTH}). In contrast, as is usually the case, the embedding of the $D=4$ scalars entails a lengthy computation. Here, we give the final result, referring to appendix \ref{subset:S6Geom} for further details on the relevant geometric structures that arise in the calculation. 

In order to express the result, it is convenient to introduce constrained coordinates $\tmu^i$, $i=1,2,3$, on the two-sphere $S^2$, 
\begin{eqnarray} \label{eq:S2}
\delta_{ij} \, \tmu^i \tmu^j = 1 \; , 
\end{eqnarray} 
and right-invariant one-forms\footnote{The right-invariant one-forms $\rho^i$ on $S^3$ shouldn't be confused with the $D=4$ pseudoscalar $\rho$.} $\rho^i$, $i=1,2,3$, on the three-sphere $S^3$. These are subject to the Maurer-Cartan equations
\begin{eqnarray} \label{eq:MC}
d\rho^i = - \tfrac12 \epsilon^i{}_{jk} \,  \rho^j \wedge \rho^k \; .
\end{eqnarray}
It is also convenient to introduce the following combinations of $D=4$ scalars \cite{Guarino:2015qaa}
\begin{eqnarray}
X = 1+ e^{2\varphi} \chi^2 \; , \qquad Y = 1 + e^{2\phi} \rho^2 \; , \qquad Z = e^{2\varphi} \chi \big( e^\phi \rho - e^\varphi \chi \big)  \; ,
\end{eqnarray}
and the following functions of $D=4$ scalars and an angle $\alpha$ on the internal $S^6$,
\begin{eqnarray} \label{deltas}
&& \Delta_1 = e^{\phi}  \sin^2 \alpha  + e^{\varphi }  \cos^2 \alpha  \;  , \nonumber \\[5pt]
&& \Delta_2 = e^{\varphi} X \sin^2 \alpha  + e^{2\varphi -\phi} Y \cos^2 \alpha  \;  , \nonumber \\[5pt]
&& \Delta_3 = X \Delta_1 \Delta_2 - Z^2  \sin^2 \alpha \,  \cos^2 \alpha    \;  .
\end{eqnarray}

Using these definitions, the complete nonlinear embedding of the SO(4)--invariant field content (\ref{fieldContentHierarchy}) of ISO(7) supergravity into type IIA reads,
{\setlength\arraycolsep{1pt}
\begin{eqnarray} \label{KKSO4sectorinIIA}
d\hat{s}^2_{10} & = & e^{\frac18 \varphi} X^{1/4} \Delta_1^{1/8} \Delta_3^{1/4} \Big[ \, ds_4^2  \nonumber \\[6pt]
&& \;\; +g^{-2} X \Delta_1 \Delta_3^{-1} \cos^2 \alpha \, \delta_{ij}    D \tilde{\mu}^i D \tilde{\mu}^j  + g^{-2} e^{-\varphi} X^{-1}  d\alpha^2+ g^{-2}  X^{-1}  \Delta_1^{-1} \sin^2 \alpha \, d\tilde{s}^2 (S^3)  \Big] ,   \nonumber \\[12pt]
e^{\hat{\phi}} &=&  e^{\frac{11}{4} \varphi} X^{-1/2} \Delta_1^{3/4} \Delta_3^{-1/2}   \; , \nonumber \\[12pt]
\hat{A}_\3 &=& C^1 \cos^2\alpha  + C^2 \sin^2\alpha  -g^{-1} \, \sin \alpha \cos \alpha \, B \wedge d\alpha   \nonumber \\[5pt] 
&& + \tfrac12 \, g^{-3} \, \chi \, \sin \alpha \cos \alpha \, d\alpha \wedge \epsilon_{ijk} \,  \tmu^i D \tmu^j \wedge \rho^k  \nonumber \\[5pt]
&& -\tfrac14 \, g^{-3} \, e^{\varphi} \chi X  \Delta_3^{-1} \big( X \Delta_1 + Z \cos^2 \alpha \big)  \sin^2 \alpha \cos^2 \alpha  \, \epsilon_{ijk} \, D \tmu^i \wedge D \tmu^j \wedge \rho^k \nonumber \\[5pt]
&& +\tfrac14 \, g^{-3} \, e^{\varphi} \chi \Delta_1^{-1} \, \sin^2 \alpha \cos^2 \alpha  \,\tmu_i  D \tmu_j \wedge \rho^i \wedge \rho^j \nonumber \\[5pt]
&& +\tfrac{1}{48} \, g^{-3} \, X^{-1}  \Delta_1^{-2} \big( e^\phi \rho X \Delta_1 + e^\varphi \chi Z \cos^2 \alpha  \big)  \sin^4 \alpha  \, \epsilon_{ijk} \,  \rho^i \wedge \rho^j \wedge \rho^k \; , \nonumber \\[12pt]
\hat{B}_\2 &=& -\tfrac12 \, g^{-2} \, e^{2\varphi} \chi X^{-1} \, \sin \alpha \, d\alpha \wedge \tmu_i \, \rho^i \nonumber \\[5pt]
&& -\tfrac12 \, g^{-2} \, e^{2\varphi+\phi} \Delta_3^{-1} \big( \rho X \Delta_1 - \chi Z \sin^2 \alpha \big) \cos^3 \alpha   \, \epsilon_{ijk} \,  \tmu^i  D \tmu^j \wedge D \tmu^k \nonumber \\[5pt]
&& +\tfrac12 \, g^{-2} \, e^{2\varphi+\phi} \chi X^{-1}  \Delta_1^{-1} \, \sin^2 \alpha \cos \alpha  \, D \tmu_i \wedge \rho^i \nonumber \\[5pt]
&& + \tfrac18 \, g^{-2} \, e^{2\varphi} \chi X^{-2}  \Delta_1^{-2} \big( e^\varphi X \Delta_1 - e^\phi Z \sin^2 \alpha  \big)  \sin^2 \alpha \cos \alpha  \, \epsilon_{ijk} \,  \tmu^i \rho^j \wedge \rho^k \, ,  \nonumber \\[12pt]
\hat{A}_\1 &=& -\tfrac12 \, g^{-1} \, e^{-2\varphi} Z \Delta_1^{-1} \, \sin^2 \alpha \cos\alpha \, \tmu_i \, \rho^i \; ,
\end{eqnarray}
where we use the ten-dimensional Einstein frame conventions of appendix A of \cite{Guarino:2015vca}.
}Indices $i,j$ are raised and lowered with $\delta_{ij}$, and $d\tilde{s}^2 (S^3)$ is the round metric on the $S^3$ on which the $\rho^i$ are defined. We have also introduced the following covariant derivative and one-form ${\cal A}^i$,
\begin{eqnarray} \label{covDerKKGen}
D \tilde{\mu}^i =  d\tilde{\mu}^i + \epsilon^i{}_{jk}  {\cal A}^j \tilde{\mu}^k \; , \qquad  \textrm{with} \qquad 
{\cal A}^i = - \tfrac12 Z X^{-1} \Delta_1^{-1} \sin^2 \alpha \, \rho^i \; .
\end{eqnarray}
These embedding formulae depend on the (non-vanishing) $D=4$ electric gauge coupling $g$, but not on the magnetic coupling $m$. Thus, they simultaneously describe the embedding of the dynamical SO(4)--invariant sector of the purely electric, $m = 0$, and dyonic, $m \neq 0$, ISO(7) gauging of $D=4$ $\cN=8$ supergravity into massless and massive, respectively, type IIA supergravity. 

The consistent embedding formulae (\ref{KKSO4sectorinIIA}) are valid in full generality for $D=4$ dynamical fields. However, being expressed in terms of the tensor hierarchy (\ref{fieldContentHierarchy}), they contain redundant degrees of freedom. As discussed in general in \cite{Guarino:2015vca}, these redundancies  can be eliminated by expressing the consistent embedding in terms of the IIA field strengths and using the $D=4$ duality relations. In the case at hand, the only contributions from the $D=4$ form field strengths (\ref{eq:FieldStrengths}) happen to occur in the IIA four-form $\hat F_\4$,
\begin{eqnarray} \label{F4D=4FS}
\hat F_\4 = H_\4^1 \cos^2 \alpha  + H_\4^2 \sin^2 \alpha +g^{-1} \sin \alpha \cos \alpha  \, d\alpha \wedge H_\3  + \cdots
\end{eqnarray}
where the dots stand for $D=4$ scalar and derivative-of-scalar contributions without Hodge dualisations. Equation (\ref{F4D=4FS}) follows from (\ref{KKSO4sectorinIIA}) after using the $D=4$ definitions (\ref{eq:FieldStrengths}). It thus provides a ten-dimensional cross-check on the four-dimensional calculation of section \ref{sec:SO(4)-sector}. More importantly, equation (\ref{F4D=4FS}) now expresses the consistent embedding in terms of the independent metric and scalar degrees of freedom contained in the $D=4$ Lagrangian (\ref{L_SO4}), (\ref{VSO4}), when the duality relations (\ref{eq:FieldStrengthDuality}) are employed. A simpler example will be presented in section \ref{sec:dilatons}.

A long calculation allows us to compute the scalar contributions to the IIA field strengths. For simplicity, we present the result for constant $D=4$ scalars\footnote{The complete, dynamical IIA field strengths contain the contributions in (\ref{F4D=4FS}), (\ref{KKfieldstrengths}), plus omitted contributions from $d\varphi$, $d\phi$, $d\chi$, $d\rho$ with no Hodge dualisations.}
{\setlength\arraycolsep{1pt}
\begin{eqnarray} \label{KKfieldstrengths} 
 \hat{F}_\4 & = & U \,  \textrm{vol}_4   \nonumber \\[8pt]
&& + \tfrac14 \,  \Big[ m g^{-4} \, e^{4\varphi +\phi } \chi  X^{-1} \Delta_3^{-1}  \,  \big[ \rho X \Delta_1 - \chi  Z \sin^2 \alpha \big]  \cos^2 \alpha 
  -2g^{-3} \, \chi   \nonumber \\
 && \qquad   + 2 g^{-3} \,  e^{-\phi} \Delta_1^{-1} \Delta_3^{-2} \sin^2 \alpha \cos^2 \alpha \nonumber \\
 && \qquad\quad\; \times    \Big( \big( e^\phi X -  e^{\varphi} Y \big) e^\varphi X \Delta_1  + \big( e^\phi -  e^{\varphi} \big) e^\phi X \Delta_2   - e^\phi Z^2 \big(\cos^2 \alpha - \sin^2 \alpha \big)  \Big)   \nonumber \\
&& \qquad\quad\;  \times \Big( e^\phi Z \big[ \rho X \Delta_1 - \chi Z \sin^2 \alpha \big] \cos^2 \alpha +e^\varphi \chi X \Delta_1 \big[ X \Delta_1 +  Z \cos^2\alpha  \big]   \Big) \nonumber \\
&& \qquad   - g^{-3} \,  e^{\phi} Z \Delta_1^{-1} \Delta_3^{-1} \sin^2 \alpha \cos^2 \alpha \nonumber \\
 && \qquad\quad\quad \times    \Big( 2\big[ (e^\phi -e^\varphi) \rho X - \chi Z \big] \cos^2 \alpha -3\big[ \rho X \Delta_1 - \chi Z \sin^2 \alpha \big]   \Big)   \nonumber \\
&& \qquad   + 2 g^{-3} \,  e^{\varphi} \chi X \Delta_3^{-1} \nonumber \\
 && \qquad\quad\; \times    \Big( \big[ X \Delta_1 - Z \cos^2 \alpha \big] - \big[ 2 X \Delta_1 - 3 Z \sin^2 \alpha \big] \cos^2 \alpha  -\big( e^\phi -  e^{\varphi} \big)  X \sin^2 \alpha \cos^2 \alpha   \Big)  \Big] \nonumber \\
 && \qquad\qquad \times   \sin \alpha \cos \alpha  \, d\alpha \wedge \epsilon_{ijk} \, D \tmu^i \wedge  D \tmu^j \wedge \rho^k  \nonumber \\[8pt]
&& - \tfrac18 \, \chi e^\varphi \, \Delta_1^{-1}\Delta_3^{-1}  \Big[ m g^{-4} \, e^{3\varphi +\phi } X^{-2} \Delta_1^{-1}  \,  \Big( \chi e^{ \phi  } \Delta_3 \sin^2 \alpha  \nonumber \\
&& \qquad  \qquad  \qquad  \qquad  \qquad  \qquad  \quad  +  \big( e^\varphi X \Delta_1 - e^\phi Z \sin^2 \alpha \big)  \big( \rho X \Delta_1 - \chi Z \sin^2 \alpha \big)  \cos^2 \alpha \Big)  \nonumber \\
 &&  \qquad  \qquad \qquad\quad \;\;\; - 2 g^{-3} \,   \Big( \Delta_3  +  \big( X \Delta_1 + Z \cos^2 \alpha \big)  \big(X \Delta_1 +Z \sin^2 \alpha \big)   \Big)     \Big] \nonumber \\
 && \qquad\qquad \times   \sin^2 \alpha \cos^2 \alpha  \, D \tmu_i \wedge  D \tmu_j \wedge \rho^i \wedge \rho^j   \nonumber \\[8pt]
&& + \tfrac14 \, X^{-1} \Delta_1^{-1}   \Big[ m g^{-4} \, e^{4\varphi +\phi } \chi^2 X^{-1} \sin^2 \alpha   \nonumber \\
 &&  \qquad  \qquad  \quad \;\;\; + 2 g^{-3} \, e^\varphi  Z  \Delta_1^{-2} \Delta_3^{-1}   \Big(e^\varphi \chi X \Delta_1  \big[ X \Delta_1 + Z \cos^2 \alpha \big] \nonumber \\
&&  \qquad  \qquad  \qquad  \qquad  \qquad  \qquad  \qquad  \quad  + e^\phi Z   \big[ \rho X \Delta_1 -\chi Z \sin^2 \alpha \big] \cos^2 \alpha  \Big) \sin^2\alpha \cos^2\alpha   \nonumber \\
 &&  \qquad  \qquad  \quad \;\;\; - g^{-3} \, \chi   \Delta_1^{-2}   \Big( 2 \, \Delta_1^2  \big[ X \Delta_1 + Z \sin^2 \alpha \big] - 2 \, e^\varphi  \big[ e^\varphi X \Delta_1 -e^\phi  Z \sin^2 \alpha \big] \cos^2\alpha   \nonumber \\
&&  \qquad  \qquad  \qquad  \qquad  \qquad  \quad \;\;    + e^\varphi \Delta_1   \big[ 2 X \Delta_1 + Z \cos^2 \alpha \big] \sin^2 \alpha  \Big)   \Big] \nonumber \\
 && \qquad\qquad \times   \sin \alpha \cos \alpha   \, d\alpha \wedge \tmu_i \,  D \tmu_j \wedge \rho^i \wedge \rho^j    \nonumber \\[8pt]
&&  - \tfrac{1}{48} \, X^{-2} \Delta_1^{-2}   \Big[ m g^{-4} \, e^{4\varphi } \chi^2 X^{-1} \big[ e^\varphi X \Delta_1 -e^\phi  Z \sin^2 \alpha \big]     \nonumber \\
 &&  \qquad  \qquad  \quad \;\;\; - g^{-3} \,  \Delta_1^{-1}  \Big( X \Delta_1  \big[ 2 e^\phi \rho X  \Delta_1 -3 e^\varphi \chi  Z \sin^2 \alpha \big] \nonumber \\
 && \qquad \qquad \qquad  \qquad \qquad \quad  +2 e^\varphi X  \big[ e^\phi \rho X  \Delta_1 + e^\varphi \chi  Z \cos^2 \alpha \big]  \nonumber \\
&&  \qquad \qquad \qquad  \qquad \qquad \quad   -2 \chi Z \Delta_1  \big[ 3 X \Delta_1 +2  Z \sin^2 \alpha \big] + e^\phi \chi Z^2 \sin^4 \alpha  \Big) \Big] \nonumber \\
 && \qquad\qquad \times   \sin^3 \alpha \cos \alpha  \, d\alpha \wedge  \epsilon_{ijk} \,  \rho^i \wedge \rho^j \wedge \rho^k \, ,  \nonumber \\[25pt] 
 \hat{H}_\3 & = &  \tfrac12 \, g^{-2} \, e^{2\varphi } \Delta_3^{-2} \,  \Big[ 2 \Big( \big( e^\phi X -  e^{\varphi} Y \big) e^\varphi X \Delta_1  + \big( e^\phi -  e^{\varphi} \big) e^\phi X \Delta_2  \nonumber \\
 && \qquad \qquad \qquad \qquad \quad - e^\phi Z^2 \big(\cos^2 \alpha - \sin^2 \alpha \big)  \Big)  \big( \rho X \Delta_1 - \chi Z \sin^2 \alpha \big) \cos^2 \alpha  \nonumber \\
 &&  \qquad \qquad \qquad \quad \; -e^\phi \Delta_3  \Big( 2 \big[ \big( e^\phi -  e^{\varphi} \big) \rho X - \chi Z \big] \cos^2 \alpha   \nonumber \\
 && \qquad \qquad \qquad \qquad \quad -3 \big( \rho X \Delta_1 - \chi Z \sin^2 \alpha \big)  \Big)    \Big]   \sin \alpha  \cos^2 \alpha \, d\alpha \wedge \epsilon_{ijk} \, \tmu^i  D \tmu^j \wedge D \tmu^k \nonumber \\[5pt]
&& - \tfrac12 \, g^{-2} \, e^{2\varphi } X^{-1} \Delta_1^{-2} \Delta_3^{-1} \,  \Big[ 2 e^{\varphi + \phi  } Z    \big( \rho X \Delta_1 - \chi Z \sin^2 \alpha \big)  \cos^2 \alpha \nonumber \\
 && \qquad \qquad \qquad \qquad \qquad \quad - \chi \, \Delta_3 \, e^\varphi \big( \Delta_1 +2 e^\phi \big)     \Big]   \sin \alpha  \cos^2 \alpha   \, d\alpha \wedge D \tmu_i \wedge  \rho^i  \nonumber \\[5pt]
&& + \tfrac18 \, g^{-2} \, X^{-2} \Delta_1^{-2} \Delta_3^{-1} \,  \Big[ e^{3\varphi }  \chi X \Delta_1 \Delta_3    + e^{2\varphi + \phi  }   \big( 2 X \Delta_1 + Z \sin^2 \alpha \big)  \Big( \chi \Delta_3  \nonumber \\
 &&  \qquad\qquad\qquad\qquad \qquad  - Z  \big( \rho X \Delta_1 - \chi Z \sin^2 \alpha \big)  \cos^2 \alpha  \Big)   \Big]   \sin^2 \alpha  \cos \alpha   \, \epsilon_{ijk} \, D \tmu^i \wedge \rho^j \wedge \rho^k  \nonumber \\[5pt]
&& + \tfrac18 \, g^{-2} \,  e^{2\varphi }  \chi  X^{-2} \Delta_1^{-2}  \,  \Big[ 2 \, e^{2\varphi } X \cos^2 \alpha  - 2 \, \Delta_1 \big( X \Delta_1 + Z \sin^2\alpha \big)   \nonumber \\
 &&  \qquad\qquad\qquad\qquad \qquad   - \big( e^\varphi X \Delta_1 - e^\phi Z \sin^2 \alpha  \big) \sin^2 \alpha  \Big] \sin \alpha   \, d\alpha \wedge  \epsilon_{ijk} \, \tmu^i \rho^j \wedge \rho^k   \; ,   \nonumber \\[25pt]
 \hat{F}_\2 & = &  \tfrac12 \, m  g^{-2} \, e^{2\varphi+\phi} \Delta_3^{-1} \big( \chi Z \sin^2 \alpha -\rho X \Delta_1 \big) \cos^3 \alpha   \, \epsilon_{ijk} \, \tmu^i  D \tmu^j \wedge D \tmu^k \nonumber \\[5pt] 
&& 
+\tfrac12 \, \Big[  m g^{-2} \, e^{2\varphi+\phi} \chi X^{-1}   - g^{-1} \, e^{-2\varphi} Z   \Big]  \Delta_1^{-1} \, \sin^2 \alpha \cos \alpha   \, D \tmu_i \wedge \rho^i \nonumber \\[5pt] 
&& -\tfrac12 \, \Big[  m g^{-2} \, e^{2\varphi} \chi X^{-1}   + g^{-1} \, e^{- \varphi} Z \Delta_1^{-2}  \big( 2\cos^2 \alpha - e^{- \varphi}  \sin^2\alpha \, \Delta_1 \big) \Big]  \, \sin \alpha   \, d\alpha \wedge \tmu_i \, \rho^i \nonumber \\[5pt] 
&& +\tfrac18 \, X^{-1} \Delta_1^{-2} \,  \Big[  m g^{-2} \, e^{2\varphi} \chi X^{-1} \big( e^\varphi X \Delta_1 - e^\phi Z \sin^2 \alpha  \big)    + 2 g^{-1} \, e^{- 2\varphi} Z\big( X \Delta_1 + Z \sin^2 \alpha \big)  \Big]  \nonumber \\
&&\qquad\qquad\qquad \;  \times  \sin^2 \alpha \cos \alpha    \, \epsilon_{ijk} \, \tmu^i  \rho^j \wedge \rho^k \; ,  
\end{eqnarray}
}together with $\hat F_\0 = m$ \cite{Guarino:2015jca}. In agreement with the discussions in \cite{Guarino:2015vca,Varela:2015uca}, the field strengths (\ref{KKfieldstrengths}) now do depend on the magnetic gauge coupling $m$ of the $D=4$ supergravity, unlike the gauge potentials (\ref{KKSO4sectorinIIA}). By the consistency of the truncation, the metric and dilaton in (\ref{KKSO4sectorinIIA}), together with the constant-scalar field strengths (\ref{KKfieldstrengths}), solve the field equations of massive type IIA supergravity at any critical point of the $D=4$ scalar potential (\ref{VSO4}). We will make this explicit for the $\cN=3$ critical point in section \ref{sec:AdSsolutions}.

The Freund--Rubin term $U \, \textrm{vol}_4$ in $\hat F_\4$ follows from the general SL(7)--covariant four-form  expression given in \cite{Guarino:2015vca}. It can be written in terms of the SO(4)--invariant four-form field strengths  $H_\4^1$ and $H_\4^2$ as
\begin{equation} \label{eq:FR1}
U \, \textrm{vol}_4 = H_\4^1 \cos^2 \alpha +H_\4^2 \sin^2 \alpha \; 
\end{equation}
or, using the dualisation equations (\ref{eq:FieldStrengthDuality}), as
{\setlength\arraycolsep{2pt}
\begin{eqnarray}
\label{USO4}
U &=& \Big[ g \,  e^{\varphi}  \big(1+e^{2\varphi} \chi^2 \big)  \Big(  4 - 4 \, e^{\phi+3 \varphi } \rho \chi^3  + e^{\varphi - \phi } \big(1- 3 e^{2\varphi} \chi^2 \big) \big(1+  e^{2\phi} \rho^2 \big)   \Big) \nonumber \\[4pt]
&& \quad + m\, e^{\phi + 6\varphi } \rho \chi^2 \Big] \cos^2\alpha  \nonumber \\[4pt]
&& +  \Big[ g \, \big(1+e^{2\varphi} \chi^2 \big)  \Big( 3 e^{\varphi}  - 3 \, e^{\phi+4 \varphi } \rho \chi^3  +6 e^{\phi} \big(1+e^{2\varphi} \chi^2 \big)   -4 e^{\phi} \big(1+e^{2\varphi} \chi^2 \big)^2    \Big) \nonumber \\[4pt]
&& \quad + m\, e^{\phi + 6\varphi } \chi^3 \Big]  \sin^2\alpha  \; ,
\end{eqnarray}
}in terms of the $D=4$ scalars. Note that, while the IIA field strengths (\ref{KKfieldstrengths}) are evaluated for constant scalars, the Freund--Rubin term (\ref{USO4}) is valid beyond that assumption: it takes on the same form also for dynamical scalars. Some calculation reveals that $U$ is related to the $D=4$ scalar potential (\ref{VSO4}) and its derivatives via
\begin{equation} \label{UintermsofV}
g \, U = -\tfrac{1}{3} \, V +\tfrac13 \Big(  \partial_\phi V -\rho \, \partial_\rho V  \Big) \, \cos^2\alpha    +\tfrac{1}{12} \Big(  \partial_\varphi V -2 \partial_\phi V - \chi \, \partial_\chi V +2 \rho \,  \partial_{\rho} V \Big) \, \sin^2\alpha \; .
\end{equation}
At the critical points of the potential, recorded in table \ref{Table:SO4Points}, this expression reduces to
{\setlength\arraycolsep{2pt}
\begin{eqnarray} \label{UintermsofVCritical}
g \, U_0 &=& -\tfrac{1}{3} \, V_0 \; , 
\end{eqnarray}
}in agreement with the general $\cN=8$ discussion of \cite{Guarino:2015vca}. See respectively \cite{Varela:2015uca} and \cite{Godazgar:2015qia,Varela:2015ywx} for related discussions in the massive IIA on $S^6$ and $D=11$ on $S^7$ contexts.

\subsection{Local and global structure} \label{sec:Regularity}

For arbitrary values of the $D=4$ scalars, the six-dimensional internal local geometry in (\ref{KKSO4sectorinIIA}) can be regarded as the warped product of an interval $I$, on which $\alpha$ takes values, and a family of five-dimensional spaces parametrised by $\alpha$. At fixed $\alpha$, the five-dimensional space corresponds to an $S^2$ bundle over $S^3$, with connection one-forms $ {\cal A}^i$ given in (\ref{covDerKKGen}). All such bundles are trivial. In the present case,  this can be seen by the fact that, at fixed $\alpha$, the curvature of the connection $ {\cal A}^i$ is identically zero by the Maurer-Cartan equations (\ref{eq:MC}). This local characterisation is useful to discuss the global extension of the geometry, to which we now turn. It is not the only possible local description, though. A different local characterisation will be given below.

Globally, the internal geometry extends smoothly into $S^6$. This is expected from the fact that the $D=4$ theory  (\ref{L_SO4}), (\ref{VSO4}) arises upon consistent Kaluza--Klein truncation of massive type IIA on $S^6$ via (\ref{KKSO4sectorinIIA}), and the Kaluza--Klein deformations are not supposed to change the internal topology. That the topology of the compactification space is indeed $S^6$ is most easily seen by continuously deforming the geometry into the G$_2$--invariant locus (\ref{eq:SO4toG2}). On this locus, the internal metric in (\ref{KKSO4sectorinIIA}) reduces to the usual, round Einstein metric (\ref{RoundS6}) on $S^6$. The local line element   (\ref{RoundS6}) is adapted to the topological construction of $S^6$ as the `join' of $S^2$ and $S^3$, provided the $S^6$ angle $\alpha$ is restricted to the interval
\begin{equation} \label{anglealpha}
\alpha \in I \equiv [ 0  , \tfrac{\pi}{2} ]  \; .
\end{equation}
On the G$_2$--invariant locus and at $\alpha = 0$, the $S^2$ remains finite and the $S^3$ collapses; at the other endpoint, $\alpha =  \frac{\pi}{2}$, the opposite happens. 

The expression (\ref{KKSO4sectorinIIA}) makes it straightforward to continuously deform the internal geometry to the round metric on $S^6$, since it is given as a function of the $D=4$ scalar manifold (\ref{ScalManN=1}). However, once the scalars are fixed to their specific values at some critical point of the potential (\ref{VSO4}), as {\it e.g.}~in the explicit $\cN=3$ solution (\ref{SO4SolN=3}) below, tracking down the deformation into the round $S^6$ geometry is no longer  obvious. In such cases, it is more useful to directly characterise the internal $S^6$  by verifying that it still corresponds to the join of $S^2$ and $S^3$. Namely, that the shrinking patterns of $S^2$ and $S^3$ at each endpoint of the interval $I$ remain valid away from the G$_2$--invariant locus. To see this, we use the definitions (\ref{deltas}) to compute the behaviour of the relevant metric functions at both endpoints of $I$. At the lower end,
\begin{eqnarray} \label{lowerend}
&& e^{ \varphi} X^2 \Delta_1 \Delta_3^{-1} \, \cos^2\alpha \,  \xrightarrow[\alpha \rightarrow 0]{} \, e^{ \phi -  \varphi}  XY^{-1} - e^{ 2\phi - 2 \varphi}  ( X^2 -e^{ -2 \varphi} Z^2 ) Y^{-2} \,  \alpha^2 \; + \ {\cal O}(\alpha^4) \; , \nonumber \\[4pt]
&& e^{\varphi} \Delta_1^{-1} \, \sin^2\alpha \,  \xrightarrow[\alpha \rightarrow 0]{} \,  \alpha^2 +{\cal O}(\alpha^4) \; .
\end{eqnarray}
Thus, $S^2$ remains finite and $S^3$ shrinks to zero size for all values of the $D=4$ scalars. At the upper end,
\begin{eqnarray}  \label{upperend}
&& e^{ \varphi} X^2 \Delta_1 \Delta_3^{-1} \, \cos^2\alpha \,  \xrightarrow[\alpha \rightarrow \frac{\pi}{2}]{} \,  (\tfrac{\pi}{2} -\alpha)^2 +{\cal O}( (\tfrac{\pi}{2}-\alpha)^4) \; , \nonumber \\[4pt]
&& e^{\varphi} \Delta_1^{-1} \, \sin^2\alpha \,  \xrightarrow[\alpha \rightarrow  \frac{\pi}{2}]{} \, e^{ -\phi +  \varphi}   +  e^{ -2\phi + 2 \varphi}   (\tfrac{\pi}{2} -\alpha)^2 +{\cal O}( (\tfrac{\pi}{2}-\alpha)^4) \; ,
\end{eqnarray}
and the opposite happens: $S^2$ shrinks and $S^3$ remains finite for all $D=4$ scalar values.

An alternate local characterisation of the internal geometry in (\ref{KKSO4sectorinIIA}) may be given as follows. The local internal geometry may  also be regarded as an $S^2$ bundle over the four-dimensional local geometry $M_4 \equiv I \times S^3$, where $I$ is the interval (\ref{anglealpha}) parametrised by $\alpha$. This local construction is a generalisation of the twistor fibration over a four-dimensional Riemannian space $M_4$. In the usual twistor construction, the metric $ds^2 (M_4)$ on $M_4$ is taken to be Einstein with (anti)self-dual Weyl tensor. The local metric on the six-dimensional twistor bundle is 
\begin{eqnarray} \label{eq:twistormetric}
ds^2_6 = \tfrac14 \delta_{ij} D\tmu^i D\tmu^j + \tfrac12 ds^2 (M_4) \; , 
\end{eqnarray}
see {\it e.g.}~\cite{Gibbons:1989er}. Here, $\tmu^i$ parametrise an $S^2$ as in (\ref{eq:S2}), and the covariant derivatives $D\tmu^i$ are defined as in the left most equation in (\ref{covDerKKGen}), in terms of a $M_4$--valued connection ${\cal A}^i$. Being four-dimensional and Einstein, $M_4$ is automatically quaternionic-K\"ahler. The curvature of the connection,
\begin{equation} \label{eq:curvature}
{\cal F}^i = d{\cal A}^i + \tfrac12 \epsilon^i{}_{jk} {\cal A}^j \wedge {\cal A}^k  \; , 
\end{equation}
is proportional to the quaternionic-K\"ahler forms $J^i$ on $M_4$. The self-duality or antiself-duality of the Weyl tensor on $M_4$ devolves in the antiselfduality or self-duality of ${\cal F}^i$ with respect to the metric $ds^2 (M_4)$. For example, the twistor bundle on $M_4 = S^4$ coincides with the three-dimensional complex projective space, $\mathbb{CP}^3$. Taking $ds^2 (M_4)$ to be the usual round metric on $S^4$,
\begin{eqnarray}
ds^2 (S^4) = d\alpha^2 + \sin^2\alpha \,  ds^2 (S^3) 
\end{eqnarray}
(with $\alpha$ here ranging in $ 0 \leq \alpha \leq \pi$) and
\begin{eqnarray}
{\cal A}^i = \tfrac12 ( 1- \cos\alpha) \, \rho^i ,
\end{eqnarray}
where the $\rho^i$ are the right-invariant Maurer-Carten one-forms on $S^3$, subject to (\ref{eq:MC}), the twistor bundle metric (\ref{eq:twistormetric}) becomes the homogeneous nearly-K\"ahler metric on $\mathbb{CP}^3$. 

The local internal metric in (\ref{KKSO4sectorinIIA}) is a generalisation of the twistor construction. In our case,  $M_4 \equiv I \times S^3$ is the upper $S^4$ hemisphere, given the range (\ref{anglealpha}) of $\alpha$. The metric $ds^2_4 (M_4)$ induced on it is not selfdual Einstein for any values of the $D=4$ scalars. On the G$_2$--invariant  locus (\ref{eq:SO4toG2}) the $S^2$ fibration trivialises, ${\cal A}^i = 0$, and the local geometry becomes locally a warped product of $S^2$ and $I \times S^3$. Away from the G$_2$--invariant locus, the $S^2$ is warped (unlike in (\ref{eq:twistormetric})), and non-trivially fibered through (\ref{covDerKKGen}) over $I \times S^3$. The curvature (\ref{eq:curvature}) of the connection ${\cal A}^i$ is
\begin{equation} \label{eq:ConnectionFS}
 {\cal F}^i = -e^{\varphi} Z X^{-1} \Delta_1^{-2} \sin \alpha \cos \alpha \, d \alpha \wedge \rho^i  + \tfrac18 Z X^{-2} \Delta_1^{-2} (2 X \Delta_1 + Z \sin^2 \alpha ) \sin^2 \alpha \,  \epsilon^i{}_{jk} \rho^j \wedge \rho^k \; , 
\end{equation}
and its Hodge dual with respect to the metric induced on $I \times S^3$,
{\setlength\arraycolsep{2pt}
\begin{eqnarray} \label{eq:ConnectionDualFS}
* {\cal F}^i &=& \tfrac12 e^{-\frac12 \varphi} Z X^{-2} \Delta_1^{-\frac32} (2 X \Delta_1 + Z \sin^2 \alpha ) \sin \alpha  \, d \alpha \wedge \rho^i \nonumber \\[4pt]
&&  -\tfrac14 e^{\frac32 \varphi}  Z X^{-1} \Delta_1^{-\frac52} \sin^2 \alpha \cos \alpha \,  \epsilon^i{}_{jk} \rho^j \wedge \rho^k \; . 
\end{eqnarray}
}The non-trivial connection $ {\cal A}^i$ is neither selfdual nor antiself-dual for any values of the $D=4$ scalars, as nowhere on the scalar manifold (\ref{ScalManN=1}) do (\ref{eq:ConnectionFS}), (\ref{eq:ConnectionDualFS}) obey $* {\cal F}^i = \pm  {\cal F}^i$. 

Massive type IIA supergravity admits a class of $\cN=1$ direct product solutions AdS$_4 \times M_6$ where $M_6$ is nearly-K\"ahler \cite{Behrndt:2004km,Behrndt:2004mj} or half-flat \cite{Lust:2004ig,Tomasiello:2007eq,Koerber:2008rx}. For example, $M_6$ can be taken to be the round $S^6$ equipped with its homogeneous nearly-K\"ahler structure, see appendix \ref{subset:S6Geom} for a review in the present context. On topologies different from $S^6$, a natural way to engineer nearly-K\"ahler geometries or half-flat geometries of the required type is via the usual twistor fibration \cite{Tomasiello:2007eq}. For example, $M_6$ can be taken to be $\mathbb{CP}^3$ with metric (\ref{eq:twistormetric}). Our local geometry (\ref{KKSO4sectorinIIA}) restricted to the G$_2$--invariant locus (\ref{eq:SO4toG2}) reduces to the round, homogeneous nearly-K\"ahler structure on $S^6$. Away from the G$_2$ locus, as in the $\cN=3$ solution of section \ref{sec:AdSsolutions}, the geometry can be locally described by the generalised twistor fibration discussed above.

On the G$_2$--invariant locus (\ref{eq:SO4toG2}), the symmetry of the configuration (\ref{KKSO4sectorinIIA}) is enhanced to a homogeneously acting G$_2$. See section \ref{subsec:G2fromSO4} for further details. Away from the G$_2$ locus, the isometry of the internal geometry is the SO(4) subgroup of SO(7) defined in either (\ref{embedding_SO4}) or (\ref{embedding_SO4_2}). The group SO(4) acts by isometries with cohomogeneity one, and is also preserved by the supergravity forms. The $ \textrm{SO}(3)_{\textrm{d}}$ subgroup of SO(4) rotates the $S^2$ fibers, and $ \textrm{SO}(3)_{\textrm{R}}$ acts on the $S^3$ base. The supersymmetry of the $\cN=3$ solution will be discussed in section \ref{sec:N=3susy}.

Some generalisations can be envisaged. When the $D=4$ scalars are restricted to the G$_2$--invariant locus (\ref{eq:SO4toG2}), the type IIA solution (\ref{KKSO4sectorinIIA}) depends only on the homogeneous nearly-K\"ahler structure on $S^6$. In this case, the $S^6$ can be replaced with any other nearly-K\"ahler manifold. This situation was discussed in \cite{Varela:2015uca}. Away from the G$_2$ locus, the solution can be also generalised. Now, the generalisation entails replacing $S^3$ with the cyclic lens space $S^3/\mathbb{Z}_p$, with the identification acting on the Hopf fiber. While $S^3/\mathbb{Z}_p$ is a smooth manifold, the total six-dimensional geometry corresponding to this generalisation displays orbifold singularities.


\section{Further truncations} \label{sec:urthertruncs}


It is useful to obtain particular cases of the uplifting formulae derived above. Here, we will discuss the truncations to the sectors of the $D=4$ supergravity with G$_2$ and $\textrm{SO}(3)^\prime \times \textrm{SO}(4)^\prime$ symmetry.

\subsection{Truncation to the G$_2$ sector} 
\label{subsec:G2fromSO4}

The sector of $D=4$ ISO(7) supergravity that retains singlets under the G$_2$ subgroup of SO(7) was analysed in detail in \cite{Guarino:2015qaa}, and its explicit ten-dimensional embedding worked out in \cite{Guarino:2015vca}. Its consistent IIA embedding was recovered from that of the SU(3)--invariant sector in \cite{Varela:2015uca}. Here, we will recover the embedding of the G$_2$-sector from the SO(4)--invariant consistent truncation formulae of section \ref{subsec:SU3UpliftSubsec}.

The $D=4$ G$_2$--invariant sector is recovered from the SO(4) sector by imposing the identifications (\ref{eq:SO4toG2}). Bringing these relations to the consistent embedding formulae (\ref{KKSO4sectorinIIA}), we find that the connection (\ref{covDerKKGen}) trivialises, ${\cal A}^i =0$, and that the scalar dependence of the internal metric factorises in front of the round Einstein metric $ds^2(S^6)$ on $S^6$ foliated as in (\ref{RoundS6}). The internal $S^6$ dependence drops out from the dilaton. Finally, all the dependence of the IIA potentials on the internal $S^6$ combines into the homogeneous nearly-K\"ahler structure ${\cal J}$, $\Upomega$ on $S^6$, through the expressions (\ref{NKintermsofmus}). More concretely, (\ref{KKSO4sectorinIIA}) reduces to 
{\setlength\arraycolsep{0pt}
\begin{eqnarray} \label{G2embeddingGeom}
&& d \hat{s}_{10}^2 = e^{\frac{3}{4} \varphi} \big( 1+e^{2 \varphi} \chi^2 \big)^{\frac{3}{4}}  ds^2_4 + g^{-2} e^{-\frac{1}{4} \varphi} \big( 1+e^{2 \varphi} \chi^2 \big)^{-\frac{1}{4}}   ds^2(S^6) \; , \nonumber \\[5pt]
&& e^{\hat \phi} = e^{\frac{5}{2} \varphi} \big( 1+e^{2 \varphi} \chi^2 \big)^{-\frac32}  \; , \nonumber \\[5pt]
&&  \hat A_\3 = C + g^{-3} \chi \, \textrm{Im} \, \Upomega  \; , \qquad
 \hat B_\2 = g^{-2} \, e^{2 \varphi} \chi  \big( 1+e^{2 \varphi} \chi^2 \big)^{-1}   \,  {\cal J}  \; , \qquad   
 \hat A_\1 = 0 \; ,
\end{eqnarray}
}in agreement with the formulae for the consistent truncation to the G$_2$--invariant sector given in (4.3) of  \cite{Guarino:2015vca}. Similarly, the constant-scalar field strengths (\ref{KKfieldstrengths}) reduce to the corresponding contributions in (4.4) of \cite{Guarino:2015vca}.


\subsection{Dilatons}
\label{sec:dilatons}

According to (\ref{eq:SO4toSO3pSO4p}), the $\textrm{SO}(3)^\prime \times \textrm{SO}(4)^\prime$ --invariant sector of $\cN=8$ ISO(7) supergravity retains only the dilatons $\phi$, $\varphi$, along with the two- and three-forms in the tensor hierarchy (\ref{fieldContentHierarchy}). From (\ref{KKSO4sectorinIIA}), (\ref{eq:SO4toSO3pSO4p}), it is apparent that the field strengths in this subsector will not contain terms in $d \varphi$ or $d\phi$, prior to imposing the dualisation (\ref{eq:FieldStrengthDuality}). In other words, equation (\ref{F4D=4FS}) for $\hat F_\4$ is exact (the dots can be disregarded) and $\hat H_\3$ and $\hat F_\2$ are zero. Using the dualisation conditions (\ref{eq:FieldStrengthDuality}), the full non-linear embedding of the $D=4$ metric plus dilaton sector into massive type IIA reads, at the level of the field strengths,
{\setlength\arraycolsep{1pt}
\begin{eqnarray} \label{KKSO4sectorinIIADilatons}
d\hat{s}^2_{10} & = & e^{\frac18 \varphi} \Delta_1^{1/8} \Delta_3^{1/4} \Big[ \, ds_4^2  \nonumber \\[6pt]
&& \;\; +g^{-2} \Delta_1 \Delta_3^{-1} \cos^2 \alpha \, d\tilde{s}^2 (S^2) + g^{-2} e^{-\varphi} d\alpha^2+ g^{-2}   \Delta_1^{-1} \sin^2 \alpha \, d\tilde{s}^2 (S^3)  \Big] ,   \nonumber \\[12pt]
e^{\hat{\phi}} &=&  e^{\frac{11}{4} \varphi} \Delta_1^{3/4} \Delta_3^{-1/2}   \; , \nonumber \\[12pt]
\hat F_\4 & =& \Big[ g \,   \big(  4 \, e^{\varphi}   + e^{2\varphi - \phi }   \big)  \cos^2\alpha + \big( 3 e^{\varphi}   + 2 e^{ \phi }   \big)  \sin^2\alpha    \Big] \, \textrm{vol}_4   +g^{-1} \sin \alpha \cos \alpha  \, d\alpha \wedge * \big( d\phi - d\varphi  \big) \nonumber \\[12pt]
\hat H_\3 & =& \hat F_\2 = 0 \; ,
\end{eqnarray}
}with $\Delta_1$, $\Delta_3$ given by (\ref{deltas}) with $\chi = \rho=0$. In this sector, the fibration of $S^2$ over $ I \times S^3$ also trivialises, ${\cal A}^i=0$. Accordingly, the symmetry preserved by the configuration (\ref{KKSO4sectorinIIADilatons}) is the  $\textrm{SO}(3)^\prime \times \textrm{SO}(4)^\prime$ subgroup of SO(7) defined in (\ref{embedding_SO4}), with $\textrm{SO}(3)^\prime$ and $\textrm{SO}(4)^\prime$ respectively acting on the $S^2$ and the $S^3$. By using the $D=4$ duality hierarchy (\ref{eq:FieldStrengthDuality}), the consistent embedding (\ref{KKSO4sectorinIIADilatons}) becomes expressed in terms of independent four-dimensional degrees of freedom only: the dilatons and their derivatives, and the metric, explicitly and through the Hodge dual.

\section{$\cN=3$ SO(4)--invariant AdS$_4$ solution of massive type IIA} 
\label{sec:AdSsolutions}

By the consistency of the embedding, the ten-dimensional metric and dilaton in (\ref{KKSO4sectorinIIA}), along with the field strengths that follow from the potentials given in that equation, satisfy the equations of motion of  massive type IIA supergravity provided the equations of motion that follow from the $D=4$ Lagrangian (\ref{L_SO4}), (\ref{VSO4}) are imposed. In particular, (\ref{KKSO4sectorinIIA}), (\ref{KKfieldstrengths}) evaluated on the critical points of the scalar potential (\ref{VSO4}) summarised in table \ref{Table:SO4Points} give rise to AdS$_4$ solutions of massive type IIA. The $\cN=1$ and $\cN=0$ critical points with G$_2$ symmetry  uplift to the solutions respectively found in \cite{Behrndt:2004km} and \cite{Lust:2008zd}. The non-supersymmetric SO(7)--invariant critical point gives rise to a solution given in \cite{Romans:1985tz}. See \cite{Varela:2015uca} for these solutions in our conventions. In all these solutions with at least G$_2$ symmetry, the fibration trivialises,  ${\cal A}^i = 0$, and the metric becomes the round,  SO(7)--symmetric Einstein metric on $S^6$. In the G$_2$--invariant solutions, the symmetry is reduced by the supergravity forms, which take values along the homogeneous nearly-K\"ahler structure on $S^6$. 

Here we are interested in the uplift of the $\cN=3$ critical point of ISO(7) supergravity \cite{Gallerati:2014xra}. Bringing the vacuum expectation values of the $D=4$ scalars recorded in table \ref{Table:SO4Points} to the formulae (\ref{KKSO4sectorinIIA}), (\ref{KKfieldstrengths}), and rescaling the external $D=4$ metric and the Freund--Rubin term with the cosmological constant recorded in the table so that AdS$_4$ is unit radius, as in \cite{Varela:2015uca}, we find the massive type IIA uplift of the $\cN=3$ solution. In Einstein frame it reads,
{\setlength{\arraycolsep}{1pt}
\begin{eqnarray} \label{SO4SolN=3}
d\hat{s}^2_{10} & = & L^2 \, \big( 3 + \cos 2\alpha  \big)^{1/8} \Big( 3 \cos^4 \alpha + 3 \cos^2 \alpha +2 \Big)^{1/4}  \Big[ \, ds^2(\textrm{AdS}_4) \nonumber  \\[4pt]
&& \qquad \quad  + \frac{2 \big( 3 + \cos 2\alpha  \big) \cos^2 \alpha}{ 3 \cos^4 \alpha +3  \cos^2 \alpha +2 } \, \delta_{ij}    D \tilde{\mu}^i D \tilde{\mu}^j  + 2 \,   d\alpha^2+ \frac{8 \sin^2 \alpha }{3 + \cos 2\alpha} \, d\tilde{s}^2 (S^3)  \Big] ,  \nonumber \\[12pt]
\label{DilatonN=3} e^{\hat{\phi}} &=&  e^{\phi_0}  \, \frac{\big( 3 + \cos 2\alpha  \big)^{3/4}}{\big( 3 \cos^4 \alpha +3  \cos^2 \alpha +2 \big)^{1/2}} \; , \nonumber \\[12pt]
L^{-3} e^{\frac{1}{4} \phi_0} \, \hat{F}_\4 & = & 3\sqrt{2} \,  \textrm{vol} (\textrm{AdS}_4 )  \nonumber \\[5pt]
&& -\frac{ 4\sqrt{6} \, \big(  2 \cos^4\alpha + 3 \cos^2 \alpha +3 \big)  \sin \alpha  \cos^3 \alpha   }{  \big( 3  + \cos 2\alpha \big) \big( 3 \cos^4\alpha + 3\cos^2 \alpha +2 \big) }  \, d\alpha \wedge \epsilon_{ijk} \, D \tmu^i \wedge  D \tmu^j \wedge \rho^k  \nonumber \\[5pt]
&& +\frac{ \sqrt{6} \,  \big( 5  + 3 \cos 2\alpha \big)   \sin^2 \alpha  \cos^2  \alpha   }{ 2 \, \big( 3 \cos^4\alpha + 3\cos^2 \alpha +2 \big) }    \, D \tmu_i \wedge  D \tmu_j \wedge \rho^i \wedge \rho^j   \nonumber \\[5pt]
&& -\frac{ 4 \sqrt{6} \,  \sin^5 \alpha \cos \alpha }{  \big( 3  + \cos 2\alpha \big)^2  }   \, d\alpha \wedge \tmu_i \,  D \tmu_j \wedge \rho^i \wedge \rho^j    \nonumber \\[5pt]
&&  -\frac{  2 \sqrt{2} \, \big( 5  + 3 \cos 2\alpha \big)  \sin^3 \alpha \cos \alpha}{ \sqrt{3} \, \big( 3  + \cos 2\alpha \big)^3 }   \, d\alpha \wedge  \epsilon_{ijk} \,  \rho^i \wedge \rho^j \wedge \rho^k \, ,  \nonumber \\[12pt] 
L^{-2} e^{-\frac{1}{2} \phi_0} \, \hat{H}_\3 & = &  -\frac{ 2\sqrt{3} \, \big( 3 \cos^6\alpha+ 8 \cos^4\alpha + 11 \cos^2 \alpha +2 \big)   }{ \big( 3 \cos^4\alpha + 3\cos^2 \alpha +2 \big)^2  }   \sin \alpha  \cos^2 \alpha \, d\alpha \wedge \epsilon_{ijk} \, \tmu^i  D \tmu^j \wedge D \tmu^k \nonumber \\[5pt]
&& +\frac{ 8\sqrt{3} \, \big(  \cos^4\alpha + \cos^2 \alpha +2 \big)  \sin \alpha  \cos^2 \alpha   }{  \big( 3  + \cos 2\alpha \big) \big( 3 \cos^4\alpha + 3\cos^2 \alpha +2 \big) }   \, d\alpha \wedge D \tmu_i \wedge  \rho^i  \nonumber \\[5pt]
&& +\frac{ \sqrt{3} \,  \big( 3  + \cos 2\alpha \big)   \sin^2 \alpha  \cos  \alpha   }{ 2 \, \big( 3 \cos^4\alpha + 3\cos^2 \alpha +2 \big) }    \, \epsilon_{ijk} \, D \tmu^i \wedge \rho^j \wedge \rho^k  \nonumber \\[5pt]
&& -\frac{ 2 \sqrt{3} \,  \sin^5 \alpha }{  \big( 3  + \cos 2\alpha \big)^2  }   \, d\alpha \wedge  \epsilon_{ijk} \, \tmu^i \rho^j \wedge \rho^k   \; ,   \nonumber \\[12pt]
L^{-1} e^{\frac{3}{4} \phi_0} \, \hat{F}_\2 & = &  \frac{ \sqrt{2} \,  \big( 5  + 3 \cos 2\alpha \big)  \cos^3 \alpha  }{ 4 \,   \big( 3 \cos^4\alpha + 3\cos^2 \alpha +2 \big) }  \, \epsilon_{ijk} \, \tmu^i  D \tmu^j \wedge D \tmu^k 
 +\frac{ 2 \sqrt{2} \,  \sin^2 \alpha \cos \alpha}{ 3  + \cos 2\alpha  }  \, D \tmu_i \wedge \rho^i \nonumber \\[5pt]  
&& -\frac{ 4 \sqrt{2} \,  \sin^3 \alpha }{ \big( 3  + \cos 2\alpha \big)^2 }  \, d\alpha \wedge \tmu_i \, \rho^i 
+\frac{ 3  \sin^4 \alpha \cos \alpha }{ \sqrt{2} \,  \big( 3  + \cos 2\alpha \big)^2 }  \, \epsilon_{ijk} \, \tmu^i  \rho^j \wedge \rho^k \; , \nonumber \\[12pt]
L \, e^{\frac{5}{4} \phi_0} \, \hat{F}_\0 & = &  \frac{ \sqrt{3}}{2\sqrt{2}} \;,
\end{eqnarray}
in the IIA conventions of appendix A of \cite{Guarino:2015vca}.
}The covariant derivative of $\tmu^i$ and the corresponding connection ${\cal A}^i$ are, from (\ref{covDerKKGen}),
\begin{eqnarray}
D \tilde{\mu}^i =  d\tilde{\mu}^i + \epsilon^i{}_{jk} {\cal A}^j \tilde{\mu}^k \; , \qquad  \textrm{with} \qquad 
{\cal A}^i =  \frac{\sin^2 \alpha }{3 + \cos 2\alpha} \, \rho^i \; ,
\end{eqnarray}
and we have defined $L^2 \equiv 2^{-\frac{31}{12}} \, 3^{\frac{3}{8}} \, g^{-\frac{25}{12}} \,  m^{\frac{1}{12}}$ and $e^{\phi_0} \equiv  2^{-\frac{1}{6}} \, 3^{\frac{1}{4}} \,  g^{\frac{5}{6}} m^{-\frac{5}{6}}$.  A set of gauge potentials for the (internal) field strengths in (\ref{SO4SolN=3}) follows from (\ref{KKSO4sectorinIIA})\footnote{This set of gauge potentials and the metric in (\ref{SO4SolN=3}) are related to the expressions given in \cite{Pang:2015vna} by identifying their $S^6$ angle $\xi_{\textrm{PR}}$ with our $\alpha$, $\xi_{\textrm{PR}}=\alpha$, relating their $S^6$ embedding coordinates $\mu_{\textrm{PR}}^{\hat{i}}$, $\nu_{\textrm{PR}}^i$ with our $\mu^{\hat{i}}$, $\mu^i$ through 
$\mu_{\textrm{PR}}^{\hat{i}}=\sin\alpha \, \tilde{\mu}^{\hat{i}}$, $\nu_{\textrm{PR}}^1=\cos\alpha \,  \tilde{\mu}^3$, $\nu_{\textrm{PR}}^2=\cos\alpha \,  \tilde{\mu}^2$, $\nu_{\textrm{PR}}^3=-\cos\alpha \,  \tilde{\mu}^1$,
letting $\frac{2^{\frac{7}{8}}}{9\sqrt{2}} L^2_{\textrm{PR}}= L^2$, and rearranging significantly. The explicit expressions (\ref{muS4coords}), (\ref{LIandRIMCforms}) for the $S^3$ embedding coordinates $\tilde{\mu}^{\hat{i}}$ and the right-invariant forms $\rho^i$ are also useful for this comparison. Note, however, that our expressions for the $\cN=3$ solution follow directly from the uplifting formulae (\ref{KKSO4sectorinIIA}) for the dynamical SO(4)--invariant sector of $\cN=8$ ISO(7) supergravity, which were not given in \cite{Pang:2015vna}.}:

{\setlength\arraycolsep{1pt}
\begin{eqnarray} \label{KKformpotentialsN=3}
L^{-3} e^{\frac{1}{4} \phi_0} \, \hat{A}_\3 &=& \frac{2\sqrt{2}}{\sqrt{3}} \, \sin \alpha \cos \alpha \, d\alpha \wedge \epsilon_{ijk} \,  \tmu^i D \tmu^j \wedge \rho^k  \nonumber \\[5pt]
&& -\frac{ 2 \sqrt{2} \,   \sin^2 \alpha \cos^2 \alpha  }{ \sqrt{3} \,  \big( 3 \cos^4\alpha + 3\cos^2 \alpha +2 \big) }  \, \epsilon_{ijk} \, D \tmu^i \wedge D \tmu^j \wedge \rho^k \nonumber \\[5pt]
&& +\frac{4 \sqrt{2} \,  \sin^2 \alpha \cos^2 \alpha}{\sqrt{3} \, \big( 3  + \cos 2\alpha \big) }  \,\tmu_i  D \tmu_j \wedge \rho^i \wedge \rho^j \nonumber \\[5pt]
&& -\frac{2 \sqrt{2} \,  \big( 2  + \cos 2\alpha \big)  \sin^4 \alpha }{3 \sqrt{3} \, \big( 3  + \cos 2\alpha \big)^2 }    \, \epsilon_{ijk} \,  \rho^i \wedge \rho^j \wedge \rho^k \; , \nonumber \\[12pt]
L^{-2} e^{-\frac{1}{2} \phi_0} \,  \hat{B}_\2 &=& -\frac{2}{\sqrt{3}}  \, \sin \alpha \, d\alpha \wedge \tmu_i \, \rho^i 
+ \frac{ \big( 5 +3 \cos 2\alpha \big)  \cos^3 \alpha }{ \sqrt{3} \,  \big( 3 \cos^4\alpha + 3\cos^2 \alpha +2 \big) }    \, \epsilon_{ijk} \,  \tmu^i  D \tmu^j \wedge D \tmu^k \nonumber \\[4pt]
&& +\frac{4 \sin^2 \alpha \cos \alpha}{\sqrt{3} \, \big( 3  + \cos 2\alpha \big) }   \, D \tmu_i \wedge \rho^i 
 +\frac{\big( 7  + \cos 2\alpha \big)  \sin^2 \alpha \cos \alpha}{\sqrt{3} \, \big( 3  + \cos 2\alpha \big)^2 }   \, \epsilon_{ijk} \,  \tmu^i \rho^j \wedge \rho^k \, ,  \nonumber \\[12pt]
L^{-1} e^{\frac{3}{4} \phi_0} \, \hat{A}_\1 &=& \sqrt{2} \; \frac{\sin^2 \alpha \cos\alpha }{3 + \cos 2\alpha } \, \tmu_i \, \rho^i \; . 
\end{eqnarray}
}

All the comments made in section \ref{sec:SO4sectorinIIA} for the generic solution away from the G$_2$-locus  apply to the specific $\cN=3$ solution (\ref{SO4SolN=3}). The internal metric and supergravity forms extend smoothly on $S^6$. Locally, the solution can be regarded as a (trivial) $S^2$ bundle over $S^3$ foliated by $\alpha$ or, alternatively, as the warped generalisation of the twistor fibration discussed in section \ref{sec:Regularity}. The angle $\alpha$ has range (\ref{anglealpha}), $\tmu^i$ parametrise $S^2$ via (\ref{eq:S2}) and $\rho^i$ are the right-invariant Maurer-Cartan one-forms on $S^3$, subject to (\ref{eq:MC}). The solution displays a cohomogeneity-one isometry group $\textrm{SO}(4) \equiv  \textrm{SO}(3)_{\textrm{d}} \times \textrm{SO}(3)_{\textrm{R}}$, where $\textrm{SO}(3)_{\textrm{d}}$ and $\textrm{SO}(3)_{\textrm{R}}$ respectively act on the $S^2$ fibers and the $S^3$ base. The solution can be generalised by replacing $S^3$ with the cyclic Lens space $S^3/\mathbb{Z}_p$, a generalisation that introduces orbifold singularities. The $\cN=3$ supersymmetry of the solution is shown in the next section.

\section{Supersymmetry of the $\cN=3$ solution} \label{sec:N=3susy}

The gravitini of the $D=4$ $\cN=8$ ISO(7) supergravity lie in the spinor representation of SO(7). Under (\ref{embedding_SO4}), this branches as\footnote{More precisely, here and below we refer to the Spin groups, $\mathrm{SU}(2)^\prime$, $\mathrm{SU}(2)_{\mathrm{L}}$, $\mathrm{SU}(3)_{\mathrm{R}}$ and $\mathrm{SU}(2)_{\mathrm{d}}$.}
\begin{eqnarray}
  \label{eq:8ofSO7branching}
  \mathbf{8}\;\stackrel{\mathrm{SO}(3)^\prime \times \mathrm{SO}(3)_{\mathrm{L}} \times \mathrm{SO}(3)_{\mathrm{R}} }{\longrightarrow} \;  ( \mathbf{2} , \mathbf{2}, \mathbf{1}) +  ( \mathbf{2} , \mathbf{1}, \mathbf{2}) 
   \;\stackrel{  \mathrm{SO}(3)_{\mathrm{d}} \times \mathrm{SO}(3)_{\mathrm{R}} }{\longrightarrow} \;
   ( \mathbf{1} , \mathbf{1} ) +    ( \mathbf{3} , \mathbf{1} ) +    ( \mathbf{2} , \mathbf{2} )   \; .
\end{eqnarray}
At the $\cN=1$ G$_2$--invariant AdS critical point, only the  $( \mathbf{1} , \mathbf{1} )$ gravitino remains massless, while all others pick up masses \cite{Guarino:2015qaa}. The full symmetry of this solution within the $D=4$ $\cN=8$ supergravity is $\textrm{OSp}(4|1) \times \textrm{G}_2$. At the $\cN=3$, SO(4)--invariant critical point, it is the $ ( \mathbf{3} , \mathbf{1} )$ gravitini that remain massless \cite{Guarino:2015qaa}. While the $\cN=3$ critical point is invariant under $\mathrm{SO}(4) \equiv \mathrm{SO}(3)_{\mathrm{d}} \times \mathrm{SO}(3)_{\mathrm{R}}$, the massless gravitini are only invariant under the second factor, and transform as a triplet under the first factor. The symmetry of the $\cN=3$ solution within the $\cN=8$ theory  is thus $\textrm{OSp}(4|3) \times \mathrm{SO}(3)_{\mathrm{R}}$, with $\mathrm{SO}(3)_{\mathrm{d}} \subset \textrm{OSp}(4|3)$ identified as the R-symmetry group.

These (super)symmetry groups are preserved by the ten-dimensional uplift, so the above considerations should allow us to identify the $G$-structures carried by the family of type IIA configurations (\ref{KKSO4sectorinIIA}). The $\mathbb{R}^7$ that furnishes the fundamental representation of the semisimple, SO(7), part of the $D=4$ gauge group is to be identified with the ambient space of the uplifting $S^6$. In other words, this SO(7) can be regarded as the generic structure group of the ambient $\mathbb{R}^7$, with the internal supersymmetry parameters transforming in the $\mathbf{8}$. On the G$_2$--invariant locus (\ref{eq:SO4toG2}), the type IIA configuration (\ref{KKSO4sectorinIIA}) is $\cN=1$. The G$_2$--invariant supersymmetry parameter corresponds, via (\ref{embedding_SO4_2}), to the  $( \mathbf{1} , \mathbf{1} )$ singlet in (\ref{eq:8ofSO7branching}). The structure of $\mathbb{R}^7$ gets reduced to G$_2$ (holonomy), which in turn descends into $S^6$ as a nearly-K\"ahler SU(3)-structure. 

Away from the G$_2$ locus, as in the solution (\ref{SO4SolN=3}), the IIA configuration  (\ref{KKSO4sectorinIIA}) is $\cN=3$. The supersymmetry parameter transforms under $\mathrm{SO}(3)_{\mathrm{d}} \times \mathrm{SO}(3)_{\mathrm{R}}$ as the $( \mathbf{3} , \mathbf{1} )$ in (\ref{eq:8ofSO7branching}).  The ambient $\mathbb{R}^7$ is thus equipped with an SU(2)--structure, with $\mathrm{SU}(2) \equiv \mathrm{SO}(3)_{\mathrm{R}}$ and R-symmetry $\mathrm{SO}(3)_{\mathrm{d}}$. Recall that an SU(2)--structure in seven dimensions is characterised by a real one-form and a real two-form, transforming as triplets of the R-symmetry group, see {\it e.g.}~\cite{DallAgata:2003txk}. Denoting the $\mathbb{R}^7$ coordinates by $x^I$, $I=1, \ldots, 7$, and splitting $I= (i, \hat{i})$, $i=1,2,3$, $\hat{i}=0,1,2,3$ as in appendix \ref{subset:S6Geom}, the one- and two-forms of our seven-dimensional $\mathrm{SO}(3)_{\mathrm{R}}$--structure can be identified as $dx^i$ and $\tfrac12 (J^i)_{\hat{i}\hat{j}} \, dx^{\hat{i}} \wedge dx^{\hat{j}} $, with $(J^i)_{\hat{i}\hat{j}}$ defined in (\ref{eq:QK4}). These indeed transform as triplets under the $\mathrm{SO}(3)_{\mathrm{d}}$ R-symmetry. The SU(2)-structure on $\mathbb{R}^7$ descends on $S^6$ as an identity structure. The latter is characterised by an $\mathrm{SO}(3)_{\mathrm{d}}$ triplet of scalars, of one-forms, and of two-forms, that can be constructed as spinor bilinears.

Rather than characterising the identity structure, we will directly contruct the $\mathrm{SO}(3)_{\mathrm{d}}$ triplet of Killing spinors, focusing on the $\cN=3$ solution (\ref{SO4SolN=3}) for definiteness. In principle, one would expect that the consistency of the uplift should determine the relevant Killing spinors from combinations of those of the round $S^6$. In practice, however, such formulae have never been worked out (although see {\it e.g.}~\cite{Nicolai:2011cy} for a discussion). It then turns out to be more efficient, though still a rather demanding exercise, to construct the Killing spinors by direct integration of the type IIA Killing spinor equations on the background (\ref{SO4SolN=3}). Here we give the end result and sketch the main steps to derive it. Further details can be found in appendices \ref{app:KillingSpinors} and \ref{app:doublets}.

Let $\hat \epsilon$ be the ten-dimensional Majorana supersymmetry parameter, let $\zeta^i_{\pm}$, $i=1,2,3$, be three of the chiral and antichiral Killing spinors of AdS$_4$, and let $\chi^i$ be an $\mathrm{SO}(3)_{\mathrm{d}}$ triplet of Dirac spinors on the internal six-dimensional geometry corresponding to the solution (\ref{SO4SolN=3}). We take 
\begin{equation} \label{eps10D}
\hat{\epsilon} = \zeta_{i+}  \otimes \chi^i+\zeta_{i -}  \otimes \chi^{ic} \; , 
\end{equation}
with the $\mathrm{SO}(3)_{\mathrm{d}}$ indices contracted, and raised and lowered with $\delta_{ij}$. The superscript $c$ denotes Majorana conjugation. The ten-dimensional spinor $\hat{\epsilon}$ given by (\ref{eps10D}) is manifestly Majorana, by the second relation in (\ref{eq:KSEAdS4}). We require that (\ref{eps10D}) annihilates the supersymmetry variations of the type IIA fermions. Using the AdS$_4$ Killing spinor equations (\ref{eq:KSEAdS4}) obeyed by $\zeta^i_{\pm}$, this turns out to be  equivalent to the following set of equations for $\chi^i$ and $\chi^{ic}$, defined on the six-dimensional internal geometry:
\begin{subequations}
\begin{align}
& e^{-\tA} \chi^{ic}+ \bigg[ d \tilde{\slashed{A} }+ \tfrac14 e^{\hat{\phi}} \left( \hat{F}_\0 +  \slashed{\hat{F}}_\2 \hat\gamma+  \slashed{\hat{G}}_\4 - i \hat{G}_\0\right)\bigg]\chi^i=0 \; ,\label{eq:6dSUSY1}\\[4pt]
&\bigg[ d \hat{\slashed{\phi}} + \tfrac{1}{2} \slashed{\hat{H}}_\3 \hat \gamma+ \tfrac14 e^{\hat{\phi}} \left(5  \hat{F}_\0 + 3  \slashed{\hat{F}}_\2  \hat\gamma+   \slashed{\hat{G}}_\4 +i \hat{G}_\0 \right)\bigg]\chi^i=0 \; ,\label{eq:6dSUSY2}\\[4pt]
&\bigg[\nabla_{\underline{M}} +\tfrac{1}{4} \slashed{\hat{H}}_{\underline{M}} \hat\gamma+ \tfrac18 e^{\tA+\hat{\phi}} \left( \hat{F}_\0 -  \slashed{\hat{F}}_\2  \hat\gamma+   \slashed{\hat{G}}_\4 + i \hat{G}_\0 \right)\gamma_{\underline{M}} \bigg]\chi^i=0\label{eq:6dSUSY3} \; .
\end{align}
\end{subequations}
Here, 
\begin{equation} \label{eq:stringframeWF}
e^{2\tA} = e^{\frac12 \phi_0} L^2 \, (3 + \cos 2\alpha)^{1/2} \; , 
\end{equation}
is the string frame, for convenience, warp factor of the solution (\ref{SO4SolN=3}), $\hat{\phi}$ the dilaton therein, $\hat{F}_\0$, $\hat{F}_\2$, $\hat{H}_\3$, the IIA field strengths, $\hat{G}_\4$ the internal component of $\hat{F}_\4$ and $\hat{G}_\0 \equiv 3\sqrt{2} \, L^{3} e^{-\frac{1}{4} \phi_0} \, e^{-4\tA}$. Also, $\gamma_{\underline{M}}$ are the six-dimensional gamma matrices, with $\underline{M} = 1, \ldots , 6$ tangent-space indices, $\hat{\gamma}$ is the six-dimensional chirality matrix, $\slashed{\hat{H}}_\3 \equiv \frac{1}{3!} \hat{H}_{\underline{M} \underline{N} \underline{P}} \, \gamma^{\underline{M} \underline{N} \underline{P}}$, and $\slashed{\hat{H}}_{\underline{M}} \equiv \frac{1}{2!} \hat{H}_{\underline{M} \underline{N} \underline{P}} \, \gamma^{ \underline{N} \underline{P}}$, etc., with $\gamma^{\underline{M_1} \ldots \underline{M_n} } \equiv \gamma^{[\underline{M_1} }  \cdots \gamma^{\underline{M_n]} } $. 

As argued above, the spinor $\chi^i$ must transform in the $( \mathbf{3} , \mathbf{1} )$ of the $\mathrm{SO}(4) \equiv \mathrm{SO}(3)_{\mathrm{d}} \times \mathrm{SO}(3)_{\mathrm{R}}$ symmetry group of the solution (\ref{SO4SolN=3}). As shown in appendix \ref{app:KillingSpinors}, the most general such spinor may be written as 
\begin{equation}\label{eq:6d triplets}
\chi^i = \tfrac{1}{2} \, e^{\frac{\tA}{2}} \, \bigg[\left(\begin{array}{c} f_{1+}\\ f_{1-}  \end{array}\right )\otimes\eta^i_1+\left(\begin{array}{c} f_{2+}\\ f_{2-}  \end{array}\right )\otimes\eta^i_2+\left(\begin{array}{c} f_{3+}\\ f_{3-}  \end{array}\right )\otimes\eta^i_3+\left(\begin{array}{c} f_{4+}\\ f_{4-}  \end{array}\right )\otimes\eta^i_4\bigg] \; .
\end{equation}
The factor of $\tfrac{1}{2} \, e^{\frac{\tA}{2}}$ is chosen for convenience, $f_{1 \pm}$, etc., are functions of $\alpha$, and $\eta^i_1 , \ldots ,  \eta^i_4$ are independent triplets of spinors on $S^2 \times S^3$ built as tensor products of the Killing spinors of $S^2$ and $S^3$. Specifically, let $\psi^\alpha$, $\alpha = 1,2$, be a doublet of spinors of $S^2$, constructed from the $S^2$ Killing spinors, and $\hat{\psi}^{\alpha}=(\sigma_2)^{\alpha}_{~\beta}\psi^{\beta c}$ for $\psi^{\alpha c}$ the Majorana conjugate of $\psi^{\alpha}$. The index $\alpha$ here labels the doublet of $\textrm{SO}(3)^\prime$ in (\ref{embedding_SO4}) which rotates $S^2$. Let $\xi^\alpha$, $\alpha =1,2$, be the two Killing spinors of $S^3$ that transform as a doublet under the $\mathrm{SO}(3)_{\mathrm{L}}$ in (\ref{embedding_SO4}) and are singlets under $\mathrm{SO}(3)_{\mathrm{R}}$. Then,
\begin{eqnarray} \label{eq:triplets}
\eta^i_1 \equiv (\sigma_2 \sigma_i)_{\alpha\beta} \, \psi^{\alpha}\otimes \xi^{\beta} \; , \qquad
\eta^i_2 \equiv  (\sigma_2 \sigma_i)_{\alpha\beta} \, \hat\psi^{ \alpha}\otimes \xi^{\beta} \; , \nonumber \\[4pt]
\eta^3_i \equiv \tmu_i (\sigma_2 )_{\alpha\beta} \, \psi^{\alpha}\otimes \xi^{\beta} \; , \qquad
\eta^4_i \equiv  \tmu_i(\sigma_2 )_{\alpha\beta} \, \hat\psi^{ \alpha}\otimes \xi^{\beta} \; ,
\end{eqnarray}
where $\sigma^i$ are the Pauli matrices and $\tmu_i$ are defined in (\ref{eq:S2}). The Pauli matrix $\sigma_2$ appears as the $\mathrm{SO}(3)_{\mathrm{d}}$ charge conjugation matrix.

Inserting $\chi^i$ given by (\ref{eq:6d triplets}), (\ref{eq:triplets}) and its Majorana conjugate $\chi^{ic}$ into the Killing spinor equations (\ref{eq:6dSUSY1})--(\ref{eq:6dSUSY3}), an involved calculation produces a(n overdetermined) system of algebraic relations among the functions $f_{1 \pm}$, etc., and a differential equation on the interval (\ref{anglealpha}) for a combination of them. The details are summarised in appendix \ref{app:KillingSpinors}. Significant further massaging allows us to bring the solution of this set of algebraic and differential equations into the form
\begin{align}\label{eq:fs}
f_{1+}&=-  i f_{2-} = \cos\left(\tfrac{\beta}{2}\right) e^{ i (\Psi_+-\frac{1}{2}\Theta)} \; ,\nonumber\\[4pt]
f_{2+}&=  i f_{1-} = \sin\left(\tfrac{\beta}{2}\right) e^{- i (\Psi_-+\frac{1}{2}\Theta)} \; ,\nonumber\\[4pt]
f_{3+}&=  i f_{4-} = \frac{\sqrt{2}\cos\alpha}{\sqrt{\cos^2\alpha+1}}\cos\left(\tfrac{\beta}{2}\right) e^{ i (\frac{\pi}{3}+\Psi_+-\frac{1}{2}\Theta)} \; ,\nonumber\\[4pt]
f_{4+}&=-  i f_{3-} = \frac{\sqrt{2}\cos\alpha}{\sqrt{\cos^2\alpha+1}}\sin\left(\tfrac{\beta}{2}\right) e^{ i (\frac{\pi}{3}-\Psi_--\frac{1}{2}\Theta)} \; , 
\end{align}
up to an arbitrary overall normalisation. We have defined the following functions of $\alpha$:
\begin{align} \label{eq:auxfuncs}
\tan\Theta& \equiv \sqrt{\frac{2}{3}}\frac{1}{\cos\alpha\sqrt{\cos^2\alpha+1}} \; , \nonumber\\[4pt]
\cos\beta& \equiv  \frac{\sqrt{2} \sin\alpha}{\sqrt{\cos^2\alpha+1}\sqrt{3\cos^4\alpha+3 \cos^2\alpha+ 2}} \; ,\nonumber\\[4pt]
\tan\Psi_{\pm}& \equiv  \pm\sqrt{\frac{2}{3}}\frac{1}{\cos\alpha\sqrt{\cos^2\alpha+1}}-\frac{\sqrt{3\cos^4\alpha+3 \cos^2\alpha+ 2}\sin\alpha}{\sqrt{3}\cos\alpha \sqrt{\cos^2\alpha+1}(\sqrt{2}\cos\alpha-\sqrt{\cos^2\alpha+1})} \; .
\end{align}
The $\mathrm{SO}(3)_{\mathrm{d}}$ triplet of $\mathrm{SO}(3)_{\mathrm{R}}$--invariant spinors $\chi^i$ given by  (\ref{eq:6d triplets}) with (\ref{eq:fs}), (\ref{eq:auxfuncs}) solve the Killing spinor equations (\ref{eq:6dSUSY1})--(\ref{eq:6dSUSY3}) on the $\cN=3$ solution (\ref{SO4SolN=3}) of massive type IIA supergravity. As an additional check, we have also verified that the three independent $\mathcal{N}=1$ pure spinors that follow from (\ref{eq:6d triplets}) with (\ref{eq:fs}), (\ref{eq:auxfuncs}) solve the pure spinor supersymmetry conditions for AdS$_4$ solutions of massive IIA supergravity given in \cite{Grana:2006kf}.

Equipped with the $\cN=3$ Killing spinors, we can proceed to compute the spinor bilinear forms and the torsion classes of the corresponding identity structure. Here we will only give the scalar bilinears. We expect one $\mathrm{SO}(3)_{\mathrm{d}}$ triplet of scalar bilinears, based on the fact that the six-dimensional identity structure is inherited from an $\mathrm{SO}(3)_{\mathrm{R}}$--structure on the ambient $\mathbb{R}^7$. Let us see how this scalar triplet arises from spinor bilinears. In principle, two such real or purely imaginary scalar bilinears can be constructed out of $\chi^i$, namely, $\chi^{i\dag}\chi^{j}$ and $\chi^{i\dag}\hat\gamma\chi^{j}$. Both of these sit in principle in the $\bm{3} \times \bm{3} \rightarrow \bm{1} + \bm{3} + \bm{5}$ of $\mathrm{SO}(3)_{\mathrm{d}}$. Direct computation from (\ref{eq:6d triplets}), (\ref{eq:fs}), (\ref{eq:auxfuncs}) shows that 
\begin{equation}\label{eq:norms}
\chi^{i\dag}\hat\gamma\chi^{j}=- i e^{\tA} \, \frac{ \sqrt{2}\cos\alpha}{\sqrt{\cos^2\alpha+1}} \, \epsilon^{ijk} \tmu_k  \; ,\qquad 
\chi^{i\dag}\chi^{j}= e^{\tA} \, \delta^{ij} \; .
\end{equation}
Thus, for both bilinears, the $\bm{5}$ components vanish identically. The first bilinear is the triplet argued above, and the second one is a singlet which, however, is not independent but is algebraically related to the former. 

Equation (\ref{eq:norms}) provides a further consistency check on our Killing spinors. It was shown in \cite{Grana:2006kf} that, for $\cN=1$ supersymmetric warped product solutions of massive IIA supergravity containing AdS$_4$, the $\cN=1$ internal Killing spinor $\chi$ must satisfy $\chi^{\dag}\hat\gamma\chi=0$ and $\chi^{\dag}\chi \propto e^{\tA}$, where $e^{2\tA}$ is the string frame warp factor. It is straightforward to see from (\ref{eq:norms}) with $i=j$ that each individual $\chi^i$, $i=1,2,3$, satisfies these $\cN=1$ conditions.

\section{Outlook}

In this paper we have studied an ${\cal N}=3$ solution of massive IIA supergravity first considered in \cite{Pang:2015vna,Pang:2015rwd}. We have described in detail the sector of ISO(7) supergravity with SO(4) invariance, which includes this solution as a point in its moduli space. This has allowed us to better understand its geometry. The solution consists of a fibration over an interval $I$ of a certain $S^2$-bundle $M_5$ over $S^3$, with the $S^2$ shrinking at one endpoint of the interval and the $S^3$ at the other, so that the full topology is that of an $S^6$ (as expected for vacua of the ISO(7) supergravity).  

Moreover, we have been able to obtain the spinorial parameters $\chi^i$, $i=1,2,3$ under which it is supersymmetric, thus confirming the expectation that it has ${\cal N}=3$ supersymmetry. This expectation was based on the amount of supersymmetry of the vacuum in the four-dimensional ISO(7) supergravity; but while uplift formulas are available for all physical fields, they are not for the supersymmetry parameters, and thus so far a full proof that the solution is ${\cal N}=3$ was lacking. 

Our results open the way to several possible developments. First of all, the structure of the spinorial parameters $\chi^i$ is not completely fixed by the SO(4) invariance. The solution has cohomogeneity one: the SO(4) orbits are copies of the $S^2$ bundle over $S^3$, and thus a priori the isometry group leaves several functions of the coordinate $\alpha$ on $I$ that appear in the $\chi^i$ undetermined. For the present solution these are fixed by the Killing spinor equations, but it is easy to set up a more general Ansatz where both these functions and those in the physical fields are allowed to vary, without breaking the SO(4) invariance and in particular ${\cal N}=3$ supersymmetry (whose R-symmetry is one of the SO(3) factors in the SO(4)).
   
Several arguments lead one to suspect the existence of more general ${\cal N}=3$ solutions in massive IIA. On $\mathbb{CP}^3$, such solutions are predicted to exist by holography \cite{gaiotto-t} and found \cite{gaiotto-t2} in first approximation in a regime where the Romans mass $\hat{F}_\0$ is small. Varying $\hat{F}_\0$ beyond this regime suggests the existence of a line of solutions. Since $\mathbb{CP}^3$ can be written as a foliation of copies of $T^{1,1}$, it is plausible that such solutions might be related to the ones we are considering here, and that thus there might be a line of deformations in this case, too. A possible analogy is offered by ${\cal N}=2$ solutions: in that case, a line of solutions exists \cite{Aharony:2010af} that connects the ${\cal N}=6$ massless solution on $\mathbb{CP}^3$ to an analogue of the solution in \cite{Guarino:2015jca} obtained by replacing $\mathbb{CP}^2$ with $\mathbb{CP}^1\times \mathbb{CP}^1$.

\section*{Acknowledgements}

We thank Alberto Zaffaroni for discussions. NTM is funded by the Italian Ministry of Education, Universities and Research under the Prin project ``Non Perturbative Aspects of Gauge Theories and Strings'' (2015MP2CX4) and INFN. AT is supported in part by INFN. OV is supported by NSF grant PHY-1720364 and, partially, by grant FPA2015-65480-P (MINECO/FEDER UE) from the Spanish Government.

\appendix

\addtocontents{toc}{\setcounter{tocdepth}{1}}


\section{\mbox{Geometric structures on $S^6$}} \label{subset:S6Geom}

In this appendix we describe the relevant geometric structures on $S^6$ that arise in the consistent truncation discussed in the main text. Let $\mu^I$, $I=1, \ldots, 7$, parametrise $S^6$ as the locus $\delta_{IJ} \mu^I \mu^J = 1$ in $\mathbb{R}^7$, and let $y^M$, $M=1, \ldots, 6$,  be the $S^6$ angles. It is convenient to split the $\mu^I$ according to the $ \textrm{SO}(3)^\prime \times \textrm{SO}(4)^\prime$ defined in (\ref{embedding_SO4}) as $\mu^I = (\mu^i , \mu^{\hat{i}})$,
\begin{eqnarray} \label{mus}
\mu^i = \cos \alpha \, \tilde{\mu}^i \; , \; i=1,2,3 \; ,  \qquad \mu^{\hat{i}} = -\sin \alpha \, \tilde{\mu}^{\hat{i}} \; , \; \hat{i}=0,1,2,3 \; ,
\end{eqnarray}
where $\alpha$ is one of the $y^M$, and is taken to have range (\ref{anglealpha}). In (\ref{mus}), $\tilde{\mu}^i$, $\tilde{\mu}^{\hat{i}}$, respectively parametrise an $S^2$ and an $S^3$ as $\delta_{ij} \tilde{\mu}^i \tilde{\mu}^j = 1$ (see equation (\ref{eq:S2})) and similarly for $\tilde{\mu}^{\hat{i}}$. For convenience, the index $\hat{i}$ ranges from 0 to 3 as indicated. In terms of these, the round, homogeneous Einstein metric on $S^6$, (E.2) of \cite{Guarino:2015vca}, is 
\begin{eqnarray} \label{RoundS6}
d\mathring{s}^2_6 = g^{-2} \Big( \cos^2 \alpha \, d\tilde{s}^2 (S^2) + d\alpha^2 + \sin^2 \alpha \, d\tilde{s}^2 (S^3)  \Big) \; ,
\end{eqnarray}
This metric has of course SO(7) isometry, although only the $ \textrm{SO}(3)^\prime \times \textrm{SO}(4)^\prime$ that rotates the $S^2$ and the $S^3$ is manifest. The local line element (\ref{RoundS6}) is adapted to the topological description of $S^6$ as the join of $S^2$ and $S^3$. 

For our calculation, we need to write the Killing vectors of $S^6$ adapted to the splitting (\ref{mus}). For this purpose,  it is useful to split the local index $M$ on $S^6$ as $M=(\alpha, a, \hat {a})$, where $a=1,2$ and $\hat{a}=1,2,3$ are local indices on $S^2$ and $S^3$, respectively. With the normalisation conventions of appendix E of \cite{Guarino:2015vca}, the non-vanshing components of the Killing vectors of the round metric (\ref{RoundS6}) are 
\begin{eqnarray} \label{KillingFoliation}
& 
K^{a}_{ij} = \tilde{K}^{a}_{ij} \; , \quad 
K^\alpha_{i\hat{i}} = -\tilde \mu_i \tmu_{\hat{i}} \; , \quad
K^{a}_{i\hat{i}} =  \tan \alpha \, \tmu_{\hat{a}} \,  \tilde{g}^{ab} \partial_b \tilde{\mu}_i \; , \nonumber \\[5pt] 
& K^{\hat{a}}_{i\hat{i}} =  -\cot \alpha \, \tmu_i \,  \tilde{g}^{\hat{a}\hat{b}} \partial_{\hat{b}} \tilde{\mu}_{\hat{i}} \; , \quad 
K^{\hat{a}}_{\hat{i}\hat{j}} = \tilde{K}^{\hat{a}}_{\hat{i}\hat{j}} \; ,
\end{eqnarray}
where $\tilde{g}^{ab}$ and $ \tilde{g}^{\hat{a}\hat{b}}$ are the round inverse metrics on $S^2$ and $S^3$ and $\tilde{K}^{a}_{ij}$,  $\tilde{K}^{\hat{a}}_{\hat{i}\hat{j}}$ their corresponding Killing vectors. The derivatives of the Killing vectors with respect to the $y^M$ angles on $S^6$ are
\begin{eqnarray} \label{KillingDerFoliation}
& K^{ij}_{\alpha a} = -4 g^{-2} \sin \alpha \cos \alpha \, \tilde{\mu}^{[i} \partial_{a} \tilde \mu^{j]} \; , \quad 
K^{ij}_{ab} = 4 g^{-2}  \cos^2\alpha \, \partial_{a} \tilde{\mu}^{[i} \partial_{b} \tilde \mu^{j]} \; , \nonumber \\[5pt]
& K^{i\hat{i}}_{\alpha a} = 2 g^{-2} \cos^2 \alpha \, \tmu^{\hat{i}} \, \partial_{a} \tilde{\mu}^i \; ,  \;
K^{i\hat{i}}_{\alpha \hat{a}} = 2 g^{-2} \sin^2 \alpha \, \tmu^i \, \partial_{\hat{a}} \tilde{\mu}^{\hat{i}} \; ,   \;\nonumber \\[5pt]
&
K^{i\hat{i}}_{a \hat{a}} = -2 g^{-2} \sin \alpha \cos \alpha \, \partial_a \tmu^i \, \partial_{\hat{a}} \tilde{\mu}^{\hat{i}} \; , \nonumber \\[5pt]
& K^{\hat{i}\hat{j}}_{\alpha \hat{a}} = 4 g^{-2} \sin \alpha \cos \alpha \, \tilde{\mu}^{[{\hat{i}}} \partial_{\hat{a}} \tilde \mu^{\hat{j}]} \; , \quad 
K^{ij}_{\hat{a}\hat{b}} = 4 g^{-2}  \sin^2\alpha \, \partial_{\hat{a}} \tilde{\mu}^{[i} \partial_{\hat{b}} \tilde \mu^{j]} \; .
\end{eqnarray}
In these expressions, $\partial_M \mu^I$ means derivative of $\mu^I = \mu^I(y^M)$ with respect to $y^M$, for $M=(\alpha, a,\hat{a})$.

With these ingredients, we can calculate the consistent embedding of the $D=4$ supergravity of section \ref{sec:SO(4)-sector} into type IIA supergravity using the uplift formulae of \cite{Guarino:2015jca,Guarino:2015vca}. The ten-dimensional embedding of the tensor hierarchy forms in (\ref{fieldContentHierarchy}) proceeds uneventfully. The embedding of the scalars is much more laborious. This is achieved by bringing the SO(4)--invariant scalar matrix ${\cal M}_{\mathbb{MN}}$ of the $D=4$ supergravity, given in appendix D.3 of \cite{Guarino:2015qaa}, to the uplifting formulae (10) of \cite{Guarino:2015jca}. Manipulating these formulae with the help of (\ref{KillingFoliation}), (\ref{KillingDerFoliation}), it turns out that the Killing vectors $K^{\hat{a}}_{\hat{i}\hat{j}}$ on $S^3$ and their derivatives always appear projected as $(J^k)^{\hat{i}\hat{j}} \, \tilde{K}^{\hat{a}}_{\hat{i}\hat{j}}$, where $(J^i)_{\hat{i}\hat{j}}$ are the components of the triplet of constant $ 4 \times 4$ matrices\footnote{Indices $i=1,2,3$ and $\hat{i}=0,1,2,3$ here correspond to indices $a=2,4,6$ and $\lambda = 1,3,5,7$ in appendix D.3 of \cite{Guarino:2015qaa}. The $J$ matrices here are the negative of the gamma matrices there: $(J^i)^{\hat{i}}{}_{\hat{j}} \textrm{``$=$"} -(\gamma^a)^\lambda{}_\mu$.}
\begin{eqnarray} \label{eq:QK4}
J^ i = e^{0i} - \tfrac12 \epsilon^i{}_{jk}\,  e^{jk} \; .
\end{eqnarray}
Here, we have written $J^i$ in terms of the set of six $ 4 \times 4$ matrices $e^{\hat{i}\hat{j}} = - e^{\hat{j}\hat{i}} $ with components $(e^{\hat{i}\hat{j}})_{\hat{k}\hat{l}} = 2\delta^{\hat{i}}_{[\hat{k}} \delta^{\hat{j}}_{\hat{l}]} $, so that $(J^i)_{0j} = \delta^i_j$ and $(J^i)_{jk} = -\epsilon^i{}_{jk} $. Indices $i$ and $\hat{i}$ are raised and lowered with the SO$(3)^\prime$ and SO$(4)^\prime$ invariant metrics $\delta_{ij}$ and $\delta_{\hat{i}\hat{j}}$, respectively. The $J^i$ are antisymmetric, $(J^i)_{\hat{i}\hat{j}} = -(J^i)_{\hat{j}\hat{i}}$, anti-selfdual,  
\begin{eqnarray} \label{antiSD}
(J^i)_{\hat{i}\hat{j}} = -\tfrac12 \epsilon_{\hat{i}\hat{j}\hat{k}\hat{l}} \, (J^i)^{\hat{k}\hat{l}} \; ,
\end{eqnarray}
satisfy the quaternion algebra,
\begin{eqnarray} \label{quaternion}
(J^i)^{\hat{i}}{}_{\hat{k}} (J^j)^{\hat{k}}{}_{\hat{j}}  = -\delta^{ij} \delta^{\hat{i}}_{\hat{j}} + \epsilon^{ij}{}_k (J^k)^{\hat{i}}{}_{\hat{j}} \; ,
\end{eqnarray}
and the identity
\begin{eqnarray} \label{QuadID}
(J^i)_{\hat{i}\hat{j}} (J_i)_{\hat{k}\hat{l}}  = 2 \delta_{\hat{i}[\hat{k}}\delta_{\hat{l}]\hat{j}} - \epsilon_{{\hat{i}}{\hat{j}}\hat{k}\hat{l}} \; . 
\end{eqnarray}
Specifically, the combinations $(J^i)^{\hat{i}\hat{j}} \, \tilde{K}^{\hat{a}}_{\hat{i}\hat{j}}$ select the Killing vectors of $S^3$ that are invariant under the SO$(3)_{\textrm{R}}$ in (\ref{embedding_SO4}), namely, the right-invariant Killing vectors on $S^3$. Lowering the index with the round $S^3$ metric $\tilde{g}_{\hat{a}\hat{b}}$, we have
\begin{eqnarray} \label{RIOF}
 (J^i)_{\hat{i}\hat{j}} \, \tilde{K}_{\hat{a}}^{\hat{i}\hat{j}} = 
2 \,  (J^i)_{\hat{i}\hat{j}} \, \tmu^{\hat{i}} \partial_{\hat{a}} \tmu^{\hat{j}} = 
\rho^i_{\hat{a}} \; ,
\end{eqnarray}
where $\rho^i_{\hat{a}}$ is the $\hat{a}$-th component of the right-invariant one-form $\rho^i$, $i=1,2,3$. These close into the Maurer-Cartan equations (\ref{eq:MC}), which are invariant under SO$(3)_{\textrm{R}}$ and lie in the adjoint of both $\textrm{SO}(3)_{\textrm{L}}$ and $\textrm{SO}(3)_{\textrm{d}}$ defined in (\ref{embedding_SO4}). 

Equations (\ref{antiSD})--(\ref{RIOF}) need to be used extensively to bring the raw consistent truncation expressions obtained from the formulae in \cite{Guarino:2015jca,Guarino:2015vca} to the final form (\ref{KKSO4sectorinIIA}) presented in the main text. Other useful identities for this purpose include 
\begin{eqnarray}
&& \epsilon_{\hat{i}\hat{j}\hat{k}\hat{l}} \, \tmu^{\hat{i}} d\tmu^{\hat{j}} \wedge d\tmu^{\hat{k}} \wedge d\tmu^{\hat{l}} = \tfrac18 \, \epsilon_{ijk} \rho^i \wedge \rho^j \wedge \rho^k = \tfrac38 \, \epsilon_{ijk} \tmu^k \tmu_h \, \rho^h \wedge \rho^i \wedge \rho^j = 6  \, \textrm{vol} (S^3) \; , \nonumber\\[6pt]
&& \epsilon_{i j k} \,  \tmu^{i} \, d\tmu^{j} \wedge d\tmu^{k}  = 2  \, \textrm{vol}(S^2) \; , \qquad 
\tilde{*}_2 \, d\tmu^i = \epsilon^i{}_{jk} \, \tmu^j d\tmu^k \; , \qquad 
\tilde{*}_3 \, \rho^i = \tfrac14 \epsilon^i{}_{jk} \, \rho^j \wedge \rho^k \; , \nonumber \\[6pt]
&&  \epsilon_{ijk} \,  \tmu^i \tmu_h \, d\tmu^j \wedge \rho^k \wedge \rho^h = \tfrac12 \, \epsilon_{ijk} \, d\tmu^i \wedge \rho^j \wedge \rho^k \; , \nonumber \\[6pt]
&& \epsilon_{ijk} \,  \tmu^i \tmu_h \, d\tmu^j \wedge d\tmu^k \wedge \rho^h = \epsilon_{ijk} \, d\tmu^i \wedge d\tmu^j \wedge \rho^k \; , \nonumber \\[6pt]
&& \epsilon_{i_1 i_2 i_3} \,  \tmu^{i_1} \, d\tmu^{i_2} \wedge d\tmu^{i_3} \wedge \epsilon_{j_1j_2j_3} \,  \tmu^{j_1} \, \rho^{j_2} \wedge \rho^{j_3} = 2  \, d\tmu_i \wedge d\tmu_j \wedge \rho^i \wedge \rho^j \; ,
\end{eqnarray}
where $\textrm{vol}(S^2)$ and $\textrm{vol}(S^3)$ are the volume forms corresponding to the metrics $d\tilde{s}^2 (S^2) $ and $d\tilde{s}^2 (S^3) $ in (\ref{RoundS6}), and $\tilde{*}_2$ and $\tilde{*}_3$ the corresponding Hodge dual operators. Note that
\begin{eqnarray}
d\tilde{s}^2 (S^2)= \delta_{ij} \, d\tmu^i \, d \tmu^j \; , \qquad 
d\tilde{s}^2 (S^3)= \delta_{\hat{i}\hat{j}} \, d\tmu^{\hat{i}} \, d \tmu^{\hat{j}}  = \tfrac14 \, \delta_{ij} \, \rho^i \rho^j \; .
\end{eqnarray}

We conclude by retrieving the homogeneous G$_2$--invariant nearly-K\"ahler structure on $S^6$ from this formalism. The nearly-K\"ahler forms are given in general by 
\begin{eqnarray} \label{JOmegaintermsofmu}
{\cal J} = \tfrac12 \, \psi_{IJK} \,  \mu^I d\mu^J \wedge d\mu^K \; , \quad 
  \Upomega  = \tfrac16 \left( \psi_{JKL} -i \, \tilde{\psi}_{IJKL} \, \mu^I \right)  d \mu^J \wedge d\mu^K \wedge d\mu^L \; , 
\end{eqnarray}
in terms of the constrained $\mu^I$ that parametrise $S^6$ and the associative and co-associative forms $\psi$ and $\tilde \psi$, on the ambient $\mathbb{R}^7$, see {\it e.g.}~appendix E of \cite{Guarino:2015vca}. It turns out that the non-vanishing components of these forms can be written for the case at hand as 
\begin{eqnarray} \label{eqapp:compsR7}
\psi_{ijk} = -\epsilon_{ijk} \; , \qquad
\psi_{i\hat{i}\hat{j}} = - (J_i)_{\hat{i}\hat{j}} \; , \qquad
\tilde \psi_{ij\hat{i}\hat{j}} = \epsilon_{ijk} (J^k)_{\hat{i}\hat{j}} \; , \qquad
\tilde \psi_{\hat{i}\hat{j}\hat{k}\hat{l}} = -\epsilon_{\hat{i}\hat{j}\hat{k}\hat{l}} \; , \qquad
\end{eqnarray}
Bringing (\ref{mus}), (\ref{eqapp:compsR7}) to (\ref{JOmegaintermsofmu}), we find
\begin{eqnarray} \label{NKintermsofmus}
 {\cal J} & =&  -\tfrac12   \cos^3 \alpha \, \epsilon_{ijk} \,  \tmu^i  d \tmu^j \wedge d \tmu^k -\tfrac12  \sin \alpha \, d\alpha \wedge \tmu_i \, \rho^i 
 \nonumber \\
&&  +\tfrac18  \sin^2 \alpha \cos \alpha   \, \epsilon_{ijk} \,  \tmu^i \rho^j \wedge \rho^k 
+\tfrac12  \sin^2 \alpha \cos \alpha \, d \tmu_i \wedge \rho^i 
  \, ,  \nonumber \\[12pt] 
 \textrm{Re} \, \Upomega &=& \tfrac12    \sin \alpha  \cos^2 \alpha \, d\alpha \wedge \epsilon_{ijk} \, \tmu^i  d \tmu^j \wedge d \tmu^k 
  +\tfrac12 \sin \alpha  \cos^2 \alpha   \, d\alpha \wedge d \tmu_i \wedge  \rho^i  \nonumber \\[5pt]
&& +\tfrac18   \sin^2 \alpha  \cos \alpha  \, \epsilon_{ijk} \, d \tmu^i \wedge \rho^j \wedge \rho^k  %
 -\tfrac18 \sin^3 \alpha  \, d\alpha \wedge  \epsilon_{ijk} \, \tmu^i \rho^j \wedge \rho^k   \; ,   \nonumber \\[12pt]
 \textrm{Im} \, \Upomega & = & \tfrac12 \, \sin \alpha \cos \alpha \, d\alpha \wedge \epsilon_{ijk} \,  \tmu^i d \tmu^j \wedge \rho^k  
+\tfrac14 \sin^2 \alpha \cos^2 \alpha  \,\tmu_i  d \tmu_j \wedge \rho^i \wedge \rho^j  \nonumber \\[5pt]
&&  -\tfrac14 \sin^2 \alpha \cos^2 \alpha  \, \epsilon_{ijk} \, d \tmu^i \wedge d \tmu^j \wedge \rho^k
 -\tfrac{1}{48} \sin^4 \alpha  \, \epsilon_{ijk} \,  \rho^i \wedge \rho^j \wedge \rho^k \;   \; .
\end{eqnarray}
These forms can indeed be checked to satisfy the nearly-K\"ahler relations 
\begin{eqnarray} \label{SU3str}
\Upomega \wedge \bar \Upomega = -\tfrac{4i}{3} {\cal J} \wedge {\cal J} \wedge {\cal J} \neq 0 \; , \quad {\cal J} \wedge \Upomega = 0  \ ,
\end{eqnarray}
and
\begin{eqnarray} \label{SU3strDif}
d {\cal J} = 3 \,  \textrm{Re} \,  \Upomega   \; , \quad d \, \textrm{Im} \, \Upomega =  -2 \,  {\cal J} \wedge {\cal J} \ . 
\end{eqnarray}
The expressions (\ref{NKintermsofmus}) are needed to show that the SO(4)--invariant consistent truncation formulae of section \ref{subsec:SU3UpliftSubsec} reduce to the G$_2$--invariant formulae of section \ref{subsec:G2fromSO4}.

\section{Flux quantisation and free energies} \label{subsec:FluxQuant}

The fluxes corresponding to the generic configuration (\ref{KKSO4sectorinIIA}) can be appropriately quantised. The quantisation conditions are
\begin{eqnarray} \label{quantFlux}
&& k = 2\pi \ell_s  \,  \hat F_\0 \equiv 2\pi \ell_s  \,  m  \; , \nonumber \\[5pt]
&& N = -\frac{1}{(2\pi \ell_s )^5 }  \int_{S^6}  e^{\frac12 \hat \phi} \   \hat{*} \hat F_\4 + \hat B_\2 \wedge d\hat A_\3 + \frac16 m \,  \hat B_\2 \wedge \hat B_\2 \wedge \hat B_\2  \; , 
\end{eqnarray}
with $k$ and $N$ integers and $\ell_s$ the string length. The fields $A_\3$, $\hat B_\2$ and $\hat \phi$ have been given in (\ref{KKSO4sectorinIIA}), and from $\hat{*} \hat F_\4$, only the contribution corresponding to the Freund--Rubin term (\ref{USO4}) is relevant. The first relation in (\ref{quantFlux}) corresponds to the relation between quantised Romans mass and $D=4$ magnetic coupling \cite{Guarino:2015jca}. The second expression can be written, integrating in the range (\ref{anglealpha}) for the angle $\alpha$, as a relation between the integer $N$ and the electric $D=4$ coupling constant $g$,
\begin{eqnarray} \label{gmNk}
N = 5 \, v(S^6) \,  (2\pi \ell_s)^{-5} \, g^{-5}  \; ,
\end{eqnarray}
with $v(S^6) = \tfrac{16}{15} \, \pi^3$ the volume of the unit radius round six-sphere. The expression (\ref{gmNk}) coincides with that given in \cite{Varela:2015uca,Guarino:2016ynd} from the embedding of the SU(3)--invariant sector. The fact that the present SO(4)--invariant calculation here and the SU(3)--invariant calculation \cite{Varela:2015uca,Guarino:2016ynd} agree provides a selfconsistency check, as the relations (\ref{quantFlux}), (\ref{gmNk}) must characterise the theory, not merely particular subsectors or solutions. 

Equipped with these values of the fluxes, we can proceed to the calculation of the gravitational free energy of the configuration (\ref{KKSO4sectorinIIA}), along the lines of \cite{Guarino:2016ynd}. For this calculation, we first redefine the external $D=4$ metric with the inverse scalar potential $V$ in (\ref{VSO4}) so that the Einstein frame warp factor reads
\begin{eqnarray} \label{warping}
e^{2A} = -6\, e^{\frac18 \varphi} X^{1/4} \Delta_1^{1/8} \Delta_3^{1/4}  \, V^{-1} \; .
\end{eqnarray}
The free energy $F$ is proportional to the inverse of the effective four-dimensional Newton's constant \cite{Emparan:1999pm}. On the geometry  (\ref{KKSO4sectorinIIA}), (\ref{warping}), this evaluates to
\begin{eqnarray} \label{FreeEnergy1}
F= \frac{16 \pi^3}{(2\pi \ell_s )^8 }  \int_{S^6} e^{8A} \,  \textrm{vol}_6 = -  \frac{96 \pi^3}{(2\pi \ell_s )^8 } \, g^{-6} \, v(S^6) \, V^{-1} \, .
\end{eqnarray}
This again reproduces the expression given in \cite{Guarino:2016ynd}. In order to write the free energy in terms of quantised fluxes, we factorise the scalar potential as $V = g^2 (m/g)^{-1/3} \,  \tilde{V}$, where $V$ is the $g=m=1$ scalar potential (\ref{VSO4}), and then replace $g$ and $m$ by their values (\ref{quantFlux}), (\ref{gmNk}) in terms of $N$ and $k$. We finally obtain
\begin{equation} \label{FreeEnergy2}
F = -96 \cdot 5^{-5/3} \,  \pi^3 \, v(S^6)^{-2/3} \, \tilde{V}^{-1} \, N^{5/3} k^{1/3} \; ,
\end{equation} 
as in \cite{Guarino:2016ynd}.

At the critical points of the scalar potential (\ref{VSO4}), recorded in table \ref{Table:SO4Points}, the free energy (\ref{FreeEnergy2}) reduces to values that have been previously given in the literature. For the critical points with at least G$_2$ symmetry, (\ref{FreeEnergy2}) reproduces the corresponding values given in table 1 of \cite{Varela:2015uca}. For the $\cN=3$ point, (\ref{FreeEnergy2}) gives
\begin{equation} \label{N=3FreeEnergy}
F = \tfrac{1}{40} \, 3^{13/6} \, \pi \, N^{5/3} k^{1/3} \; ,
\end{equation}
in agreement with \cite{Pang:2015rwd}. This result also matches the field theory result, equation (8.4) of \cite{Jafferis:2011zi}, after correcting a typo there\footnote{There is a factor of $1/2$ missing in the r.h.s.~of (8.4) of \cite{Jafferis:2011zi}. We thank D.~Jafferis for confirming this.}.

The formula (\ref{FreeEnergy2}) holds for geometries with running $D=4$ scalars, not only for scalars frozen at critical points of the $D=4$ potential. It was conjectured in \cite{Guarino:2016ynd}, based on an SU(3)--invariant calculation, that the free energy (\ref{FreeEnergy2}) should further hold at any point of the 70-dimensional coset space E$_{7(7)}/\textrm{SU}(8)$ of the full $D=4$ $\cN=8$ dyonically-gauged ISO(7) supergravity, where  $\tilde{V}$ given by the full $g=m=1$, $\cN=8$ potential, normalised as in \cite{Guarino:2015qaa}. The present SO(4)--invariant calculation provides further evidence in favour of this conjecture.

\section{SO(4)--invariant AdS$_4$ solutions of massive type IIA} \label{SO4InvAdS4Sols}

The type IIA solutions that we have presented in the main text have the local form 
{\setlength\arraycolsep{0pt}
\begin{eqnarray} \label{SO4IIAconfig}
&& d\hat{s}_{10}^2 = e^{2X(\alpha)} ds^2 (\textrm{AdS}_4)  + e^{2B(\alpha)} \, \delta_{ij}    D \tilde{\mu}^i D \tilde{\mu}^j  + e^{2A(\alpha)} d\alpha^2  + e^{2C(\alpha)} \, d\tilde{s}^2 (S^3)  \; , 
 \qquad \hat \phi = \phi(\alpha) \; ,  \nonumber \\[10pt]
&& \hat F_\4 = \mu_0 \textrm{vol}_4 
+C_1(\alpha)  \, d\alpha \wedge \epsilon_{ijk} \, D \tmu^i \wedge  D \tmu^j \wedge \rho^k 
+C_2(\alpha)    \, D \tmu_i \wedge  D \tmu_j \wedge \rho^i \wedge \rho^j   \nonumber \\[5pt]
&&\qquad \; +C_3(\alpha)  \, d\alpha \wedge \tmu_i \,  D \tmu_j \wedge \rho^i \wedge \rho^j   
+C_4(\alpha)    \, d\alpha \wedge  \epsilon_{ijk} \,  \rho^i \wedge \rho^j \wedge \rho^k \, ,  
\\[10pt] 
&& \hat H_\3 =  B_1(\alpha)   \, d\alpha \wedge \epsilon_{ijk} \, \tmu^i  D \tmu^j \wedge D \tmu^k 
 +B_2(\alpha)  \, d\alpha \wedge D \tmu_i \wedge  \rho^i  +B_3(\alpha)    \, \epsilon_{ijk} \, D \tmu^i \wedge \rho^j \wedge \rho^k  \nonumber \\[5pt]
&& \qquad \;  + B_4 (\alpha)   \, d\alpha \wedge  \epsilon_{ijk} \, \tmu^i \rho^j \wedge \rho^k   \; ,   \nonumber \\[10pt]
&& \hat F_\2 =  A_1 (\alpha)  \, \epsilon_{ijk} \, \tmu^i  D \tmu^j \wedge D \tmu^k 
+A_2 (\alpha)   \, D \tmu_i \wedge \rho^i 
+A_3 (\alpha)    \, d\alpha \wedge \tmu_i \, \rho^i 
+A_4 (\alpha)   \, \epsilon_{ijk} \, \tmu^i  \rho^j \wedge \rho^k \; , \nonumber 
\end{eqnarray}
}where AdS$_4$ and $S^3$ are unit radius, $\mu_0$ is a constant, $X(\alpha)$, etc., are functions of the angle $\alpha$, the $\tmu^i$, $i=1,2,3$, are constrained coordinates that define a unit radius $S^2$ through (\ref{eq:S2}) and $\rho^i$ are the right-invariant forms on $S^3$, subject to (\ref{eq:MC}). The corresponding potentials are of the form
{\setlength\arraycolsep{2pt}
\begin{eqnarray} \label{SO4IIAconfigPotentials}
\hat{A}_\3 &=& c_1 (\alpha) \, d\alpha \wedge   \epsilon_{ijk} \,  \tmu^i D \tmu^j \wedge \rho^k 
 + c_2 (\alpha)    \, \epsilon_{ijk} \, D \tmu^i \wedge D \tmu^j \wedge \rho^k 
 +c_3 (\alpha)   \,\tmu_i  D \tmu_j \wedge \rho^i \wedge \rho^j   \nonumber \\[5pt]
&& +c_4 (\alpha)     \, \epsilon_{ijk} \,  \rho^i \wedge \rho^j \wedge \rho^k \; , \nonumber \\[10pt]
\hat{B}_\2 &=& b_1 (\alpha)  \, d\alpha \wedge \tmu_i \, \rho^i 
+ b_2 (\alpha)    \, \epsilon_{ijk} \,  \tmu^i  D \tmu^j \wedge D \tmu^k 
+ b_3 (\alpha)  \, D \tmu_i \wedge \rho^i  
+ b_4 (\alpha)    \, \epsilon_{ijk} \,  \tmu^i \rho^j \wedge \rho^k \; , \nonumber \\[10pt]
\hat{A}_\1 &=& a_1 (\alpha)   \, \tmu_i \, \rho^i \; . 
\end{eqnarray}
}In this appendix, we will denote the covariant derivative of $\tmu^i$ as 
\begin{eqnarray} \label{covderapp}
D \tilde{\mu}^i =  d\tilde{\mu}^i + \epsilon^i{}_{jk} {\cal A}^j \tilde{\mu}^k \; , \qquad  \textrm{with} \qquad 
{\cal A}^i =  A_0 (\alpha)  \, \rho^i \; .
\end{eqnarray}
The function $A_0 (\alpha)$, as well as all other functions of $\alpha$ in (\ref{SO4IIAconfig}), (\ref{SO4IIAconfigPotentials}) can be read off from the concrete expressions given in the main text.

The configuration (\ref{SO4IIAconfig})--(\ref{covderapp}) preserves the SO(4) subgroup of SO(7) defined in (\ref{embedding_SO4}). We will now work out the differential and algebraic equations that the functions $X(\alpha)$, etc., must obey for (\ref{SO4IIAconfig}) to solve the Bianchi identities and equations of motion of massive type IIA supergravity. Firstly, the functions that specify the potentials (\ref{SO4IIAconfigPotentials}) can be related to those entering the field strengths (\ref{SO4IIAconfig}) using the corresponding definitions (see (A.3) of \cite{Guarino:2015vca}). We find 
\begin{equation} \label{F2=dA1}
A_1 = m b_2 \; , \qquad 
A_2 = a_1 + m b_3 \; , \qquad 
A_3 = a_1^\prime + m  b_1 \; , \qquad 
A_4 =  -\tfrac12 (1-2A_0) a_1 + m  b_4 \; ,  
\end{equation}
and
\begin{eqnarray} \label{H3=dB2}
&& B_1 = b_2^\prime \; , \nonumber \\[4pt]
&& B_2 = b_3^\prime - b_1 -2 A_0^\prime b_2  \; ,   \\[4pt]
&& B_3 = b_4  +\tfrac12 (1-2A_0) b_3  - (1-A_0) A_0  b_2   \; , \nonumber  \\[4pt]
&& B_4 = b_4^\prime  -A_0^\prime b_3  +\tfrac12 (1-2A_0) b_1 \; , \nonumber
\end{eqnarray}
and
\begin{eqnarray} \label{F4=dA3}
&& C_1 = c_2^\prime -c_1 -a_1 b_2^\prime + m b_1 b_2 \; , \nonumber \\[4pt]
&& C_2 = c_3 -(1-2A_0) c_2  + 2 m b_2 b_4  - \tfrac12 m b_3^2  \; ,  \\[4pt]
&& C_3 = c_3^\prime + 2 c_2 A_0^\prime  -(1-2A_0) c_1 +a_1 \big(  b_3^\prime - b_1 -2 A_0^\prime b_2 \big)   - m b_1 b_3  \; , \nonumber  \\[4pt]
&& C_4 = c_4^\prime + \tfrac13 c_3 A_0^\prime  + \tfrac13 (1- A_0) A_0 c_1  -\tfrac13 a_1 \big(  b_4^\prime  -A_0^\prime b_3  +\tfrac12 (1-2A_0) b_1  \big)   +\tfrac13  m b_1 b_4   \; . \nonumber
\end{eqnarray}
We have dropped the explicit $\alpha$ dependence and have denoted with a prime the derivative with respect to it. We used the expressions (\ref{F2=dA1})--(\ref{F4=dA3}) to construct the constant scalar field strengths (\ref{KKfieldstrengths}) from the potentials (\ref{KKSO4sectorinIIA}).

Moving to the field equations (see (A.3), (A.5) of \cite{Guarino:2015vca}), some calculation shows that the type IIA Bianchi identities impose the relations
{\setlength\arraycolsep{0pt}
\begin{eqnarray} \label{SO4IIABianchis}
&& C_2^\prime -C_3 + C_1 -2 A_0 C_1 -2 A_1 B_4 -2 A_4 B_1 +A_2 B_2  =0 \;, \nonumber \\[5pt]
&& B_3^\prime - \tfrac12 (1 - 2 A_0 ) B_2  - B_4 + (1 - A_0 ) A_0  B_1  =0  \; ,  \nonumber \\[4pt]
&& A_1^\prime - m B_1 =0  \; ,  \nonumber \\[4pt]
&& A_2^\prime -A_3 -2 A_1 A_0^\prime  - m B_2  =0  \; ,  \nonumber \\[4pt]
&& A_4^\prime - A_2 A_0^\prime + \tfrac12 (1 - 2 A_0 ) A_3 - m B_4  =0  \; ,  \nonumber \\[4pt]
&& A_4 + \tfrac12 (1 - 2 A_0 ) A_2 - (1 - A_0 ) A_0  A_1 - m B_3  =0  \; .
\end{eqnarray}
}Next, a long calculation shows that the $\hat F_\4$ equation of motion gives
{\setlength\arraycolsep{0pt}
\begin{eqnarray} \label{SO4IIAF4eom}
&&  \big( e^{ \frac12 \phi +4X-A-2B+C }  C_1 \big)^\prime  +4  \, e^{ \frac12 \phi +4X + A -2B -C }  C_2 - 8  \, e^{ \frac12 \phi +4X + A -2B-C }  A_0 C_2  \nonumber \\
&& \qquad \qquad  - 4  \, e^{ \frac12 \phi +4X - A -C }  C_3 A_0^\prime + 2 \mu_0 B_4 =0 \;, \nonumber \\[10pt]
&&   e^{ \frac12 \phi +4X-A-2B+C }  C_1  +2  \, e^{ \frac12 \phi +4X - A -C }  (1-2A_0) C_3 - 48  \, e^{ \frac12 \phi +4X - A +2B- 3C }  (1-A_0 ) A_0 C_4  \nonumber \\
&& \qquad \qquad  + 2 \mu_0 B_3 =0 \;, \nonumber \\[10pt]
&&  \big( e^{ \frac12 \phi +4X-A-C }  C_3 \big)^\prime  -2  \, e^{ \frac12 \phi +4X + A -2B -C }  C_2 -24  \, e^{ \frac12 \phi +4X - A  + 2B -3C }  C_4 A_0^\prime + \tfrac12  \mu_0 B_2 =0 \;, \nonumber \\[10pt]
&&  \big( e^{ \frac12 \phi +4X-A +2B-3C }  C_4 \big)^\prime  + \tfrac{1}{24} \,   \mu_0 B_1 =0 \;,
\end{eqnarray}
}the $\hat H_\3$ equation of motion gives
{\setlength\arraycolsep{0pt}
\begin{eqnarray} \label{SO4IIAH3eom}
&&  \big( e^{ - \phi +4X-A-2B+3C }  B_1 \big)^\prime +4 e^{ - \phi +4X-A+ C }  B_2 A_0^\prime + 32  e^{ - \phi +4X +A- C }  (1-A_0) A_0  B_3  \nonumber \\
&& \qquad \qquad  -4 e^{ \frac12 \phi +4X - A -2B+C }  A_3 C_1 -32  e^{ \frac12 \phi +4X + A -2B  -C }  A_4 C_2   - m  \, e^{ \frac32 \phi +4X + A -2B +3C } A_1  \nonumber \\
&& \qquad \qquad - 24 \mu_0 C_4 =0 \;, \nonumber \\[12pt]
&&  \big( e^{ - \phi +4X-A+C }  B_2 \big)^\prime +8 e^{ - \phi +4X-A+2B- C }  B_4 A_0^\prime - 8  e^{ - \phi +4X +A- C }  (1-2A_0) B_3  \nonumber \\
&& \qquad \qquad  +8 e^{ \frac12 \phi +4X + A -2B-C }  A_2 C_2 +4  e^{ \frac12 \phi +4X - A -C }  A_3 C_3   - m  \, e^{ \frac32 \phi +4X + A +C } A_2    \nonumber \\
&& \qquad \qquad - 2 \mu_0 C_3 =0 \;, \nonumber \\[12pt]
&&  e^{ - \phi +4X-A+C }  B_2  - 4  e^{ - \phi +4X -A +2B- C }  (1-2A_0)  B_4    -2 e^{ \frac12 \phi +4X - A -2B+C }  A_1 C_1  \nonumber \\
&& \qquad \qquad +4  e^{ \frac12 \phi +4X - A -C }  A_2 C_3 -96  e^{ \frac12 \phi +4X - A +2B-3C }  A_4 C_4   - \tfrac12  m  \, e^{ \frac32 \phi +4X - A +2B +C } A_3 \nonumber \\
&& \qquad \qquad  - 2 \mu_0 C_2 =0 \;, \nonumber \\[12pt]
&&  \big( e^{ - \phi +4X-A+2B-C }  B_4 \big)^\prime -2 e^{ - \phi +4X+A - C }  B_3   -2 e^{ \frac12 \phi +4X + A -2B-C }  A_1 C_2  \nonumber \\
&& \qquad \qquad  -12  e^{ \frac12 \phi +4X - A  +2B-3C }  A_3 C_4   - m  \, e^{ \frac32 \phi +4X + A +2B -C } A_4    - \tfrac12 \mu_0 C_1 =0 \;, 
\end{eqnarray}
}the $\hat F_\2$ equation of motion gives
{\setlength\arraycolsep{0pt}
\begin{eqnarray} \label{SU3IIAF2eom}
&& \big( e^{ \frac32 \phi +4X-A+2B + C }  A_3 )^\prime -2 \,  e^{ \frac32 \phi +4X + A +C } A_2 +8 \,  e^{ \frac32 \phi +4X + A +2B -C }  (1-2 A_0)  A_4    \\
&& \qquad \quad   +4 \,  e^{ \frac12 \phi +4X - A -2B +C } B_1 C_1   -8 \,  e^{ \frac12 \phi +4X - A -C } B_2 C_3  +192 \,  e^{ \frac12 \phi +4X - A +2B -3C } B_4 C_4  = 0 \; ,  \nonumber
\end{eqnarray}
}and the dilaton equation of motion gives

{\setlength\arraycolsep{0pt}
\begin{eqnarray} \label{SO4IIADilaton}
&& \big( e^{ 4X-A+2B + 3C }  \phi^\prime )^\prime - 3  \,  e^{ \frac32 \phi +4X  } \Big( e^{ A-2B+3C  } A_1^2 + 2 e^{ A+C  } A_2^2 + e^{ -A+2B+C  } A_3^2 + 16 e^{ A+2B-C  } A_4^2   \Big)    \nonumber \\
&& \qquad \qquad  +2  \,  e^{ - \phi +4X  } \Big( e^{ -A-2B+3C  } B_1^2 + 2 e^{ -A+C  } B_2^2 +32 e^{A-C  } B_3^2 + 16 e^{ -A+2B-C  } B_4^2   \Big)     \nonumber \\
&&\qquad \qquad  -4  \,  e^{ \frac12 \phi +4X  }  \Big( e^{ -A-2B+C  } C_1^2 + 4 e^{ A -2B -C  } C_2^2 +2 e^{-A-C  } C_3^2 + 144 e^{ -A+2B-3C  } C_4^2   \Big)  \nonumber \\
&&  \qquad \qquad -\tfrac54   \, m^2 \,   e^{ \frac52 \phi +4X + A +2B +3C }  +\tfrac14   \, \mu_0^2 \,   e^{ \frac12 \phi -4X + A +2B +3C } = 0 \; . 
\end{eqnarray}
}As for the Einstein equation, we have only computed the external components, which produce the following equation for the warp factor:
{\setlength\arraycolsep{0pt}
\begin{eqnarray} \label{SO4IIAEinsteinExt}
&& \big( e^{ 4X-A+2B + 3C }  X^\prime )^\prime - \tfrac{1}{4}  \,  e^{ \frac32 \phi +4X  } \Big( e^{ A-2B+3C  } A_1^2 + 2 e^{ A+C  } A_2^2 + e^{ -A+2B+C  } A_3^2 + 16 e^{ A+2B-C  } A_4^2   \Big)    \nonumber \\
&& \qquad \qquad - \tfrac{1}{2}  \,  e^{ - \phi +4X  } \Big( e^{ -A-2B+3C  } B_1^2 + 2 e^{ -A+C  } B_2^2 +32 e^{A-C  } B_3^2 + 16 e^{ -A+2B-C  } B_4^2   \Big)     \nonumber \\
&&\qquad \qquad  -3  \,  e^{ \frac12 \phi +4X  }  \Big( e^{ -A-2B+C  } C_1^2 + 4 e^{ A -2B -C  } C_2^2 +2 e^{-A-C  } C_3^2 + 144 e^{ -A+2B-3C  } C_4^2   \Big)  \nonumber \\
&&  \qquad \qquad +\tfrac{1}{16}   \, m^2 \,   e^{ \frac52 \phi +4X + A +2B +3C } -\tfrac{5}{16}   \, \mu_0^2 \,   e^{ \frac12 \phi -4X + A +2B +3C }  + 3 \, e^{ 2X + A +2B +3C }  = 0 \; . 
\end{eqnarray}
}

We have employed equations (\ref{SO4IIABianchis})--(\ref{SO4IIAEinsteinExt}) to verify that the generic expressions (\ref{KKSO4sectorinIIA}), (\ref{KKfieldstrengths}) evaluated on the critical points of ISO(7) supergravity with at least SO(4) invariance, recorded in table \ref{Table:SO4Points} of the main text, are indeed solutions of massive type IIA supergravity. We have verified on a case-by-case basis that all solutions mentioned in section \ref{sec:AdSsolutions}, particularly the $\cN=3$ solution (\ref{SO4SolN=3}), do satisfy equations  (\ref{SO4IIABianchis})--(\ref{SO4IIAEinsteinExt}). Up to a check of the internal Einstein equations, this shows that all constant-scalar configurations presented in the main text are indeed solutions of massive type IIA supergravity.

\section{Derivation of the $\cN=3$ Killing spinors} \label{app:KillingSpinors}

Following, for convenience, the string frame conventions of \cite{Hassan:1999bv}, the supersymmetry transformations of the type IIA fermions read
\begin{align}\label{eq:hassansusy}
\delta \hat{\lambda} &= \left(d\hat{\slashed{\phi}} +\tfrac{1}{4} \slashed{\hat{H}}_\3 \hat\Gamma\right)\hat \epsilon+ \tfrac18 e^{\hat\phi} \left(5 \hat{F}_\0+ 3 \slashed{\hat{F}}_\2 \hat\Gamma+ \slashed{\hat{F}}_\4 \right) \hat\epsilon  \; , \nonumber
\\[4pt]
\delta \hat{\psi}_M  &=\left(\nabla_M+ \frac{1}{4} \slashed{\hat{H}}_M \hat\Gamma\right) \hat{\epsilon} +\tfrac18 e^{\hat\phi}  \left(\hat{F}_\0 - \slashed{\hat{F}}_\2 \hat\Gamma + \slashed{\hat{F}}_\4  \right) \hat\Gamma_M \hat{\epsilon}  \; .
\end{align}
The IIA dilatino, $\hat\lambda$, gravitino, $\hat{\psi}_M$, and supersymmetry paramater, $\hat\epsilon$, are Majorana. The ten-dimensional gamma matrices are $\hat\Gamma_{\underline{M}}$, with $\underline{M} = 0, 1, \ldots, 9$, and $M = 0, 1, \ldots, 9$ denoting here ten-dimensional tangent space and local indices, respectively. The slashed forms are defined, as usual, as  $\slashed{\hat{H}}_\3 \equiv \frac{1}{3!} \hat{H}_{\underline{M} \underline{N} \underline{P}} \, \hat\Gamma^{\underline{M} \underline{N} \underline{P}}$, and $\slashed{\hat{H}}_{\underline{M}} \equiv \frac{1}{2!} \hat{H}_{\underline{M} \underline{N} \underline{P}} \, \hat\Gamma^{ \underline{N} \underline{P}}$, etc.,  with $\hat\Gamma^{\underline{M_1} \ldots \underline{M_n} } \equiv \hat\Gamma^{[\underline{M_1} }  \cdots \hat\Gamma^{\underline{M_n]} } $. The ten-dimensional chirality matrix has been denoted by $\hat\Gamma$, and $\hat\Gamma_M$ denotes the contraction of $\hat\Gamma_{\underline{M}}$ with the ten-dimensional vielbein.

Let us show that the $\cN=3$ solution (\ref{SO4SolN=3}) obeys the IIA Killing spinor equations, $\delta \hat{\lambda} =0 $, $\delta \hat{\psi}_M =0 $, for the supersymmetry parameters given in section \ref{sec:N=3susy}. We start by reducing these to the Killing spinor equations (\ref{eq:6dSUSY1})--(\ref{eq:6dSUSY3}) on the internal six-dimensional geometry. In order to do this, we choose a basis for the ten-dimensional gamma matrices such that
\begin{equation} \label{eq:GammaMatDecomp}
\hat\Gamma_\mu = e^{\tA}\gamma^{(4)}_{\mu}\otimes \mathbb{I},~~~ \hat\Gamma_{\underline{M}} = \hat \gamma^{(4)} \otimes \gamma_{\underline{M}},~~~ \hat{B}^{(10)}= \mathbb{I}\otimes B,~~~ \hat\Gamma= \hat\gamma^{(4)}\otimes \hat\gamma
\end{equation}
with $\mu=0,1,2,3$ local indices on AdS$_4$ and ${\underline{M}}=1, \ldots , 6$ now rebranded as a tangent space index on the internal six-dimensional geometry. Here, $e^{2\tA}$ is the string frame warp factor (\ref{eq:stringframeWF}), $\gamma^{(4)}_\mu$ and $\gamma_{\underline{M}}$ four- and six-dimensional gamma matrices (the former contracted with the AdS$_4$ vielbein), $\hat\gamma^{(4)}=-i  \gamma^{(4)}_{0123}$ and $\hat\gamma = i \gamma_{123456}$ the respective chirality matrices. The six and ten-dimensional intertwiners  $B$ and $\hat{B}^{(10)}$, with $BB^*=\mathbb{I}$, are such that $\chi^c = B \chi^*$, where $c$ denotes Majorana conjugation, $\gamma_{\underline{M}}^{*}=- B^{-1}\gamma_{\underline{M}} B$, and similarly for $\hat{B}^{(10)}$. Let $\zeta^{i}_{\pm}$ be (anti)chiral Killing spinors on AdS$_4$,
\begin{equation} \label{eq:KSEAdS4}
D_{\mu} \zeta^{i}_{\pm}= \frac{1}{2}\gamma^{(4)}_{\mu}\zeta^{i}_{\mp} \; , 
\qquad 
 \zeta^{i}_{\mp}= \zeta^{ic}_{\pm} \; , 
\end{equation}
and let $\chi^i$, $i=1,2,3$, three arbitrary Dirac spinors on the internal six-dimensional geometry. Writing the ten-dimensional spinor paramater $\hat \epsilon$ in terms of $\zeta^{i}_{\pm}$ and $\chi^i$ as in (\ref{eps10D}) and making use of the decompositions (\ref{eq:GammaMatDecomp}) and the AdS$_4$ Killing spinor equation (\ref{eq:KSEAdS4}), we obtain the six-dimensional Killing spinor equations (\ref{eq:6dSUSY1})--(\ref{eq:6dSUSY3}) from the expressions (\ref{eq:hassansusy}) equated to zero.  

One method to proceed in general (as pursued in \cite{Gabella:2012rc,Passias:2017yke,Passias:2018zlm} for $\mathcal{N}=2$ AdS$_4$ solutions) is to form spinor bilinears from \eqref{eq:6dSUSY1}--\eqref{eq:6dSUSY3}, use them to show that a  geometrically realised SO(3) R-symmetry ${\rm SO}(3)_{\mathcal{R}}$ necessarily emerges and then locally determine the metric, dilaton and fluxes up to PDEs. However given that the solution of section \ref{sec:AdSsolutions} contains  $S^2$ fibred over $S^3$, we find it easier to construct all spinors on $S^2\times S^3$ that transform as triplets under ${\rm SO}(3)_{\mathcal{R}}$ and couple them to arbitrary spinors on the interval spanned by $\alpha$, thereby effectively reducing the problem from 6 to 1 dimensions.

\subsection{Constructing an  ${\rm SO}(3)_{\mathrm{d}}$ triplet in 6d}

The first thing that needs addressing is exactly how one constructs a triplet of spinors that transforms under ${\rm SO}(3)_{\mathcal{R}}$. Clearly the metric on a round $S^2$ and $S^3$ preserve ${\rm SO}(3)_{S^2}$ and ${\rm SO}(4)_{S^4}$ isometries respectively but, as explored in the Minkowski classifications of \cite{Macpherson:2016xwk, Macpherson:2017mvu,Apruzzi:2018cvq}, each independent Killing spinor on these spheres can only individually be used to form $\mathrm{SU}(2)$ doublets. We also have to consider the fact that $S^2$ is fibred over $S^3$ in terms of the $\mathrm{SU}(2)$ right-invariant 1-forms as (\ref{covderapp}). This means that if one decomposes  ${\rm SO}(4)_{S^4}=\mathrm{SO}(3)_{\mathrm{L}}\times \mathrm{SO}(3)_{\mathrm{R}}$ (with $\mathrm{L}/\mathrm{R}$ standing for left/right) it is only $\mathrm{SO}(3)_{\mathrm{R}}$ and the diagonal $\mathrm{SO}(3)_{\mathrm{d}}$ formed from $\mathrm{SO}(3)' \equiv \mathrm{SO}(3)_{S^2}$ and $\mathrm{SO}(3)_\mathrm{L}$ that are preserved by the full space: the anti-diagonal is broken. Since $\mathrm{SO}(3)_\mathrm{R}$ only involves $S^3$, the preceding discussion suggests that we should identify the R-symmetry as $\mathrm{SO}(3)_{\mathcal{R}}=\mathrm{SO}(3)_{\mathrm{d}}$, but we will need to be more explicit to construct its corresponding triplets. As we shall show, the fundamental building blocks of the triplets are actually the $\mathrm{SU}(2)$ doublets on $S^2$ and $S^3$, for which we give further details in appendix \ref{sec:DoubletsAppendix}.

In what follows, we shall parametrise the two $\mathrm{SU}(2)$ doublets on $S^2$ as $\psi^{\alpha}$ and $\hat\psi^{\alpha}$, while the single $\mathrm{SU}(2)_{\mathrm{L}}$ doublet on $S^3$, that is a singlet with respect to $\mathrm{SU}(2)_{\mathrm{R}}$, shall be $\xi^{\alpha}$ for $\alpha=1,2$. The Killing vectors of $\mathrm{SO}(3)'$ we denote by $K_i$ and of $\mathrm{SO}(3)_{\mathrm{L}/\mathrm{R}}$ by $L^i/R^i$. The key property of the doublets that we shall need is how they transform under the action of the spinorial Lie derivative along these Killing vectors, namely
\beq\label{eq:doubletcondtion}
\mathcal{L}_{K_i}\psi^{\alpha}= \frac{i}{2} (\sigma_i)^{\alpha}_{~\beta}\psi^{\beta}\,,\qquad\mathcal{L}_{L_i}\xi^{\alpha}= \frac{i}{2} (\sigma_i)^{\alpha}_{~\beta}\xi^{\beta}\,,\qquad\mathcal{L}_{R_i}\xi^{\alpha}= 0
\eeq
with a corresponding expression for the action of $K_i$ on $\hat\psi^{\alpha}$. Here $\sigma_i$ are the Pauli matrices, so that a doublet of a given $\mathrm{SU}(2)$ realises the corresponding Lie algebra under its action. To realise the R-symmetry $\mathrm{SO}(3)_{\mathrm{d}}$, then we must construct products of these doublets that transform as
\beq\label{eq:tripletcondtion}
\mathcal{L}_{K^{\mathrm{d}}_i}\eta^{j}=\epsilon_{ijk}\eta^{k}\,,\qquad\mathcal{L}_{R_i}\eta^{j}=0\,,\qquad K^{\mathrm{d}}_i=K_i+ L_i\,.
\eeq
The obvious way one might try to construct a triplet is to contract $\psi^{\alpha}$ and $\xi^{\alpha}$ with the Pauli matrices. This is almost correct, but it is actually the matrices $\sigma_2 \sigma_i$, which are symmetric, that give the correct transformation properties. It is then not hard to show that
\beq\label{eq:triplets12}
\eta^i_1 = (\sigma_2 \sigma_i)_{\alpha\beta} \psi^{\alpha}\otimes \xi^{\beta},~~~\eta^i_2 = (\sigma_2 \sigma_i)_{\alpha\beta} \hat\psi^{\alpha}\otimes \xi^{\beta}\,,\qquad\eta^{ic}_1=-\eta^i_2,
\eeq
obey \eqref{eq:tripletcondtion} using \eqref{eq:doubletcondtion} and standard Pauli matrix identities, which gives us two triplets. But this is not the whole story. It is also possible to construct two $\mathrm{SO}(3)_{\mathrm{d}}$ singlets by contracting the doublets with $\sigma_2$; so, given that the embedding coordinates $\tilde{\mu}_i$ of $S^2$  transform as a triplet under $\mathrm{SO}(3)_{\mathrm{d}}$, 
\begin{equation}\label{eq:triplets34}
\eta^3_i = \tilde{\mu}_i(\sigma_2 )_{\alpha\beta} \psi^{\alpha}\otimes \xi^{\beta}\, ,\qquad \eta^4_i = \tilde{\mu}_i(\sigma_2 )_{\alpha\beta} \hat\psi^{\alpha}\otimes \xi^{\beta}\,,\qquad \eta^{ic}_3= \eta_4^i,
\end{equation}
also obey \eqref{eq:tripletcondtion}, by Leibniz rule.

We can derive (\ref{eq:triplets12}), (\ref{eq:triplets34}) by using ambient space coordinates and group theory. On $S^3$, our spinors need to be singlets under $\mathrm{SO}(3)_{\rm R}$. The only spinors with this feature are the $\xi^\alpha$. This can be seen by going to the left-invariant frame, where the spinorial Lie derivative reduces to partial directional derivative. The $\xi^\alpha$ transform as a doublet under $\mathrm{SO}(3)_{\rm L}$; in order to produce a triplet under $\mathrm{SO}(3)_{\rm d}$ (the diagonal in $\mathrm{SO}(3)'\times \mathrm{SO}(3)_{\rm L}$) we need to tensor them with $S^2$ spinors that transform either as a doublet or as a quadruplet of $\mathrm{SO}(3)'$, since only $(s=1/2)$ and $(s=3/2)$ are such that $(s)\otimes (1/2)$ contains $(1/2)$. 

In order to produce such $S^2$ spinors, we can work in the ambient $\mathbb{R}^3$. Here it is clear that our ingredients are constant spinors and the three coordinates; going back to $S^2$, these become the $\psi^\alpha$ and the $\tilde\mu_i$. From the $\psi^\alpha$ and $m$ copies of the $\mu_i$ one obtains a representation $(m)\otimes (1/2)= (m-1/2)\oplus (m+1/2)$. This contains $(1/2)$ or $(3/2)$ for $m=0,1,2$. Defining $\tilde\mu_{\alpha \beta}\equiv \tilde\mu^i (\sigma_2 \sigma_i)_{\alpha \beta}$, the doublets one obtains this way can be written as $\psi^\alpha$, $\tilde\mu_{\alpha \beta} \psi^\beta$; it turns out that the latter is simply $\hat\psi^\alpha$. We can now tensor these two doublets with $\xi^\alpha$ on $S^3$ and extract the triplet using $(\sigma_2 \sigma_i)_{\alpha \beta}$; this gives (\ref{eq:triplets12}). On the other hand, the quadruplets one can write in this way can be written as
$\tilde\mu_{(\alpha \beta} \psi_{\gamma)}$, $\tilde\mu_{(\alpha \beta} \tilde\mu_{\gamma) \delta} \psi^\delta = \tilde\mu_{(\alpha \beta} \hat \psi_{\gamma)}$. Tensoring these with the $\xi^\gamma$ and contracting one index to extract the triplet, after some Pauli matrix algebra one obtains a linear combination of (\ref{eq:triplets12}) and (\ref{eq:triplets34}).

Thus we have obtained a set of triplets $\{\eta^1_i,\eta^2_i,\eta^3_i,\eta^4_i\}$, which is linearly independent and closed under Majorana conjugation and the action \eqref{tab:12formaction}--\eqref{tab:34formaction} of the $\mathrm{SO}(3)_{\mathrm{d}}$ invariant forms \eqref{eq:invarientforms1}--\eqref{eq:invarientforms2}. From the embedding coordinates argument we have just given, this set is also exhaustive. One can also see this in the following way. Any 5d spinor can be decomposed in a basis of four linearly independent spinors with complex functional coefficients defined in 5d. This means any additional triplet can be decomposed in a basis of our existing triplets as $\tilde{\eta}^i = \sum_{n=1}^4 (a_n)^{i}_{~j} \eta^j_n$. Since this new triplet needs to transform as \eqref{eq:tripletcondtion}, and $\eta^i_n$ already do transform in this fashion, there are only two options. i) We take $(a_n)^i_{~j}$ constant, in which case $\tilde{\eta}^i$ is not linearly independent of $\eta^i_n$ by  definition. ii) We take $(a_n)^i_{~j}$ to be proportional to the $S^2$ embedding coordinates $\tilde{\mu}_j$, which  leads to $\tilde{\eta}^i=\sum_{n=1}^4 c_n\epsilon_{ijk}\tilde{\mu}_k\eta^j_n$. At first sight this does seem to give two additional triplets (not four, because those involving $\eta^i_3,\eta^i_4$ are proportional to $\epsilon_{ijk}\tilde{\mu}_j\tilde{\mu}_k=0$). However, one can show that $\epsilon_{ijk}\tilde{\mu}_k\eta^j_1= -i \eta^i_2+\eta^i_3$ and $\epsilon_{ijk}\tilde{\mu}_k\eta^j_2= -i \eta^i_1+\eta^i_4$; so these triplets depend linearly on $\eta^i_n$.

The most general $\mathrm{SO}(3)_{\mathrm{d}}$  triplet of spinors on a fibration of $S^2$ over $S^3$ times an interval is then of the form
\begin{equation}\label{eq:6d triplets-app}
\chi^i = \frac{1}{2}e^{\frac{\tA}{2}} \bigg[\left(\begin{array}{c} f_{1+}\\ f_{1-}  \end{array}\right )\otimes\eta^i_1+\left(\begin{array}{c} f_{2+}\\ f_{2-}  \end{array}\right )\otimes\eta^i_2+\left(\begin{array}{c} f_{3+}\\ f_{3-}  \end{array}\right )\otimes\eta^i_3+\left(\begin{array}{c} f_{4+}\\ f_{4-}  \end{array}\right )\otimes\eta^i_4\bigg]
\end{equation}
where $f_{n\pm}$ are functions of $\alpha$, the coordinate on the interval, to be determined by \eqref{eq:6dSUSY1}--\eqref{eq:6dSUSY3}.  Up to this point our supersymmetry discussion has been quite general and will apply to any AdS$_4$ solution with metric, dilaton and fluxes preserving $\mathrm{SO}(3)_{\mathrm{d}}\times \mathrm{SO}(3)_{\mathrm{R}}$ with a  $S^2\times S^3$ fibration. We shall now proceed to solve the arbitrary functions of the interval for the solution of section \ref{sec:AdSsolutions}. We shall return to this system in full generality in a follow up.

\subsection{Solving for the Killing Spinor}\label{sec:Solvingthespinor}

In this subsection, we will explicitly compute the triplet of spinors preserved by the solution of section \ref{sec:AdSsolutions} by plugging \eqref{eq:6d triplets-app} into \eqref{eq:6dSUSY1}--\eqref{eq:6dSUSY3} and solving for the undetermined functions of the interval. We shall work with the following $6=1+2+3$ decomposition of the flat space gamma matrices
\begin{equation}\label{eq:6dgammas}
\gamma_{\underline{\alpha}}= \sigma_1\otimes \mathbb{I}\otimes \mathbb{I}\, ,\qquad\gamma_{\underline{a}}= \sigma_2\otimes \sigma_a\otimes \mathbb{I}\, ,\qquad\gamma_{\underline{\hat a}}= \sigma_2\otimes \sigma_3\otimes \sigma_{\hat a} \, ,\qquad B=\sigma_2\otimes \sigma_1\otimes \sigma_2
\end{equation}
for $a=1,2$ and ${\hat a}=1,2,3$, so that the six-dimensional chirality matrix is $\hat\gamma=\sigma_3\otimes \mathbb{I}\otimes \mathbb{I}$ and only acts on the interval part of the spinor.  The string frame vielbein on the internal space can be succinctly written as
\begin{equation}
\begin{split}
	e^{\alpha}&=\sqrt{2}e^{A}d\alpha \, ,\qquad e^{\hat a}= 2\sqrt{2}L^2e^{-A+\frac{1}{2}\phi_0} \sin\alpha\rho_{\hat a},\\[2mm]
	e^{a}&=\sqrt{2}L e^{\hat{\phi}-\frac{3}{2}\phi_0}\cos\alpha \left(dy^a-2L^4 e^{-4A+\phi_0}\sin^2\alpha K^{y^{a}}_{i}\rho_i\right)\,;
\end{split}	
\end{equation}
$y^a$ are coordinates on $S^2$, and $K_i$ are the $\mathrm{SU}(2)$ Killing vectors on $S^2$ given in \eqref{eq:S2SU2}.
The AdS warp factor and dilaton are respectively
\beq
e^{2A}=e^{\frac{1}{2}\phi_0} L^2 \sqrt{2(\cos^2\alpha+1)}\,,\qquad e^{\hat{\phi}}=e^{\phi_0}\frac{(2(\cos^2\alpha+1))^{\frac{3}{4}}}{\sqrt{3 \cos^4\alpha+ 3 \cos^2 \alpha+2}}.
\eeq
In (\ref{SO4SolN=3}) all possible 10d fluxes of massive IIA are turned on. However, for $\mathrm{SO}(3)_{\mathrm{d}}$ to be preserved, the fluxes should be singlets under its action. To show this is so, and because it will be helpful in what follows, we introduce a basis of $\mathrm{SO}(3)_{\mathrm{d}}$ invariant forms  on $S^2\times S^3$:
\begin{equation}\label{eq:invarientforms1}
\begin{split}
	\omega_1 &= \frac{1}{2}\rho_i \tilde{\mu}_i ,~~~ \omega^1_2= \frac{1}{2} \epsilon_{ijk}\tilde{\mu}_i D\tilde{\mu}_j\wedge D\tilde{\mu}_k,~~~ \omega^2_2=\frac{1}{2}\rho_i\wedge D \tilde{\mu}_i,\\[2mm]
	\omega^3_2&= \frac{1}{2} \epsilon_{ijk} \tilde{\mu}_i \rho_j\wedge D\tilde{\mu}_k,~~~ \omega^4_2= \frac{1}{8} \epsilon_{ijk} \tilde{\mu}_i \rho_j\wedge \rho_k\,.	
\end{split}	
\end{equation}
In terms of these, all invariant forms on $S^2\times S^3$ can be expressed \cite{Rota:2015aoa}, 
\begin{equation}\label{eq:invarientforms2}
\begin{split}
	\omega^1_3&=\omega_1\wedge \omega^1_2 \, ,\qquad\omega^2_3=\omega_1\wedge \omega^2_2 \, ,\qquad\omega^3_3=\omega_1\wedge \omega^3_2 \, ,\qquad	 \omega^4_3=\omega_1\wedge \omega^4_2,\\[2mm]
	\omega_4&=\omega^1_2\wedge\omega^4_2=-\frac{1}{2}\omega^2_2\wedge\omega^2_2=-\frac{1}{2}\omega^3_2\wedge \omega^3_2 \, ,\qquad\omega_5= \omega_1\wedge \omega_4\,.
\end{split}	
\end{equation}
This exhausts the list of forms.
The fluxes appearing in \eqref{eq:6dSUSY1}--\eqref{eq:6dSUSY3} then take the form
\begin{align}
e^{4A}\hat{G}_{(0)}&=4 \sqrt{3}s e^{\phi_0}\hat{F}_{(0)}L^4, \nonumber\\[2mm]
\hat{F}_{(2)}&=4 s e^{\frac{1}{2}\phi_0} \hat{F}_\0 L^2\bigg[-\frac{2\sin^3\alpha}{\sqrt{3}(\cos^2\alpha+1)^2}d\alpha\wedge \omega_1+\frac{\cos^3\alpha(3\cos^2\alpha+1)}{\sqrt{3}(3\cos^4\alpha+3\cos^2\alpha+2)}\omega^1_2\nonumber\\[2mm]
&~~~~~~~~~~~~~~~~~~~-\frac{2\cos\alpha \sin^2\alpha}{\sqrt{3}(\cos^2\alpha+1)}\omega^2_2+\frac{\sqrt{3}\cos\alpha \sin^4\alpha}{(\cos^2\alpha+1)^2}\omega^4_2\bigg],\nonumber\\[2mm]
\hat{G}_{(4)}&=16 e^{\phi_0} \hat{F}_\0 L^4\bigg[ \frac{\cos^2\alpha\sin^2\alpha(3\cos^2\alpha+1)}{(3\cos^4\alpha+3\cos^2\alpha+2)}\omega_4- d\alpha\wedge \bigg(\frac{\cos\alpha\sin^5\alpha}{(\cos^2\alpha+1)^2}\omega^2_3\nonumber\\[2mm]
&+\frac{2\cos\alpha\sin^3\alpha(3\cos^2\alpha+1)}{(\cos^2\alpha+1)^3}\omega^4_3
+\frac{2\cos^3\alpha\sin\alpha(2\cos^4\alpha+3\cos^2\alpha+3)}{(\cos^2\alpha+1)(3\cos^4\alpha+3\cos^2\alpha+2)}\omega^1_3\bigg)\bigg],\nonumber\\[2mm]
\hat{H}_{(3)}&=4 s e^{\frac{1}{2}\phi_0} L^2\bigg[\frac{2\sqrt{3}\cos\alpha\sin^2\alpha(\cos^2\alpha+1)}{3\cos^4\alpha+3\cos^2\alpha+2}\omega^2_3-d\alpha\wedge\bigg(\frac{\sqrt{3}\sin^5\alpha}{(\cos^2\alpha+1)^2}\omega^4_2\nonumber\\[2mm]
&~~~~~~~~~~~~~~~~~~+\frac{\sqrt{3}\cos^2\alpha \sin\alpha(3\cos^6\alpha+8 \cos^4\alpha+11 \cos^2\alpha+2)}{(3\cos^4\alpha+3\cos^2\alpha+2)^2}\omega^1_2\nonumber\\[2mm]
&~~~~~~~~~~~~~~~~~~+\frac{2\sqrt{3}\cos^2\alpha\sin\alpha(\cos^4\alpha+\cos^2\alpha+2)}{(\cos^2\alpha+1)(3\cos^4\alpha+3\cos^2\alpha+2)}\omega^2_2\bigg)\bigg],
\end{align}
which is a mild generalisation of (\ref{SO4SolN=3}) including a possible world sheet parity inversion parametrised by
\beq
s=\pm 1.
\eeq
One can check explicitly that this still solves all the Bianchi identities and flux equations of motion with the remaining equations of motion following once
\begin{equation}
3e^{-\frac{5}{2} \phi_0}=  8 \hat{F}_{(0)}^2 L^2
\end{equation}
is imposed.

In the remainder of this appendix and in appendix \ref{sec:DoubletsAppendix} we will sketch the computation of the functions $f_{n\pm}$ appearing in (\ref{eq:6d triplets-app}).

The first supersymmetry condition we shall solve is the dilatino variation (\ref{eq:6dSUSY2}). The reason to start here is that this condition contains no derivatives of the spinors, no Majorana conjugation and the $S^2\times S^3$ data only appears packaged in the invariant forms. Thus the computation just consists of writing  (\ref{eq:6dSUSY2}) in the form 
\begin{equation}\label{eq:6dcondtion}
\sum_{n=1}^4\left(\begin{array}{c} X_{n+}\\X _{n-}  \end{array}\right )\otimes\eta^i_n,
\end{equation}
using the action of the invariant forms listed in \eqref{tab:12formaction}--\eqref{tab:34formaction}. Once this is done, one knows that all eight $X_{n\pm}$ must individually vanish, because the 4 triplets $\eta^i_n$ are independent, and likewise the positive and negative chirality components of the expression. This leads to eight complex algebraic constraints that the $f_{n\pm}$ must satisfy.  We found it useful to introduce an auxiliary set of complex functions of the interval:
\begin{equation}\label{eq:ftot}
\begin{split}
	f_{1\pm}= \frac{1}{2}(t_{1\pm}+t_{2\pm})\, ,\qquad f_{3\pm}= \frac{1}{2}(-t_{1\pm}+t_{2\pm}+t_{3\pm}+t_{4\pm}),\\[2mm]
	f_{2\pm}= \frac{1}{2}(t_{1\pm}-t_{2\pm})\, ,\qquad f_{4\pm}= \frac{1}{2}(-t_{1\pm}-t_{2\pm}+t_{3\pm}-t_{4\pm})\	
\end{split}
\end{equation}
and then solve instead for $t_{n\pm}$. This is because the dilaton conditions fix several of the $t_{n\pm}$ in terms of just one other and known functions of the interval.
The simplest four of the conditions that follow take the form
\begin{align}\label{eq:X1}
X_{1\pm}+X_{2\pm}&=Z_{\pm}(t_{1\pm},~t_{1\mp}) \, ,\qquad X_{1\pm}-X_{2\pm}=Z_{\mp}(t_{2\pm},~t_{2\mp}),\nonumber\\[2mm]
Z_{\pm}(z_1,~z_2)&\propto z_1\sqrt{\tilde{\Delta}_3}\big(2i\sqrt{6}s \cos^2\alpha+\sqrt{2}\tilde{\Delta}_1(2+\cos^2\alpha)\pm \sqrt{\tilde{\Delta}_1}\cos\alpha(2+ \sqrt{3}si+\cos^2\alpha) \big)\nonumber\\[2mm]
&+z_2\sin\alpha\big(\sqrt{6}(3+\cos^2\alpha)\cos^2\alpha\pm 4si\sqrt{\tilde{\Delta}_1}(1+2 \cos^2\alpha)\big)\,.
\end{align}
$\tilde{\Delta}_i$ are defined in \eqref{eq:deltas}; we have factored out a common non-vanishing $\alpha$-dependent factor in $Z_{\pm}$.
It is easy to check that \eqref{eq:X1} contains only 2 independent expressions, namely the middle two of \eqref{eq:dilatoncondtions}. The remaining 4 $X_{n\pm}$ are more complicated and we only quote their solution. Of the eight complex conditions that follow from (\ref{eq:6dSUSY2}) only 5 are independent. All in all, the solution can be written as
\begin{equation}\label{eq:dilatoncondtions}
\begin{split}
	t_{3\pm}&=-i s t_{4\mp},\\[2mm]
	t_{1+}&= - \frac{\cos\alpha\big(\sqrt{3}\tilde{\Delta}_1+2 s i\big)+\sqrt{2}\sqrt{\tilde{\Delta}_1}\big(\sqrt{3}\cos^2\alpha+si \big)}{\big(1+3 \cos^2\alpha+2\sqrt{2}\sqrt{\tilde{\Delta}_1}\cos\alpha\big)\sqrt{\tilde{\Delta}_3}}\sin\alpha t_{1-},\\[2mm]
	t_{2+}&= - \frac{\cos\alpha\big(\sqrt{3}\tilde{\Delta}_1+2 s i\big)-\sqrt{2}\sqrt{\tilde{\Delta}_1}\big(\sqrt{3}\cos^2\alpha+si \big)}{\big(1+3 \cos^2\alpha-2\sqrt{2}\sqrt{\tilde{\Delta}_1}\cos\alpha\big)\sqrt{\tilde{\Delta}_3}}\sin\alpha t_{2-},\\[2mm]
	\sin\alpha t_{4+}&=\frac{\sqrt{3}\cos\alpha\tilde{\Delta}_1^3+4 si \cos^3\alpha \tilde{\Delta}_2+\sqrt{2}\big(2\sqrt{3}\cos^4\alpha+s i\tilde{\Delta}_2\big)\tilde{\Delta}_1^{\frac{3}{2}}}{\big(7 \cos^4\alpha+4 \cos^2\alpha+1\big)\sqrt{\tilde{\Delta}_3}} t_{4-}	
\end{split}	
\end{equation}
where
\begin{equation}\label{eq:deltas}
\tilde{\Delta}_1=\cos^2\alpha+1,~~~\tilde{\Delta}_2=(3\cos^4\alpha+2\cos^2\alpha+1),~~~\tilde{\Delta}_3=(3\cos^4\alpha+3\cos^2\alpha+2).
\end{equation}

We next  turn our attention to the AdS$_4$ gravitino (\ref{eq:6dSUSY1}), which can be dealt with in much the same fashion as the dilatino. The only additional ingredient one needs is that $\chi^{ic}$ can be expressed in a basis of $\eta^i_n$ using the relations between triplets under Majorana conjugation in \eqref{eq:triplets12}--\eqref{eq:triplets34}. Given (\ref{eq:dilatoncondtions}), (\ref{eq:6dSUSY1}) provides one additional complex and one real constraints on $t_{n\pm}$:
\begin{equation}\label{eq: AdS4condtions}
\begin{split}
	\sin\alpha t_{2-}&=i(\sqrt{\tilde{\Delta}_1}-\sqrt{2}\cos\alpha) t^*_{1-},\\[2mm]
	\text{Im}t_{4_-}&=\frac{\big(\sqrt{2}\tilde{\Delta}_2+\sqrt{3}s \sqrt{\tilde{\Delta}_3}\cos^2\alpha\big)\big(\sqrt{3}s \cos\alpha \tilde{\Delta}_1^{\frac{3}{2}}-(1+2\cos^2\alpha)\sqrt{\tilde{\Delta}_3}\big)}{2+\cos^2\alpha \tilde{\Delta}_1(9\cos^4\alpha+6 \cos^2\alpha+8)}\text{Re}t_{4_-}.
\end{split}	
\end{equation}
As these conditions are the only ones that involve complex conjugation, we shall delay solving them until we have fixed more of $t_{n\pm}$ with the internal gravitino conditions (\ref{eq:6dSUSY3}).

The gravitino conditions on $S^2\times S^3$ do require us to take covariant derivatives of spinor \eqref{eq:spin connection}, and $H_a= \frac{1}{2}H_{abc}\gamma^{bc}$ now appears; both depend on more than simply the invariant forms. Therefore the action of the forms is insufficient to solve (\ref{eq:6dSUSY3}) in these directions.

The first thing we need to know is the form of the covariant derivative entering in (\ref{eq:6dSUSY3}), which requires that we compute the spin connection on $M_6$.
The vielbein on a generic $M_6$ consisting of  $S^2$ fibred over $S^3$ times an interval preserving $\mathrm{SO}(3)_{\mathrm{d}}\times \mathrm{SO}(3)_{\mathrm{R}}$ may be expressed as
\beq\label{eq:genvielbein}
e^{\alpha}=e^{k}d\alpha\,,\qquad e^{\hat a}= \frac{1}{2}e^{C_2}\rho_{\hat a},~~~e^{a}=e^{C_1}\bigg(dy^{a}-\lambda K^{y^{ a}}_{i}\rho_i\bigg),~~~D\tilde{\mu}_i=d\tilde{\mu}_i+ \lambda\epsilon_{ijk}d\tilde{\mu}_j\tilde{\mu}_k
\eeq
for $y^a$ coordinates on $S^2$ and  $e^{K},~e^{C_1},~e^{C_2},~\lambda$ functions of the interval. The spin connection on $M_6$ defined through $de^M+\Omega^M_{~N}\wedge e^N$ enters into the definition of (\ref{eq:6dSUSY3}) as $\nabla_{M} = \partial_M +\frac{1}{4}\Omega_{M,\underline{P}\underline{Q}}\gamma^{\underline{P}\underline{Q}}$. After computing this it is possible to show that the covariant derivatives along the internal space defined by  \eqref{eq:genvielbein} decompose as
\begin{align}\label{eq:spin connection}
\nabla_{\alpha}&=\partial_{\alpha}-\frac{1}{2} e^{2C_1}\lambda'\omega^3_2,\nonumber\\[2mm]
K^{y^a}_i\nabla_{y^{a}}&=K^{y^a}_i D^{S^2}_{y^a}+\frac{1}{2}K^{y^a}_i\gamma_{y^a}dC_1+ e^{2C_1}\lambda(\lambda-1)\big(\frac{1}{4}d\rho_i- \tilde{\mu}_i \omega^4_2\big)+\frac{1}{4}e^{2C_1} d\lambda\wedge(\rho_i -2 \tilde{\mu}_i \omega^1),\nonumber\\[2mm]
L^{y^{\hat{a}}}_i\nabla_{y^{\hat{a}}}&=L^{y^{\hat{a}}}_i D^{S^3}_{y^{\hat{a}}}+\frac{1}{2}L^{y^{\hat{a}}}_i\gamma_{y^{\hat{a}}}dC_2-e^{2C_1}\lambda^2(\lambda-1)\big(\frac{1}{4}d\rho_i- \tilde{\mu}_i \omega^4_2\big)-\frac{1}{4}e^{2C_1}\lambda d\lambda\wedge(\rho_i-2 \tilde{\mu}_i \omega^1) \nonumber\\[2mm]
&-\frac{1}{4}e^{2C_1} \lambda \epsilon_{ijk} D\tilde{\mu}_j\wedge D\tilde{\mu}_k-\frac{1}{2}e^{2C_1}\epsilon_{ijk}\tilde{\mu}_kD\tilde{\mu}^j \wedge(\frac{1}{2}d\lambda+ \lambda dC_1- \lambda dC_2)\nonumber\\[2mm]
&- \frac{1}{2}e^{2C_1}\lambda(\lambda-1)(\omega^1\wedge D\tilde{\mu}_i-\tilde{\mu}_i \omega^2_2)
\end{align}
where $D^{S^i}_{y^{a/\hat{a}}}$ is the covariant derivative on $S^2/S^3$; we contract the $S^2\times S^3$ directions with the two Killing vectors that make up $\mathrm{SO}(3)_{\mathrm{d}}$. Form expressions should be understood through the Clifford map $dx^{m_1}\wedge \ldots \wedge dx^{m_k} \mapsto \gamma^{m_1\ldots m_k}$.  Elsewhere in the text we have set
\beq\label{eq:kC}
\begin{split}
&e^{k}=\sqrt{2}e^{A} \, ,\qquad e^{C_1}=\sqrt{2}L e^{\hat{\phi}-\frac{3}{2}\phi_0}\cos\alpha \, ,\\
&e^{C_2}= 4\sqrt{2}L^2e^{-A+\frac{1}{2}\phi_0} \sin\alpha \, ,\qquad\lambda=2L^4 e^{-4A+\phi_0}\sin^2\alpha\,.	
\end{split}
\eeq
The other additional object appearing in (\ref{eq:6dSUSY3}) is $H_p$, but this is not hard to compute if we contract the $S^2\times S^3$ directions of this along the Killing vectors as in \eqref{eq:spin connection} and make use of the identities in the last column of \eqref{eq:S2LieandContr} and \eqref{eq:S2LieandContr}.

 We proceed by solving these conditions spinor component by spinor component. By  making use of the rotation\footnote{Specifically, when we use the coordinates presented at the beginning of section \ref{sec:DoubletsAppendix} the rotation \eqref{eq:rotation} maps the components of each triplet to
\beq
\Lambda\chi^j=\frac{1}{2}e^{\frac{A}{2}}(-i u_j t_{1+},~i \tilde{\mu}_j t_{3+},~\tilde{\mu}_jt_{4_+},~-\overline{u}_j t_{2+},~-i u_jt_{1-},~i \tilde{\mu}_jt_{3-},~\tilde{\mu}j t_{4_-},~-\overline{u}_j t_{2-})^T
\eeq
where
\beq
u_j= s_j+ i t_j = (\cos\theta_1\cos\phi_1+ i \sin\phi_1,~\cos\theta_1\sin\phi_1- i \cos\phi_1,-\sin\theta_1)_j
\eeq
and $\mu_i$ are the embedding coordinates of $S^2$.} outlined in appendix \ref{sec:triplettandforms} it is possible to factorise the $S^2\times S^3$ data out of each of these components leaving  many expression involving only functions of the interval that must vanish. After a lengthy calculation we find that the $S^2\times S^3$  gravitino imposes just two additional complex constraints that may be expressed as
\begin{align}\label{eq:S2S3condtions}
t_{4-}&=- \left(1+ \frac{(1+i \sqrt{3} s)\cos\alpha}{\sqrt{2 }\sqrt{\tilde{\Delta}_1}}\right)t_{2_-},\nonumber\\[2mm]
\sin\alpha\sqrt{\tilde{\Delta}_3} t_{2_-}&=\big((2+ \sqrt{3}si  \tilde{\Delta}_1)\cos\alpha- \sqrt{2} \sqrt{\tilde{\Delta}_1}(1+ \sqrt{3}s i\cos^2\alpha)\big)t_{1-}
\end{align}
after using (\ref{eq:dilatoncondtions})  to tame many expressions. This leaves only one complex function, $t_{1_-}$ say, to be determined.

The last condition we need to deal with is the internal gravitino condition along the interval. For this expression the covariant derivative \eqref{eq:spin connection} and  $H_{\alpha}$ are expressed in terms on the invariant forms. We can thus once more use the action of the forms of the triplets to massage the interval component of (\ref{eq:6dSUSY3}) into the form \eqref{eq:6dcondtion}.
Once (\ref{eq:dilatoncondtions}) and (\ref{eq:S2S3condtions}) are used to eliminate the other $t_{n\pm}$ we are left with a single ODE for $t_{1_-}$:
\begin{equation}\label{eq:intervalcondtion}
\partial_{\alpha}\log(t_{1-})=\frac{\cot\alpha}{\tilde{\Delta}_1}-\frac{1}{\sqrt{2}\sin\alpha \sqrt{\tilde{\Delta}_1}} - s i\sqrt{\frac{3}{2}} \frac{1+2 \cos^2\alpha}{\tilde{\Delta}_3\sqrt{\tilde{\Delta}_1}}\sin\alpha.
\end{equation}
Although this may appear a little intimidating a closed form solution does in fact exist, and after some effort one finds that
\begin{equation}\label{eq:t1m}
t_{1-}=- c \sqrt{1+ \frac{\sqrt{2}\cos\alpha}{\sqrt{\tilde{\Delta}_1}}} e^{-\frac{1}{2}s i \Theta},~~~ \cot\Theta= \sqrt{\frac{3}{2}}\cos\alpha \sqrt{\tilde{\Delta}_1},
\end{equation}
is the general solution to \eqref{eq:intervalcondtion}, where the sign is chosen to simplify \eqref{eq:fs}.
At this point we have completely determined the spinor up to a complex constant $c$ and a sign $s=\pm 1$, but we still need to check if (\ref{eq: AdS4condtions}) actually holds. After a comparatively brief computation it is possible to show that consistency can  be achieved if either
\begin{equation}
(s=1, ~\text{Im}c=0)\qquad\text{or}\qquad(s=-1, ~\text{Re}c=0)\,.
\end{equation}
However, the fact that $c$ should either be real or purely imaginary is just a consequence of our choice of intertwiner in \eqref{eq:6dgammas}. If we include a constant phase in its definition, $c$ can be completely arbitrary and it is this phase that is fixed by the choice of $s=\pm 1$. We use this fact to allow for an arbitrary constant in the main text.

We have now completely fixed all 8 auxiliary functions $t_{n\pm}$; inverting (\ref{eq:ftot}), after significant massaging one is led to the $f_{n\pm}$ in (\ref{eq:auxfuncs}). 
 This confirms that a Killing spinor preserving $\mathcal{N}=3$ superconformal symmetry does indeed exist.

\section{Further details on doublets, invariant forms and triplets} \label{app:doublets}

In this appendix we shall first give further explicit details about the $\mathrm{SU}(2)$ doublets from which the triplets are constucted in appendix \ref{sec:DoubletsAppendix}. Later we discuss the $\mathrm{SO}(3)_{\mathrm{d}}$ invariant forms, triplets and how the former acts on the latter in appendix \ref{sec:triplettandforms}.  We will mostly work with a concrete choice of coordinates,  $y^a=(\theta_1,~\phi_1)^a$ on $S^2$ and $y^{\hat{a}}=(\theta_2,~\phi_2,~\psi_2)^{\hat{a}}$ on $S^3$.

\subsection{$\mathrm{SU}(2)$ doublets}\label{sec:DoubletsAppendix}
In this appendix we construct $\mathrm{SU}(2)$ doublets form the Killing spinors on $S^2$ and $S^3$. In principle these calculations were already performed in \cite{Macpherson:2016xwk, Macpherson:2017mvu,Apruzzi:2018cvq}; however, we are using slightly different conventions here (notably with the choice of frame on $S^3$ and sign of the Killing spinor equation on $S^2$), so we provided some additional details here.

\subsubsection{$\mathrm{SU}(2)$ doublets on $S^2$}
The one-forms 
\begin{equation}
k_i =\epsilon_{ijk} d\tilde{\mu}_j \tilde{\mu}_k
\end{equation}
are dual to the $\mathrm{SO}(3)'$ Killing vectors $K_i$ on $S^2$; under the action of $d$, they behave  as the right invariant forms on $S^3$, with the same sign.
We shall use the specific parametrisation of the $S^2$ embedding coordinates
\begin{equation}
\tilde{\mu}_i=(\sin\theta_1\cos\phi_1,~\sin\theta_1\sin\phi_1,~\cos\theta_1)_i,
\end{equation}
in terms of which the Killing vectors are
\begin{align}\label{eq:S2SU2}
K_1&= \sin\phi_1 \partial_{\theta_1}+ \cot\theta_1 \cos\phi_1 \partial_{\phi_1}, \nonumber\\[2mm]
K_2&= -\cos\phi_1 \partial_{\theta_1}+ \cot\theta_1 \sin\phi_1 \partial_{\phi_1}, \\[2mm]
K_3&= -\partial_{\phi_1}. \nonumber
\end{align}
It is not hard to show that
\beq\label{eq:S2LieandContr}
\mathcal{L}_{K_i} \tilde{\mu}_j= \iota_{K_i}d\tilde{\mu}_j= \epsilon_{ijk} \tilde{\mu}_k\,,\qquad\mathcal{L}_{K_i} k_j=\epsilon_{ijk} k_k,
\eeq
so clearly $\tilde{\mu}_i,~d\tilde{\mu}_i,~k_i$ are all charged under $\mathrm{SO}(3)$.
There exist Killing spinors on $S^2$ that solve the equation
\begin{equation}
\nabla_{\underline{a}}\psi=- \frac{i}{2}\sigma_{a} \psi\,.
\end{equation}
If the frame $e^a=(d\theta_1,~\sin\theta_1 d\phi_1)^a$, with $\gamma_{\underline{a}}= \sigma_a$, $a=1,2$ one such specific example is
\begin{equation}
\psi= e^{-\frac{i}{2}\theta_1\sigma_1}e^{\frac{1}{2}\phi_1\sigma_1\sigma_2} \left(\begin{array}{c} 0\\ 1  \end{array}\right ).
\end{equation}
Using this, one can construct two $\mathrm{SU}(2)$ doublets on $S^2$, namely
\begin{equation}
\psi^{\alpha}= \left(\begin{array}{c} \psi\\-\sigma_2 \psi^* \end{array}\right )^{\alpha},~~~\hat\psi^{\alpha}= \left(\begin{array}{c} \sigma_3\psi\\-\sigma_3\sigma_2 \psi^*\end{array} \right )^{\alpha}.
\end{equation}
Both indeed transform under the action of the spinorial Lie derivative as
\begin{equation}
\mathcal{L}_{K_i}\psi^{\alpha}= \frac{i}{2}(\sigma_i)^{\alpha}_{~\beta}\psi^{\beta}\,,
\end{equation}
as required.

\beq
\tilde{\mu}^{\hat{1}}+i\tilde{\mu}^{\hat{2}} =\cos\left(\frac{\theta_2}{2}\right) e^{\frac{i}{2}(\phi_2+\psi_2)},~~~\tilde{\mu}^{\hat{3}}+i\tilde{\mu}^{\hat{4}} =\sin\left(\frac{\theta_2}{2}\right) e^{\frac{i}{2}(\phi_2-\psi_2)}
\eeq
and reabsorb these angles into the right invariant 1-forms everywhere they appear
\beq
\rho_1+i \rho_2= ie^{-i \phi_2}(d\theta_2+ i \sin\theta_2 d\psi_2),~~~\rho_3= d\phi_2+ \cos\theta_2 d\psi_2.
\eeq

\subsubsection{$\mathrm{SU}(2)$ doublets on $S^3$}
There are two independent sets of three Killing vectors on $S^3$ that realise each of the $\mathrm{SU}(2)$ factors of $\mathrm{SO}(4)= \mathrm{SU}(2)_{\mathrm{L}}\otimes \mathrm{SU}(2)_{\mathrm{R}}$ . The one forms dual to these $R/L$ vectors are the L/R invariant 1-forms of $\mathrm{SU}(2)$, that are defined in terms of $g\in \mathrm{SU}(2)$ as
\begin{equation} \label{muS4coords}
\begin{split}
	\lambda_i&= -i\text{Tr}\left(\sigma_i g^{-1} dg \right)\,,\qquad d\lambda_i-\frac{1}{2}\epsilon_{ijk}\lambda_j\wedge \lambda_k=0,\\[2mm]
	\rho_i&= -i \text{Tr}\left(\sigma_i dg g^{-1} \right)\,,\qquad d\rho_i+\frac{1}{2}\epsilon_{ijk}\rho_j\wedge \rho_k=0\,.	
\end{split}	
\end{equation}
We shall specifically take our group element to be
\begin{equation}
g= e^{\frac{i}{2}\sigma_3 \phi_2}e^{\frac{i}{2}\sigma_2 \theta_2}e^{\frac{i}{2}\sigma_3 \psi_2}\,,
\end{equation}
for which a consistent embedding of $S^3$ into $\mathbb{R}^4$ is given by
\begin{equation}
\tilde{\mu}^{\hat{1}}+i\tilde{\mu}^{\hat{2}} =\cos\left(\frac{\theta_2}{2}\right) e^{\frac{i}{2}(\phi_2+\psi_2)},~~~\tilde{\mu}^{\hat{3}}+i\tilde{\mu}^{\hat{4}} =\sin\left(\frac{\theta_2}{2}\right) e^{\frac{i}{2}(\phi_2-\psi_2)}.
\end{equation}
This leads to the following definition of the Killing vectors
\begin{equation} \label{LIandRIMCforms}
\begin{split}
	R_1&=-\sin\psi_2 \partial_{\theta_2}+\csc\theta_2 \cos\psi_2-\cot\theta_2\cos\psi_2\partial_{\psi_2},\\[2mm]
	R_2&=\cos\psi_2 \partial_{\theta_2}+\csc\theta_2 \sin\psi_2-\cot\theta_2\sin\psi_2\partial_{\psi_2},\\[2mm]
	R_3&=\partial_{\psi_2},\\[2mm]
	L_1&=\sin\phi_2 \partial_{\theta_2}+\cot\theta_2\cos\phi_2\partial_{\phi_2}- \csc\theta_2 \cos\phi_2 \partial_{\psi_2},\\[2mm]
	L_2&=\sin\phi_2 \partial_{\theta_2}-\cot\theta_2\sin\phi_2\partial_{\phi_2}+ \csc\theta_2 \sin\phi_2 \partial_{\psi_2},\\[2mm]
	L_3&=\partial_{\phi_2},
\end{split}	
\end{equation}
with dual one forms
\begin{equation}
\begin{split}
	\rho_1&=-\cos\phi_2 \sin\theta_2d\psi_2+\sin\phi_2 d\theta_2,\\[2mm]
	\rho_2&=\cos\phi_2d\theta_2+\sin\phi_2 \sin\theta_2d\psi_2 ,\\[2mm]
	\rho_3&=d\phi_2+\cos\theta_2d\psi_2,\\[2mm]
	\lambda_1&=\cos\psi_2 \sin\theta_2d\phi_2-\sin\psi_2 d\theta_2,\\[2mm]
	\lambda_2&=\cos\psi_2d\theta_2+\sin\psi_2 \sin\theta_2d\phi_2,\\[2mm]
	\lambda_3&=d\psi_2+\cos\theta_2d\phi_2.
\end{split}	
\end{equation}
Using these it is not hard to show that the Killing vectors obey the following relations when acting of the forms
\begin{equation}\label{eq:S3LieandContr}
\begin{split}
	\mathcal{L}_{R_i} \rho_j&= 0,~~~\mathcal{L}_{R_i} \lambda_j= \epsilon_{ijk} \lambda_k\,,\qquad\iota_{R_i}\lambda_j= \delta_{ij},\\[2mm]
	\mathcal{L}_{L_i} \lambda_j&= 0,~~~\mathcal{L}_{L_i} \rho_j= \epsilon_{ijk} \rho_k\,,\qquad\iota_{L_i}\rho_j= \delta_{ij},	
\end{split}	
\end{equation}
so that $(\rho/\lambda)^i$ are triplets under $\mathrm{SO}(3)_{\mathrm{L}/\mathrm{R}}$ and singlets under $\mathrm{SO}(3)_{\mathrm{R}/\mathrm{L}}$.
  
Working in the canonical frame of an $S^3$ spanned by $\rho_i$ (rather than the Hopf frame that \cite{Macpherson:2017mvu} uses) and with $\gamma_{\underline{\hat {a}}}= \sigma_{\hat {a}}$, $\hat {a}=1,2,3$, the spinors on $S^3$ that are charged under $\mathrm{SU}(2)_{\mathrm{L}}$ are solutions to
\begin{equation}
\nabla_{\underline{\hat{a}}}\xi= -\frac{i}{2}\sigma_{\hat{a}}\xi 
\end{equation}
which one can show is solved by any constant spinor; we choose
\begin{equation}
\xi= \left(\begin{array}{c} 0\\ -i  \end{array}\right ).
\end{equation}
From this we can construct a doublet of $\mathrm{SU}(2)_{\mathrm{L}}$
\begin{align}
\xi^{\alpha}=\left(\begin{array}{c} \xi\\ \sigma_1 \xi^* \end{array}\right )^{\alpha},
\end{align}
which transforms as
\begin{equation}
\mathcal{L}_{L_i}\xi^{\alpha}= \frac{i}{2}(\sigma_i)^{\alpha}_{~\beta}\xi^{\beta}\,,\qquad\mathcal{L}_{R_i}\xi^{\alpha}=0
\end{equation}
under the action of the spinorial Lie derivative.

\subsection{Invariant forms and triplets}\label{sec:triplettandforms}
In this subsection  we will focus on the five-dimensional manifold spanned by the unwarped fibration of $S^2$ over $S^3$; we give some additional details about the $\mathrm{SO}(3)_{\mathrm{d}}$ invariant forms, triplets and how the former acts on the latter. Our flat space 5d gamma matrices are
\beq
\gamma^{(5)}_{\underline{a}}= \sigma_a\otimes \mathbb{I},~~~\gamma^{(5)}_{\underline{\hat{a}}}= \sigma_3\otimes \sigma_{\hat{a}},~~B^{(5)}= \sigma_1\otimes \sigma_2.
\eeq
Five-dimensional Majorana conjugation is defined as
\beq
\eta^c= B_5 \eta^*\,.
\eeq
If one views the invariant forms as spinor bilinears, one can factor out all their dependence on the $S^2$ angles using the matrix
\begin{equation}\label{eq:rotation}
\Lambda = \mathbb{I}\otimes \mathbb{I}\otimes e^{\frac{i}{2}\sigma_2 \theta_1} e^{-\frac{1}{2}\sigma_2\sigma_1 \phi_1}\,.
\end{equation}
Using this, it is not hard to show that \eqref{eq:invarientforms1} becomes
\begin{equation}
\begin{split}
	\label{eq:omegaslash}
	\omega^{(5)}_1 &= \Lambda^{-1} \gamma^{(5)}_{\hat{3}}\Lambda \, ,\qquad \omega^{1,(5)}_2= - \Lambda^{-1} \gamma^{(5)}_{12}\Lambda \, ,\qquad\omega^{4,(5)}_2=  \Lambda^{-1} \gamma^{(5)}_{\hat{1}\hat{2}}\Lambda,\\[2mm]
	\omega^{2,(5)}_2&= - \Lambda^{-1} (\gamma^{(5)}_{1\hat{1}}+\gamma^{(5)}_{2\hat{2}})\Lambda \, ,\qquad \omega^{3,(5)}_2=  \Lambda^{-1} (\gamma^{(5)}_{1\hat{2}}+\gamma^{(5)}_{\hat{1}2})\Lambda,
\end{split}	
\end{equation}
under the 5 dimensional Clifford map on the unwarped $S^2\times S^3$.
Likewise the 4 dimensional components of the $\mathrm{SO}(3)_{\mathrm{d}}$ triplets in \eqref{eq:triplets12}, \eqref{eq:triplets34}   undergo a simplification when acted on by $\Lambda$. One can show that they can be expressed  as
\begin{equation}
\begin{split}
	\label{eq:tripletrot}
	\Lambda\eta^1_i&= s_i \eta^1_0- i t_i\eta^{1c}_0+ \tmu_i \eta^2_0\,,\qquad\qquad\Lambda\eta^3_i = - \tmu_i \eta^{2c}_0,\\[2mm]
	\Lambda\eta^2_i&= -(s_i \eta^{1c}_0- i t_i\eta^{1}_0+ \tmu_i \eta^{2c}_0)\,,\qquad\Lambda\eta^4_i = \tmu_i \eta^{2}_0	
\end{split}	
\end{equation}
in terms of two orthogonal constant spinors
\beq\label{eq:constatnspinors}
\eta^1_0=\left(\begin{array}{c}-i\\0\\0\\1\end{array}\right)\,,\qquad\eta^2_0=\left(\begin{array}{c}0\\i\\-1\\0\end{array}\right),
\eeq
and their Majorana conjugates. We have introduced
\beq\label{eq:st}
s_i=(\cos\theta_1\cos\phi_1~,\cos\theta_1\sin\phi_1~,-\sin\theta_1)_i\,,\qquad t_i=(\sin\phi_1~,-\cos\phi_1~,0)_i.
\eeq

Together, \eqref{eq:omegaslash} and \eqref{eq:tripletrot} can be used to greatly simplify the 8 dimensional spinorial components one needs to solve when plugging \eqref{eq:6d triplets} into \eqref{eq:6dSUSY1}--\eqref{eq:6dSUSY3}. This was extremely useful for deriving the system of sufficient supersymmetry conditions in \eqref{eq:dilatoncondtions}, \eqref{eq: AdS4condtions}, \eqref{eq:S2S3condtions}.  Using these conditions, it is also not hard to establish that the triplets satisfy the following relations under Majorana conjugation:
\begin{equation}
\eta^{ic}_1=-\eta^i_2\,,\qquad \eta^{ic}_3= \eta_4
\end{equation}
and under inner product
\begin{align}\label{eq:etaInP}
\eta^{i\dag}_1\eta^j_1&=\eta^{i\dag}_2\eta^j_2= 2\delta^{ij},~~~\eta^{i\dag}_1\eta^j_2=-2i \epsilon_{ijk}\tmu_k,\nonumber\\[2mm]
\eta^{i\dag}_3\eta^j_3&=\eta^{i\dag}_4\eta^j_4=\eta^{i\dag}_1\eta^j_4=\eta^{i\dag}_2\eta^j_3=2\tmu_i\tmu_j,\\[2mm]
\eta^{i\dag}_3\eta^j_4&=\eta^{i\dag}_1\eta^j_3=\eta^{i\dag}_2\eta^j_4=0. \nonumber
\end{align}
(\ref{eq:etaInP}) is useful to derive \eqref{eq:norms}.

Finally, let us work out how the invariant forms act on the 4 triplets. It is quite easy to show that the various gamma matrix combinations appearing in \eqref{eq:omegaslash} act on the constant spinors of \eqref{eq:constatnspinors} in the following fashion:
\begin{equation}\label{tab:12slashformaction}
\begin{array}{c||c|c|c|c|c|c|c}
~       &\gamma^{(5)}_{\hat{3}}          &\gamma^{(5)}_{12}   &\gamma^{(5)}_{\hat{1}\hat{2}}    &\gamma^{(5)}_{1\hat{1}}    & \gamma^{(5)}_{2\hat{2}} & \gamma^{(5)}_{1\hat{2}} & \gamma^{(5)}_{\hat{1}2}\\
\hline
\hline
\eta^1_0&\eta^1_0& -i \eta^{1c}_0  &-i \eta^{1c}_0  &  i \eta^1_0 &  -i \eta^1_0 &- \eta^{1c}_0 &  \eta^{1c}_0\\
\hline
\eta^2_0&-\eta^2_0& -i \eta^{2c}_0 &i \eta^{2c}_0  & - i\eta^2_0& - i \eta^2_0 &- \eta^{2c}_0 &- \eta^{2c}_0\,.\\
\end{array}
\end{equation}
Using this table, and the rotated form of the triplets in \eqref{eq:tripletrot}, it is relatively simple to establish the action of the invariant forms.
For the one-form and two-forms one can compute
\begin{equation}\label{tab:12formaction}
\begin{array}{c||c|c|c|c|c}
~       &\omega^{(5)}_1          &\omega^{1,(5)}_2   &\omega^{2,(5)}_2    &\omega^{3,(5)}_2    & \omega^{4,(5)}_2\\
\hline
\hline
\eta^i_1&\eta^i_1-2\eta^i_4& i \eta^i_2  &2 i \eta^i_4   &  2\eta^i_3  &  i\eta^i_2-2i\eta^i_3\\
\hline
\eta^i_2&\eta^i_2-2\eta^i_3& i \eta^i_1	 &-2i \eta^i_3    & - 2\eta^i_4 & i\eta^i_1-2i\eta^i_4\\
\hline
\eta^i_3&-\eta^i_3         & i \eta^i_4  & -2i\eta^i_3    & - 2\eta^i_4 &- i \eta^i_4\\
\hline
\eta^i_4&-\eta^i_4				 & i \eta^i_3  & 2i \eta^i_4    & 2\eta^i_3   &- i \eta^i_3\,.\\
\end{array}
\end{equation}
For the three- and four-forms one finds
\begin{equation}
\begin{array}{c||c|c|c|c|c}\label{tab:34formaction}
~       &\omega^{1,(5)}_3          &\omega^{2,(5)}_3   &\omega^{3,(5)}_3    &\omega^{4,(5)}_3    & \omega^{(5)}_4\\
\hline
\hline
\eta^i_1 &i \eta^i_2-2i \eta^i_3& -2 i \eta^i_4&-2\eta^i_3& i\eta^i_2&-\eta^i_1+2 \eta^i_4\\
\hline
\eta^i_2&i \eta^i_1-2i \eta^i_4&2 i \eta^i_3&2\eta^i_4&i\eta^i_1&-\eta^i_2+2 \eta^i_3\\
\hline
\eta^i_3&-i\eta^i_4&2 i \eta^i_3&2\eta^i_4&i\eta^i_4&\eta^i_3\\
\hline
\eta^i_4&-i\eta^i_3 &-2 i \eta^i_4&-2\eta^i_3&i\eta^i_3&\eta^i_4\,.\\
\end{array}
\end{equation}
The 5-form just flips the sign of every triplet, but does not appear in the lower form RR sector in IIA and so does not concern us.

The actions \eqref{tab:12formaction}--\eqref{tab:34formaction} are appropriate for the forms on the five-dimensional $S^2$-fibration over $S^3$. A little extra care needs to be taken when applying them to the main text and \eqref{eq:spin connection}. This is because the six-dimensional gamma matrices \eqref{eq:6dgammas} contain a factor that acts on the interval; moreover, the $S^2$ and $S^3$ are multiplied in the main text by functions $e^{C_1}$, $e^{C_2}$. In terms of bilinears, the relation between the full forms in six dimensions and their five-dimensional counterpart is 
\begin{equation}
\begin{split}
	e^{C_2}\omega_{1} &=\sigma_2\otimes \omega^{(5)}_{1} \, ,\qquad e^{2C_1}\omega^{1}_{2} =\mathbb{I}\otimes \omega^{1,(5)}_{2} \, ,\qquad e^{C_1+C_2}\omega^{2}_{2} =\mathbb{I}\otimes \omega^{2,(5)}_{2},\\[2mm]
	&e^{C_1+C_2}\omega^{3}_{2}= \mathbb{I}\otimes \omega^{3,(5)}_{2} \, ,\qquad e^{2C_2}\omega^{4}_{2} =\mathbb{I}\otimes \omega^{4,(5)}_{2}\,.	
\end{split}	
\end{equation}

\bibliography{references}
\end{document}